\documentclass[reqno]{amsart}
\usepackage{hyperref}
\usepackage{amsmath,amssymb} 

\usepackage{euscript}
\usepackage[T2A]{fontenc}
\usepackage[utf8]{inputenc}
\usepackage{textcomp}

\allowdisplaybreaks

\theoremstyle{plain}
\newtheorem{theorem}{Theorem}[section]

\theoremstyle{definition}

\theoremstyle{remark}
\newtheorem{remark}[theorem]{Remark}
\numberwithin{equation}{section}

\begin{document}
	
	\title[ Disorder solutions  of the two-dimensional Ising-like models ]
	{Disorder solutions for the partition functions of the two-dimensional Ising-like models	}
	
	\author[P. Khrapov]{Pavel Khrapov}
	\address{Pavel Khrapov, Department of Mathematics, Bauman Moscow State Technical University (5/1 2-nd
Baumanskaya St., Moscow 105005, Russia)
}  
	\email{khrapov@bmstu.ru }	



	\begin{abstract}
	For the generalized Ising models with all possible interactions within a face of the square lattice	the formulas for finding partition function and free energy per lattice site in the thermodynamic limit  were derived on a certain, in the general case, 8-dimensional  subset of exact disordered solutions of 10-dimensional set of the Hamiltonian's parameters.  When a part of parameters are set to zero, as a consequence, the disorder solutions were got for the models with nearest, next-nearest-neighbor  and the interaction of four spins  in an external field and without an external field,  triangular  and "checkerboard-triangular" Ising models with triple interactions in an external magnetic field.			
\end{abstract}
	\maketitle

		\section{Introduction}
		\label{introduction}
		
		At the present time there are a lot of different anisotropic models like Ising or Potts models with different coupling constants in the different directions, for which the wonderful subsets   called "disorder solutions" are found in the space of parameters, where the partition function can be calculated and represented in the simple form .  Remarkable examples were provided in the case of anisotropic models. Stephenson J. \cite{Stephenson} explicitly researched pair correlations between spins at the sites of the anisotropic triangular lattice along the axes, Enting I.G. \cite{Enting}  showed that the ratios of certain triplet order parameters to magnetisation in
		honeycomb and diamond lattice Ising models can be easily calculated  ,
		Baxter R.J. \cite{Baxter_1984} analyzed the disorder varieties of 
		the Ising model with all possible interactions around a face of
		the square lattice,    "interactions-round-a-face" (IRF) model
		on the square lattice. For the disorder solutions expressions obtained for the free energy and intra-row correlations. These are applied to the checkerboard Potts model. Ruj$\acute a$n P. \cite{RujAn_1}, \cite{RujAn_2}, \cite{RujAn_3} researched the 
		IRF model on the square lattice. He examined the general eight-vertex model thoroughly. Wu F.Y. \cite{Wu1985} found  the disorder solutions for the "checkerboard-triangular" lattice. 
		M. T. Jaekel and J. M. Maillard \cite{Jaekel_Maillard_1} found a local criterion which characterizes  disorder varieties for any dimensionality and explains
		the effective dimensional reduction occurring in the
		model. 	Dhar D., Maillard J.M. \cite{Dhar_Maillard} ,   
		Georges A., Hansel D., Doussal P. L., Maillard J. M. \cite{Georges_Hansel_Doussal_Maillard_1987} used this local criterion to calculate
		correlation functions on the disorder varieties of Ising and Potts models.
		Meyer H., Angl\`{e}s  d’Auriac J.-C.,  Maillard J.-M.  \cite{Meyer_et_al_1997} studied the disorder varieties
		of the eight vertex model in the framework of a random
		matrix theory approach to the transfer matrix.
		Various methods were used to obtain these solutions:
		methods related to crystal growth (Enting I.G.  \cite{Enting},  Welberry T.R., Galbraith R. \cite{Welberry_Galbraith}, Welberry T.R., Miller G.H. \cite{Welberry_Miller}), to Markov
		processes (Verhagen A.M.W. \cite{Verhagen}) and 
		transfer matrix technique ( Ruj$\acute a$n P. \cite{RujAn_2}, \cite{RujAn_3}, Baxter, R.J. \cite{Baxter_1984}). In the most cases, the ploblem of calculating the partition function was compared with an equivalent one in another area, where appropriate methods were available to find a solution.
		Reviews on this theme can be found in Wu F.Y.  \cite{Wu2009exactly} , Baxter R.J.  \cite{Baxter2016}, Pelizzola A. \cite{Pelizzola_2000}, \cite{Pelizzola_2005}.
		This article is a logical continuation of the author’s works  \cite{Khrapov3}, \cite{Khrapov4},  \cite{Khrapov5},  \cite{Khrapov6} which outlines a general methodology for finding such disordered solutions for generalized Ising and Potts models, and explicitly some of these solutions for a square and three-dimensional generalized Ising model are obtained. 
		In this work the IRF model on the square lattice is given in "interaction representation" rather than in the "weight (Boltzmann) representation"
		in the articles \cite{RujAn_2}, \cite{RujAn_3},  \cite{Baxter_1984}.  The disorder solutions were found on the 8-dimensional subset of 10-dimensional space of all independent parameters.  This is the generalization of results from \cite{Khrapov6}, where 	     the solutions were found on the 7-dimensional subset of the 10-dimensional space of all independent parameters. When a part of parameters are set to zero, as a consequence, the disorder solutions were also found for models   with nearest, next-nearest-neighbor and quadruple interactions in an external field and without an external field, for models on the triangular lattice with all possible interactions and for "checkerboard-triangular" lattice. The last solutions is compared with solutions from   Wu F.Y.  \cite{Wu1985} for "checkerboard-triangular" Ising model, their coincidence is obtained on numerical examples . 				
		At zero magnetic field, the Hamiltonian with nearest, next-nearest-neighbor and quadruple interactions becomes invariant under the change of signs of all spins, the elementary transfer matrix acquires central symmetry. These facts made possible to reduce the number of equations from 8 to 4 and to find the exact value of free energy in the thermodynamic limit on the 4-dimensional subset of the 5-dimensional space of parameters of model in the classical interpretation. This is the generalization of results from \cite{Khrapov3}, where the dimensional of the subset of the disorder solutions is equal to 3. 
		
		This work has the following stucture.
		In the section (\ref{Model_description}) the lattice model with helical boundary conditions were described.
		Toroidal boundary conditions are with a shift by one (similar to helical ones), and a cyclic closure of the set of all points (in natural ordering) \cite{Khrapov3}.
		For these models the Hamiltonian and the partition function were written. Sparse elementary transfer matrices  \cite{Khrapov3} with non-negative elements were constructed to calculate them.
		Section (\ref{largest_eigenvalue}) is devoted to finding disorder solutions for the partition function and free energy of model with the Hamiltonian (\ref{Hamiltonian1}).
		In the section  (\ref{largest_eigenvalue}) for a special kind of eigenvector (\ref{eigenvector1})  the system of equations (\ref{main_system1}) is written for the parameters of the Hamiltonian (\ref{Hamiltonian1}) , and parametric (depended on 11 parameters, not all of them are independent) full resuls of its solution in the form of sequentially calculated remaining parameters. The system of equations (\ref{main_system1}) and the value of free energy per lattice site in the thermodynamic limit remain the same for two-dimensional models with Hamiltonians in which the values of two (out of four) neighboring maximal spins in the natural ordering are replaced by the values of the spins at any other two lattice points neighboring in the natural ordering, this is significantly expands the set of models having disordered exact solutions. 	    
		The parameters are chosen in such way that the denominators of the considered formulas were deliberately positive, and so that there are as few branching decisions as possible. Firstly the quadratic equation is solved  (\ref{quadratic_equation}), then in the subsections (\ref{CaseAneq0Bneq0Cneq0}) , (\ref{CaseA0Bneq0Cneq0}),  (\ref{CaseA0B0C0}) various ratios for the coefficients of the quadratic equation are considered in sequence. In all cases numerical solutions are given to show that the set of solutions is nonempty. All numerical solutions are verified by directly finding the highest eigenvalue and eigenvector of the transfer matrix numerically by the power method. The system solution output  (\ref{main_system1}) and its equivalent system (\ref{main_system2}) can be found in the section  (\ref{solution_main_system}). For a numerical example section (\ref{CaseAneq0Bneq0Cneq0}) the rank of the Jacobi matrix of the mapping from the set of parameters of the solution to the set of parameters of the classical Hamiltonian (\ref{Hamiltonian_classic}), (\ref{J_classic}) calculated. It is equal to 8 with tolerance $0.0001 $, and it shows the dimension of the set of exact solutions in this case. 
		Considering that the spectrum of the transposed matrix coincides with the spectrum of the original matrix, in the section (\ref{largest_eigenvalue_transpon}) a similar eigenvector  (\ref{transpon_eigenvector1}) and similar system of equations (\ref{transpon_main_system1})  are written out for the transposed elementary transfer matrix. By some transformations, the system of equations (\ref{largest_eigenvalue_transpon}) is reduced to the system of equations (\ref{transpon_main_system4new}), between the solutions of which and the solutions of the system (\ref{main_system4})  and equation (\ref{main_system2} $\; $a) a bijective correspondence is found for the original transfer matrix. Hence  we will not get new solutions for the transposed matrix.
		In the section   (\ref{without_triple}) by vanishing the coefficients, responsible for triple interactions, formulas were written for finding the parameters of the exact disorder solutions for Hamiltonian with nearest, next-nearest-neighbor and quadruple interactions in an external field. The solution splits into two branches, for each of which an example of a numerical solution is given. 
		In the section (\ref{even_model}) formulas for finding the parameters are derived for models with Hamiltonians, invariant under inversion of all signs of the spins in the considered volume, i.e. for the models with nearest, next-nearest-neighbor and quadruple interactions  without an external field. A numerical example shows that the dimension of the set of the disorder solutions is equal to 4 in the 5-dimensional space of the classical Hamiltonian's parameters. 
		In the section   (\ref{triangles}) , by vanishing some of the parameters  $K_i$ in two different ways, exact solutions are obtained for models with Hamiltonians on a triangular lattice. It is shown by a numerical example that in the general case the disorder solutions form a 4-dimensional subset in the 6-dimensional space of parameters of the Hamiltonian (\ref{Hamiltonian_classic}) considering the conditions  (\ref{triangles_zero} (a)). 
		In the work \cite{Verhagen}
		an anisotropic triangular Ising model, in which the first- and second-order
		parameters and the field parameters are functionally related, is solved exactly
		by representing the distribution of the atom patterns in terms of a suitably
		constructed Markov process. The considered Hamiltonian does not contain triple interactions, and the disorder solutions form the 2-dimensional subset in the 3-dimensional space of the Hamiltonian's parameters. 	   
		With additional vanishing  some parameters and reduction of the Hamiltonian to the "checkerboard-triangular" case, formulas for finding the partition function and free energy are derived for each particular case, when the solution splits into multiple branches.  An example of a numerical solution is given for each branch of the solution. In the cases of "checkerboard-triangular" lattice the exact appropriate solutions from \cite{Wu1985} were found for all numerical solutions. Along the way, this shows that in the thermodynamic limit the model from \cite{Wu1985} coincides with the models in this paper.

		\section{Model description} 
		\label{Model_description}
		Let us consider two-dimensional lattice (more detailed description of this lattice can be found in \cite{Khrapov3}). Let us assume
		
		\begin{equation}\label{a}
			\mathcal {L}_2=\{t=(t_1,t_2),t_i=0,1,...,L_i, i=1,2\}.
		\end{equation}
		Then let us make the following identification
		
		\begin{equation}\label{b}
			(L_1,t_2)\equiv (0,t_2+1), (L_1,L_2-1)\equiv (0,L_2) \equiv (0,0).
		\end{equation}
		
		Due to the process of  identifying points, the total number of lattice sites $\mathcal {L}_2 $    is equal to 
		$L=L_1  L_2$   . Thus, special boundary toroidal cyclic helical (with a shift) conditions are set on $\mathcal {L}_2$. We renumber all points $ \mathcal {L}_2 : $
		\begin{equation}\label{c} 
			\tau^0=(0,0) ,
			\tau^1=(1,0) , \tau^2=(2,0) ,...,\tau^{L_1}= (L_1,0)\equiv (0,1)
			,\tau^{L}=(0,0)\equiv \tau^0 .
		\end{equation}

		This numeration determines the nature cyclic detour at every point (in the positive direction) and local (cyclic) ordering. 
		Let us consider that there is a particle in each site $t=(t_1,t_2)$. The state of particle is defined by spin $\sigma_t  $  , which at every site of lattice $t=(t_1,t_2)$ can take two values: $\sigma_t \in X=\{+1,-1\} $  . 
		The Hamiltonian of generalized two-dimensional Ising model has the form 
		
		\begin{equation}\label{Hamiltonian1}
			\begin{gathered}
				\mathcal {H} (\sigma )= -\sum_{i=0}^{L-1} (
				J_{01}\sigma_{	\tau^i}\sigma_{	\tau^{i+1}}+J_{02}\sigma_{	\tau^i}\sigma_{	\tau^{i+L_1}} +J_{23}\sigma_{	\tau^{i+L_1}}\sigma_{\tau^{i+L_1+1}} +\\
				J_{13}\sigma_{	\tau^{i+1}}\sigma_{\tau^{i+L_1+1}} +J_{03}\sigma_{	\tau^{i}}\sigma_{\tau^{i+L_1+1}}+
				J_{12}\sigma_{\tau^{i+1}}\sigma_{\tau^{i+L_1}}+J_{012}\sigma_{	\tau^{i}}\sigma_{\tau^{i+1}}\sigma_{\tau^{i+L_1}}+\\
				J_{023}\sigma_{	\tau^{i}}\sigma_{\tau^{i+L_1}}\sigma_{\tau^{i+L_1+1}}+
				J_{123}\sigma_{	\tau^{i+1}}\sigma_{\tau^{i+L_1}}\sigma_{\tau^{i+L_1+1}}+
				J_{013}\sigma_{	\tau^{i}}\sigma_{\tau^{i+1}}\sigma_{\tau^{i+L_1+1}}+\\
				J_{0123}\sigma_{\tau^{i}}\sigma_{\tau^{i+1}}\sigma_{\tau^{i+L_1}}\sigma_{	\tau^{i+L_1+1}}+
				J_0 \sigma_{	\tau^{i}} +
				J_1 \sigma_{	\tau^{i+1}} +
				J_2 \sigma_{\tau^{i+L_1}} +
				J_3 \sigma_{\tau^{i+L_1+1}}).
			\end{gathered} 
		\end{equation}
		
		where $ J_i $, $ i \in I=\{0, 1, 2, 3, 01,02,23,13,03,12, 012, 023,123, 013,0123\} $ are corresponding coefficients of multi-spin interaction. The lattice model with the Hamiltonian (\ref{Hamiltonian1}) is equal to the model with the Hamiltonian  
		
		\begin{equation}\label{Hamiltonian_classic}
			\begin{gathered}
				\mathcal {H'} (\sigma )= -\sum_{i=0}^{L-1} (
				J'_{01}\sigma_{	\tau^i}\sigma_{	\tau^{i+1}}+J'_{02}\sigma_{	\tau^i}\sigma_{	\tau^{i+L_1}} +J_{03}\sigma_{	\tau^{i}}\sigma_{\tau^{i+L_1+1}}+\\
				J_{12}\sigma_{\tau^{i+1}}\sigma_{\tau^{i+L_1}}+J_{012}\sigma_{	\tau^{i}}\sigma_{\tau^{i+1}}\sigma_{\tau^{i+L_1}}+
				J_{023}\sigma_{	\tau^{i}}\sigma_{\tau^{i+L_1}}\sigma_{\tau^{i+L_1+1}}+\\
				J_{123}\sigma_{	\tau^{i+1}}\sigma_{\tau^{i+L_1}}\sigma_{\tau^{i+L_1+1}}+
				J_{013}\sigma_{	\tau^{i}}\sigma_{\tau^{i+1}}\sigma_{\tau^{i+L_1+1}}+\\
				J_{0123}\sigma_{\tau^{i}}\sigma_{\tau^{i+1}}\sigma_{\tau^{i+L_1}}\sigma_{	\tau^{i+L_1+1}}+
				H' \sigma_{	\tau^{i}},
			\end{gathered} 
		\end{equation}
		where
		
		\begin{equation} \label{J_classic}
			\begin{gathered}
				J'_{01}=J_{01}+J_{23} , \\
				J'_{02}=J_{02}+J_{13}, \\
				H'=J_0+J_1+J_2+J_3.   
			\end{gathered}
		\end{equation}

		Partition function of model with the Hamiltonian (\ref{Hamiltonian1})  can be written in the following form
		
		\begin{equation}\label{Partition_function1} 
			\begin{split}
				Z_{{L}}=Z_{{L_1}{L_2}}=\sum_{\sigma}\exp ( -\mathcal {H}(\sigma )/{(k_B T)}), 
			\end{split} 
		\end{equation}

		where summation perfomed over all spins.
		
		Let us assume
		
		\begin{equation}\label{K}
			\begin{split}
				K =\{ K_i=J_i/{(k_B T)} , i \in I \}	
			\end{split} 
		\end{equation}
		- the set of parameters of the considerable model. Here 
		$ T $ is temperature, $k_B $ is Boltzmann's constant. Then partition function (\ref{Partition_function1}) can be represented in the form 
		
		\begin{equation}\label{Partition_function2}
			\begin{gathered}
				Z_{{L}}=Z_{{L_1}{L_2}}=\sum_{\sigma}\exp ( \sum_{i=0}^{L-1} (
				K_{01}\sigma_{	\tau^i}\sigma_{	\tau^{i+1}}+K_{02}\sigma_{	\tau^i}\sigma_{	\tau^{i+L_1}} +K_{23}\sigma_{	\tau^{i+L_1}}\sigma_{\tau^{i+L_1+1}} +\\
				K_{13}\sigma_{	\tau^{i+1}}\sigma_{\tau^{i+L_1+1}} +K_{03}\sigma_{	\tau^{i}}\sigma_{\tau^{i+L_1+1}}+
				K_{12}\sigma_{\tau^{i+1}}\sigma_{\tau^{i+L_1}}+K_{012}\sigma_{	\tau^{i}}\sigma_{\tau^{i+1}}\sigma_{\tau^{i+L_1}}+\\
				K_{023}\sigma_{	\tau^{i}}\sigma_{\tau^{i+L_1}}\sigma_{\tau^{i+L_1+1}}+
				K_{123}\sigma_{	\tau^{i+1}}\sigma_{\tau^{i+L_1}}\sigma_{\tau^{i+L_1+1}}+
				K_{013}\sigma_{	\tau^{i}}\sigma_{\tau^{i+1}}\sigma_{\tau^{i+L_1+1}}+\\
				K_{0123}\sigma_{\tau^{i}}\sigma_{\tau^{i+1}}\sigma_{\tau^{i+L_1}}\sigma_{	\tau^{i+L_1+1}}+
				K_0 \sigma_{	\tau^{i}} +
				K_1 \sigma_{	\tau^{i+1}} +
				K_2 \sigma_{\tau^{i+L_1}} +
				K_3 \sigma_{\tau^{i+L_1+1}}). 
			\end{gathered} 
		\end{equation}

		For using model we write the elementary transfer matrix $ \Theta=\Theta_{p,q} $ of size $ 2^{L_1+1}  \times 2^{L_1+1}$ in the same way, as in \cite{Khrapov3}, \cite{Khrapov5}.  Nonzero elements of the transfer matrix    $ \Theta=\Theta_{p,q} $ are specified by all sorts of pairs of sets of spins
		$ \{(\sigma_{\tau^0},\sigma_{\tau^{1}},...,\sigma_{\tau^{L_1}}),$ 
		$(\sigma_{\tau^1},\sigma_{\tau^{2}},...,\sigma_{\tau^{L_1+1}})  \} $:

		\begin{multline} \label{Theta1}
			\Theta_{p,q}=\Theta_{ \{ 
				(\sigma_{\tau^0},\sigma_{\tau^{1}},...,\sigma_{\tau^{L_1}}),
				(\sigma_{\tau^1},\sigma_{\tau^{2}},...,\sigma_{\tau^{L_1+1}})  \}}=
			\exp (
			K_{01}\sigma_{	\tau^i}\sigma_{	\tau^{i+1}}+\\K_{02}\sigma_{	\tau^i}\sigma_{	\tau^{i+L_1}} +
			K_{23}\sigma_{	\tau^{i+L_1}}\sigma_{\tau^{i+L_1+1}} +
			K_{13}\sigma_{	\tau^{i+1}}\sigma_{\tau^{i+L_1+1}} +K_{03}\sigma_{	\tau^{i}}\sigma_{\tau^{i+L_1+1}}+\\
			K_{12}\sigma_{\tau^{i+1}}\sigma_{\tau^{i+L_1}}+
			K_{012}\sigma_{	\tau^{i}}\sigma_{\tau^{i+1}}\sigma_{\tau^{i+L_1}}+
			K_{023}\sigma_{	\tau^{i}}\sigma_{\tau^{i+L_1}}\sigma_{\tau^{i+L_1+1}}+\\
			K_{123}\sigma_{	\tau^{i+1}}\sigma_{\tau^{i+L_1}}\sigma_{\tau^{i+L_1+1}}+
			K_{013}\sigma_{	\tau^{i}}\sigma_{\tau^{i+1}}\sigma_{\tau^{i+L_1+1}}+\\
			K_{0123}\sigma_{\tau^{i}}\sigma_{\tau^{i+1}}\sigma_{\tau^{i+L_1}}\sigma_{	\tau^{i+L_1+1}}+
			K_0 \sigma_{	\tau^{i}} +
			K_1 \sigma_{	\tau^{i+1}} +\\
			K_2 \sigma_{\tau^{i+L_1}} +
			K_3 \sigma_{\tau^{i+L_1+1}}) ,   
		\end{multline}

		wherein
		\begin{equation}\label{p}
			\begin{split}
				p= \sum_{k=0}^{L_1}{((1-\sigma_{\tau^k})/2)2^k}, p=0,1,\ldots,2^{L_1+1}-1,	
			\end{split} 
		\end{equation}
		
		\begin{equation}\label{q}
			\begin{split}
				q= \sum_{k=0}^{L_1}{((1-\sigma_{\tau^{1+k}})/2)2^k}, q=0,1,\ldots,2^{L_1+1}-1.	
			\end{split} 
		\end{equation}

		Then 
		
		\begin{equation}\label{Partition_function3}
			\begin{gathered}
				Z_L=Z_{{L_1}{L_2}}=\sum_{ \{\sigma_{\tau^0},\sigma_{\tau^{1}},\ldots,\sigma_{\tau^{L-1}}  \}}
				\Theta_{\{ (\sigma_{\tau^0},\sigma_{\tau^{1}},...,\sigma_{\tau^{L_1}}),
					(\sigma_{\tau^1},\sigma_{\tau^{2}},...,\sigma_{\tau^{L_1+1}})  \} }\\
				\Theta_{\{ (\sigma_{\tau^1},\sigma_{\tau^{2}},...,\sigma_{\tau^{L_1+1}}),
					(\sigma_{\tau^2},\sigma_{\tau^{3}},...,\sigma_{\tau^{L_1+2}})  \} }\ldots \\
				\Theta_{\{ (\sigma_{\tau^{L-1}},\sigma_{\tau^{0}},...,\sigma_{\tau^{L_1-1}}),
					(\sigma_{\tau^0},\sigma_{\tau^{1}},...,\sigma_{\tau^{L_1}})  \} } 
				=
				Tr({	\Theta}^L). 
			\end{gathered} 	
		\end{equation}
		
		Now the free energy at one site of lattice $ f $ in the thermodynamic limit can be written in the following form  \cite{Baxter2016} :
		\begin{equation}\label{e}
			\begin{split}
				f(T,	K)=-kT \lim\limits_{L\to {\infty}} {\ln(\lambda_{\max}(L,T,K))} , 
			\end{split} 
		\end{equation}
		where $\lambda_{\max}$ is the largest transfer matrix  $ \Theta=\Theta_{p,q} $ eigenvalue.
		
		By the Perron–Frobenius theorem \cite{Perron} the only one largest eigenvalue $\lambda_{\max}$ of transfer matrix $ \Theta=\Theta_{p,q} $ will correspond to a matrix with positive elements (all matrix elements $ \Theta^{L_1+1} $ will strictly be greater than zero, the structure filled with nonzero elements becomes clear already for $ \Theta^{2} $ . Actually, at first the Perron–Frobenius theorem is used for matrix $ \Theta^{L_1+1} $).
		Let us assume
		
		\begin{equation}\label{G} 
			\begin{gathered} 
				G(\sigma_{\tau^0},\sigma_{\tau^1},\sigma_{\tau^{L_1}},\sigma_{\tau^{L_1+1}},K)=\\
				\exp (
				K_{01}\sigma_{	\tau^i}\sigma_{	\tau^{i+1}}+K_{02}\sigma_{	\tau^i}\sigma_{	\tau^{i+L_1}} +K_{23}\sigma_{	\tau^{i+L_1}}\sigma_{\tau^{i+L_1+1}} +\\
				K_{13}\sigma_{	\tau^{i+1}}\sigma_{\tau^{i+L_1+1}} +K_{03}\sigma_{	\tau^{i}}\sigma_{\tau^{i+L_1+1}}+
				K_{12}\sigma_{\tau^{i+1}}\sigma_{\tau^{i+L_1}}+K_{012}\sigma_{	\tau^{i}}\sigma_{\tau^{i+1}}\sigma_{\tau^{i+L_1}}+\\
				K_{023}\sigma_{	\tau^{i}}\sigma_{\tau^{i+L_1}}\sigma_{\tau^{i+L_1+1}}+
				K_{123}\sigma_{	\tau^{i+1}}\sigma_{\tau^{i+L_1}}\sigma_{\tau^{i+L_1+1}}+
				K_{013}\sigma_{	\tau^{i}}\sigma_{\tau^{i+1}}\sigma_{\tau^{i+L_1+1}}+\\
				K_{0123}\sigma_{\tau^{i}}\sigma_{\tau^{i+1}}\sigma_{\tau^{i+L_1}}\sigma_{	\tau^{i+L_1+1}}+
				K_0 \sigma_{	\tau^{i}} +
				K_1 \sigma_{	\tau^{i+1}} +
				K_2 \sigma_{\tau^{i+L_1}} +
				K_3 \sigma_{\tau^{i+L_1+1}}).  
			\end{gathered} 
		\end{equation}
		
		Let us define the values $ a_{ij} $  ,$ j=0,1,...,7 $  , $ i=0,1 $:

		\begin{equation}\label{aij}   
			\begin{split} 
				a_{i0}=G(+1,+1,+1,1-2i,K),\\
				a_{i1}=G(-1,+1,+1,1-2i,K),\\  
				a_{i2}=G(+1,-1,+1,1-2i,K),\\
				a_{i3}=G(-1,-1,+1,1-2i,K),\\  
				a_{i4}=G(+1,+1,-1,1-2i,K),\\
				a_{i5}=G(-1,+1,-1,1-2i,K),\\  
				a_{i6}=G(+1,-1,-1,1-2i,K),\\
				a_{i7}=G(-1,-1,-1,1-2i,K).\\ 
			\end{split} 
		\end{equation}
		Then nonzero elements of matrix $ \Theta $ can be written in the form:
		\begin{equation}\label{Theta2}  
			\begin{gathered} 
				\Theta_{2r+l+4k+s2^{L_1},r+2k+s2^{L_1-1}+i2^{L_1}}=a_{i,2r+l+4s}, \\
				r=0,1,  l=0,1,s=0,1,i=0,1, k=0,1,\ldots,2^{L_1-2}-1.
			\end{gathered} 
		\end{equation}

		%\section{Section title}
		%\label{sec:1}
		%\subsection{Subsection title}
		%\label{sec:2}
		
		%%%%%%%%%%%%%%%%%%%%%%%%%%%%%%%%%%%%%%%%%%%%%%%%%%%%%%%%%%%%%%%%%%%%%%%%%%%%%%%%%%%%%%%%%%%
		\section{Disorder solutions for the partition function and free energy of model with the Hamiltonian (\ref{Hamiltonian1})} \label{largest_eigenvalue}
		
		The eigenvector of the transfer matrix $ \Theta=\Theta_{p,q} $, corresponding to
		the largest eigenvalue $ F $, we represent in the following form: \\
		
		\begin{equation}\label{eigenvector1}   
			\overrightarrow{x}=
			(\underbrace{1,b_2,\dots,1,b_2 }_{2^{L_1}};\underbrace{b_3,b_4,\dots,b_3,b_4}_{2^{L_1}})^T
			.
		\end{equation}

		Then by the Perron–Frobenius theorem \cite{Perron} this eigenvector with all positive elements will correspond to the single maximal eigenvalue $F$ of the transter matrix $ \Theta=\Theta_{p,q} $. Let us denote 
		\begin{equation}\label{R_i1}  
			\begin{split} 
				R_i = \exp(K_i), \\
				u_i = \exp(2 K_i),\\
				s_i = \exp(4 K_i), i \in I.
			\end{split}. 
		\end{equation}

		From (\ref{R_i1}),  the form of the elementary transfer matrix  $ \Theta $ and the form of the eigenvector (\ref{eigenvector1}), we have at  
		
		\begin{equation}\label{R_i2} 
			F>0, \;\; b_2>0, \;\; b_3>0, \;\; b_4>0 , 
		\end{equation}
		
		the following system of equations 
		\begin{equation}\label{main_system1}  
			\begin{split} 
				\left\{  
				\begin{array}{rcl}  
					F &= & a_{00}+ b_3 a_{10},  \\  
					b_2 F &= & a_{01}+b_3a_{11}, \\ 
					F &= & b_2 a_{02}+b_4 a_{12}, \\ 
					b_2 F &= & b_2 a_{03}+ b_4 a_{13}, \\ 
					b_3 F &= &  a_{04}+ b_3 a_{14}, \\
					b_4 F &= &  a_{05}+ b_3 a_{15}, \\
					b_3 F &= &  b_2 a_{06}+ b_4 a_{16}, \\
					b_4 F &= &  b_2 a_{07}+ b_4 a_{17}, \\
				\end{array}   
				\right.
			\end{split} 
		\end{equation}
		where  $ a_{ij}$ is from (\ref{aij}).
		
		These eight equations (\ref{main_system1}) will repeat.
		Then , using the form of the Hamiltonian (\ref{Hamiltonian1}), after changing variable (\ref{R_i1}) , we have

		\begin{multline} \label{main_system2}
			F   =   ((b_3 R_0 R_1 R_2 R_{01} R_{012} R_{02} R_{12})/
			(R_3 R_{0123} R_{013} R_{023} R_{03} R_{123} R_{13} R_{23}) + \\
			R_0 R_1 R_2 R_3 R_{01} R_{012} R_{0123} R_{013} R_{02} R_{023} R_{03} R_{12} R_{123} R_{13} R_{23}),   \;\; \;\;  \;\; \;\;(a) \\
			b_2 F   =   ((b_3 R_1 R_2 R_{0123} R_{013} R_{023} R_{03} R_{12})/
			(R_0 R_3 R_{01} R_{012} R_{02} R_{123} R_{13} R_{23}) + \\
			(R_1 R_2 R_3 R_{12} R_{123} R_{13} R_{23})/
			(R_0 R_{01} R_{012} R_{0123} R_{013} R_{02} R_{023} R_{03})),   \;\;  \;\; \;\;(b)\\
			F   =   ((b_4 R_0 R_2 R_{0123} R_{013} R_{02} R_{123} R_{13})/
			(R_1 R_3 R_{01} R_{012} R_{023} R_{03} R_{12} R_{23}) +\\
			(b_2 R_0 R_2 R_3 R_{02} R_{023} R_{03} R_{23})/
			(R_1 R_{01} R_{012} R_{0123} R_{013} R_{12} R_{123} R_{13})),   \;\; \;\;  \;\; \;\;(c)   \\
			b_2 F   =   ((b_4 R_2 R_{01} R_{012} R_{023} R_{03} R_{123} R_{13})/
			(R_0 R_1 R_3 R_{0123} R_{013} R_{02} R_{12} R_{23}) +\\
			(b_2 R_2 R_3 R_{01} R_{012} R_{0123} R_{013} R_{23})/
			(R_0 R_1 R_{02} R_{023} R_{03} R_{12} R_{123} R_{13})) ,   \;\; \;\;  \;\; \;\;(d) \\
			b_3 F  =   ((R_0 R_1 R_3 R_{01} R_{013} R_{03} R_{13})/
			(R_2 R_{012} R_{0123} R_{02} R_{023} R_{12} R_{123} R_{23}) +\\
			(b_3 R_0 R_1 R_{01} R_{0123} R_{023} R_{123} R_{23})/
			(R_2 R_3 R_{012} R_{013} R_{02} R_{03} R_{12} R_{13})) ,   \;\; \;\;  \;\; \;\;(e)\\
			b_4 F   =   ((R_1 R_3 R_{012} R_{0123} R_{02} R_{023} R_{13})/
			(R_0 R_2 R_{01} R_{013} R_{03} R_{12} R_{123} R_{23}) + \\
			(b_3 R_1 R_{012} R_{013} R_{02} R_{03} R_{123} R_{23})/
			(R_0 R_2 R_3 R_{01} R_{0123} R_{023} R_{12} R_{13}))  ,   \;\; \;\;  \;\; \;\;(f)  \\
			b_3 F   =  ((b_2 R_0 R_3 R_{012} R_{0123} R_{03} R_{12} R_{123})/
			(R_1 R_2 R_{01} R_{013} R_{02} R_{023} R_{13} R_{23}) +\\
			(b_4 R_0 R_{012} R_{013} R_{023} R_{12} R_{13} R_{23})/
			(R_1 R_2 R_3 R_{01} R_{0123} R_{02} R_{03} R_{123}))   ,   \;\; \;\;  \;\; \;\;(g) \\
			b_4 F   =   ((b_2 R_3 R_{01} R_{013} R_{02} R_{023} R_{12} R_{123})/
			(R_0 R_1 R_2 R_{012} R_{0123} R_{03} R_{13} R_{23}) + \\
			(b_4 R_{01} R_{0123} R_{02} R_{03} R_{12} R_{13} R_{23})/
			(R_0 R_1 R_2 R_3 R_{012} R_{013} R_{023} R_{123})) .   \;\; \;\;  \;\; \;\;(h)       
		\end{multline}

		\begin{remark}\label{remark1} 
			The system of eight equations will remain the same (\ref{main_system1})-(\ref{main_system2}), if in the formula of the Hamiltonian (\ref{Hamiltonian1}) we change $ \sigma_{\tau^{i+L_1}} $ and $ \sigma_{\tau^{i+L_1+1}}  $ to $ \sigma_{\tau^{i+L_1+m}} $ and $ \sigma_{\tau^{i+L_1+m+1}}  $ respectively with random integer value $ m>0 $. The  transfer matrix size will only change. This result helps to extend the class of models which disorder solutions are found for. 		
		\end{remark}
		The solution of the system (\ref{main_system2}) is in the section \ref{solution_main_system}. Below we write the resulting formulas of solution of the system (\ref{main_system2}), depending on  parameters
		
		$b_3, K_{012}, K_{013}, K_{03}, K_{123}, K_{13}, K_{23}, K_3, K_{0123}, K_{023}, K_{13}.$
		
		Let us consider quadratic equation 
		\begin{equation}\label{quadratic_equation}
			A x^2 + B x+  C=0,
		\end{equation}
		where $ x=s_{02}, $
		
		\begin{equation}\label{coefficient_A} 
			\begin{array}{l}  
				A =-b_3^2 u_{012}^4 u_{013} u_{03} u_{123} u_{13}^2 u_{23} (u_3 u_{0123} u_{023} u_{13} + 
				b_3 u_{013} u_{03} u_{123} u_{23})^2 \\
				(b_3 + 
				u_3 u_{0123} u_{013} u_{023} u_{03} u_{123} u_{13} u_{23})^2 (b_3 u_3 u_{012}^4 u_{0123} u_{023}^3 \
				u_{13} - \\
				b_3 u_3 u_{0123}^3 u_{013}^2 u_{023} u_{03}^2 u_{13} + 
				b_3^2 u_{012}^4 u_{013} u_{023}^2 u_{03} u_{123} u_{23} - \\
				b_3^2 u_{0123}^4 u_{013} u_{023}^2 u_{03} u_{123} u_{23} - 
				u_3^2 u_{0123}^2 u_{013} u_{03} u_{123} u_{13}^2 u_{23} + \\
				u_3^2 u_{012}^4 u_{0123}^2 u_{013} u_{023}^4 u_{03} u_{123} u_{13}^2 u_{23} - 
				b_3 u_3 u_{0123}^3 u_{023} u_{123}^2 u_{13} u_{23}^2 + \\
				b_3 u_3 u_{012}^4 u_{0123} u_{013}^2 u_{023}^3 u_{03}^2 u_{123}^2 u_{13} u_{23}^2), 
			\end{array}  
		\end{equation}
		
		\begin{equation}\label{coefficient_B}  
			\begin{array}{l}  
				B = -b_3 u_3 u_{012}^2 u_{0123} u_{023} u_{13} (u_3 u_{013} u_{03} u_{13} + 
				b_3 u_{0123} u_{023} u_{123} u_{23}) \\(u_3 u_{0123} u_{023} u_{13} + 
				b_3 u_{013} u_{03} u_{123} u_{23}) (b_3 u_{0123} u_{013} u_{023} u_{03} + 
				u_3 u_{123} u_{13} u_{23}) \\
				(b_3 + 
				u_3 u_{0123} u_{013} u_{023} u_{03} u_{123} u_{13} u_{23}) (b_3 u_3 u_{012}^4 u_{0123} u_{023} \
				u_{03}^2 u_{123}^2 u_{13} - \\
				b_3 u_3 u_{0123} u_{013}^4 u_{023} u_{03}^2 u_{123}^2 u_{13} - 
				b_3^2 u_{0123}^2 u_{013}^3 u_{023}^2 u_{03} u_{123}^3 u_{23} + \\
				b_3^2 u_{012}^4 u_{013} u_{03}^3 u_{123}^3 u_{23} - 
				u_3^2 u_{013}^3 u_{03} u_{123}^3 u_{13}^2 u_{23} + \\
				u_3^2 u_{012}^4 u_{0123}^2 u_{013} u_{023}^2 u_{03}^3 u_{123}^3 u_{13}^2 u_{23} + 
				b_3 u_3 u_{012}^4 u_{0123} u_{013}^2 u_{023} u_{13} u_{23}^2 - \\
				b_3 u_3 u_{0123} u_{013}^2 u_{023} u_{03}^4 u_{13} u_{23}^2 - 
				b_3 u_3 u_{0123} u_{013}^2 u_{023} u_{123}^4 u_{13} u_{23}^2 + \\
				b_3 u_3 u_{012}^4 u_{0123} u_{013}^2 u_{023} u_{03}^4 u_{123}^4 u_{13} u_{23}^2 + 
				b_3^2 u_{012}^4 u_{013}^3 u_{03} u_{123} u_{23}^3 - \\
				b_3^2 u_{0123}^2 u_{013} u_{023}^2 u_{03}^3 u_{123} u_{23}^3 + 
				u_3^2 u_{012}^4 u_{0123}^2 u_{013}^3 u_{023}^2 u_{03} u_{123} u_{13}^2 u_{23}^3 - \\
				u_3^2 u_{013} u_{03}^3 u_{123} u_{13}^2 u_{23}^3 - 
				b_3 u_3 u_{0123} u_{023} u_{03}^2 u_{123}^2 u_{13} u_{23}^4 + \\
				b_3 u_3 u_{012}^4 u_{0123} u_{013}^4 u_{023} u_{03}^2 u_{123}^2 u_{13} u_{23}^4),
			\end{array}  
		\end{equation}

		\begin{equation}\label{coefficient_C}  
			\begin{array}{l}  
				C = -u_3^2 u_{013} u_{03} u_{123} u_{23} (u_3 u_{013} u_{03} u_{13} + 
				b_3 u_{0123} u_{023} u_{123} u_{23})^2 \\(b_3 u_{0123} u_{013} u_{023} u_{03} + 
				u_3 u_{123} u_{13} u_{23})^2 (b_3 u_3 u_{012}^4 u_{0123}^3 u_{023} u_{13} - \\
				b_3 u_3 u_{0123} u_{013}^2 u_{023}^3 u_{03}^2 u_{13} + 
				b_3^2 u_{012}^4 u_{0123}^2 u_{013} u_{03} u_{123} u_{23} - \\
				b_3^2 u_{0123}^2 u_{013} u_{023}^4 u_{03} u_{123} u_{23} - 
				u_3^2 u_{013} u_{023}^2 u_{03} u_{123} u_{13}^2 u_{23} + \\
				u_3^2 u_{012}^4 u_{0123}^4 u_{013} u_{023}^2 u_{03} u_{123} u_{13}^2 u_{23} - 
				b_3 u_3 u_{0123} u_{023}^3 u_{123}^2 u_{13} u_{23}^2 + \\
				b_3 u_3 u_{012}^4 u_{0123}^3 u_{013}^2 u_{023} u_{03}^2 u_{123}^2 u_{13} u_{23}^2).
			\end{array}  
		\end{equation}
		
		We consider different relations between $A, B, C$ below.
		\subsection{Case 1 }\label{CaseAneq0Bneq0Cneq0}
		Let us $ A \neq 0 , B \neq 0 , C \neq 0  $.
		
		Hence $ s_{02}=x_1 $ or $ s_{02}=x_2 $  ($ s_{02}>0 $), where $x_1, x_2 $ - the roots of the square equation (\ref{quadratic_equation}) ,
		$  u_{02} = s_{02}^{1/2} $,
		$ R_{02} = u_{02}^{1/2}$, 
		$K_{02} = \log(R_{02})  .$
		Then
		\begin{equation} \label{s12} 
			\begin{gathered}
				s_{12} = (u_{012}^2 (u_3 u_{0123} u_{023} u_{13} + 
				b_3 u_{013} u_{03} u_{123} u_{23}) \\
				(b_3^2 u_3 u_{012}^2 u_{0123} u_{02}^2 u_{023}^2 u_{03} \
				u_{123} u_{13}^2 + b_3 u_3^2 u_{0123}^2 u_{013}^3 u_{023} u_{03}^2 u_{13} u_{23} + \\
				b_3^3 u_{012}^2 u_{013} u_{02}^2 u_{023} u_{03}^2 u_{123}^2 u_{13} u_{23} + 
				b_3 u_3^2 u_{012}^2 u_{0123}^2 u_{013} u_{02}^2 u_{023}^3 u_{03}^2 u_{123}^2 u_{13}^3 u_{23} \
				+\\ b_3^2 u_3 u_{0123}^3 u_{013}^2 u_{023}^2 u_{03} u_{123} u_{23}^2 + 
				u_3^3 u_{0123} u_{013}^2 u_{03} u_{123} u_{13}^2 u_{23}^2 + \\
				b_3^2 u_3 u_{012}^2 u_{0123} u_{013}^2 u_{02}^2 u_{023}^2 u_{03}^3 u_{123}^3 u_{13}^2 \
				u_{23}^2 + b_3 u_3^2 u_{0123}^2 u_{013} u_{023} u_{123}^2 u_{13} u_{23}^3))/\\
				((b_3 u_{0123} u_{013} \
				u_{023} u_{03} + 
				u_3 u_{123} u_{13} u_{23}) \\(b_3 u_3^2 u_{0123} u_{013}^3 u_{023}^2 u_{03}^2 u_{123} u_{13} + 
				b_3^2 u_3 u_{0123}^2 u_{013}^2 u_{023}^3 u_{03} u_{123}^2 u_{23} + \\
				b_3^2 u_3 u_{012}^2 u_{0123}^2 u_{02}^2 u_{023} u_{03} u_{13}^2 u_{23} + 
				u_3^3 u_{013}^2 u_{023} u_{03} u_{123}^2 u_{13}^2 u_{23} + \\
				b_3^3 u_{012}^2 u_{0123} u_{013} u_{02}^2 u_{03}^2 u_{123} u_{13} u_{23}^2 + 
				b_3 u_3^2 u_{0123} u_{013} u_{023}^2 u_{123}^3 u_{13} u_{23}^2 + \\
				b_3 u_3^2 u_{012}^2 u_{0123}^3 u_{013} u_{02}^2 u_{023}^2 u_{03}^2 u_{123} u_{13}^3 u_{23}^2 \
				+ b_3^2 u_3 u_{012}^2 u_{0123}^2 u_{013}^2 u_{02}^2 u_{023} u_{03}^3 u_{123}^2 u_{13}^2 u_{23}^3)) ,  
			\end{gathered}
		\end{equation} 
		$u_{12} = s_{12}^{1/2}, $ $ R_{12} = s_{12}^{1/4},$
		
		\begin{equation} \label{s01} 
			\begin{gathered}
				s_{01} = ((u_3 u_{0123} u_{023} u_{13} + 
				b_3 u_{013} u_{03} u_{123} u_{23}) \\(b_3^2 u_3 u_{012}^2 u_{0123}^2 u_{013} u_{02}^2 u_{023} \
				u_{123} u_{13}^2 + b_3 u_3^2 u_{0123} u_{013}^2 u_{023}^2 u_{03}^3 u_{13} u_{23} +\\ 
				b_3^3 u_{012}^2 u_{0123} u_{013}^2 u_{02}^2 u_{03} u_{123}^2 u_{13} u_{23} + 
				b_3 u_3^2 u_{012}^2 u_{0123}^3 u_{013}^2 u_{02}^2 u_{023}^2 u_{03} u_{123}^2 u_{13}^3 u_{23} \
				+\\
				b_3^2 u_3 u_{0123}^2 u_{013} u_{023}^3 u_{03}^2 u_{123} u_{23}^2 + 
				u_3^3 u_{013} u_{023} u_{03}^2 u_{123} u_{13}^2 u_{23}^2 + \\
				b_3^2 u_3 u_{012}^2 u_{0123}^2 u_{013}^3 u_{02}^2 u_{023} u_{03}^2 u_{123}^3 u_{13}^2 \
				u_{23}^2 + b_3 u_3^2 u_{0123} u_{023}^2 u_{03} u_{123}^2 u_{13} u_{23}^3))/\\
				(u_{12}^2 (b_3 + 
				u_3 u_{0123} u_{013} u_{023} u_{03} u_{123} u_{13} u_{23}) (b_3 u_3^2 u_{0123} u_{013}^3 \
				u_{023}^2 u_{03}^2 u_{123} u_{13} + \\b_3^2 u_3 u_{0123}^2 u_{013}^2 u_{023}^3 u_{03} u_{123}^2 u_{23} +
				b_3^2 u_3 u_{012}^2 u_{0123}^2 u_{02}^2 u_{023} u_{03} u_{13}^2 u_{23} + \\
				u_3^3 u_{013}^2 u_{023} u_{03} u_{123}^2 u_{13}^2 u_{23} + 
				b_3^3 u_{012}^2 u_{0123} u_{013} u_{02}^2 u_{03}^2 u_{123} u_{13} u_{23}^2 + \\
				b_3 u_3^2 u_{0123} u_{013} u_{023}^2 u_{123}^3 u_{13} u_{23}^2 + 
				b_3 u_3^2 u_{012}^2 u_{0123}^3 u_{013} u_{02}^2 u_{023}^2 u_{03}^2 u_{123} u_{13}^3 u_{23}^2 \
				+\\ b_3^2 u_3 u_{012}^2 u_{0123}^2 u_{013}^2 u_{02}^2 u_{023} u_{03}^3 u_{123}^2 u_{13}^2 u_{23}^3)) ,  
			\end{gathered}
		\end{equation} 
		
		$u_{01} = s_{01}^{1/2}, $ $ R_{01} = s_{01}^{1/4},$
		
		\begin{equation} \label{b_2} 
			\begin{gathered}
				b_2 = ((b_3 b_4 u_3 u_{0123} u_{013}^2 u_{023} u_{03}^2 u_{13} + 
				b_3^2 b_4 u_{0123}^2 u_{013} u_{023}^2 u_{03} u_{123} u_{23} + \\
				b_4 u_3^2 u_{013} u_{03} u_{123} u_{13}^2 u_{23} + 
				b_3 b_4 u_3 u_{0123} u_{023} u_{123}^2 u_{13} u_{23}^2)/\\
				(b_3 u_{012}^2 u_{02}^2 (b_3 u_3 u_{0123} u_{023} u_{13} + b_3^2 u_{013} u_{03} u_{123} u_{23} + \\
				u_3^2 u_{0123}^2 u_{013} u_{023}^2 u_{03} u_{123} u_{13}^2 u_{23} + 
				b_3 u_3 u_{0123} u_{013}^2 u_{023} u_{03}^2 u_{123}^2 u_{13} u_{23}^2)))    ,  
			\end{gathered}
		\end{equation} 
		
		\begin{equation} \label{u1} 
			\begin{gathered}
				u_1 = (b_2 u_3 u_{01} u_{013} u_{023} u_{12} u_{123} + b_4 u_{01} u_{0123} u_{03} u_{12} u_{13} u_{23})/\\
				(
				u_{012} (u_3 u_{0123} u_{023} u_{13} + 
				b_3 u_{013} u_{03} u_{123} u_{23}))  ,  
			\end{gathered}
		\end{equation} 
		
		\begin{equation} \label{u2} 
			\begin{gathered}
				u_2 = (u_3 u_{013} u_{03} u_{13} + b_3 u_{0123} u_{023} u_{123} u_{23})/\\(
				b_3 u_{012} u_{02} u_{12} (b_3 + 
				u_3 u_{0123} u_{013} u_{023} u_{03} u_{123} u_{13} u_{23})) ,  
			\end{gathered}
		\end{equation}

		\begin{equation} \label{u0}
			\begin{gathered}
				u_0 = ((u_3 u_{0123} u_{023} u_{13} + b_3 u_{013} u_{03} u_{123} u_{23})/\\(
				b_4 u_2 u_{01} u_{12} (b_3 + 
				u_3 u_{0123} u_{013} u_{023} u_{03} u_{123} u_{13} u_{23}))),  
			\end{gathered}
		\end{equation} 
		
		$R_i=u_i, i \in I$,
		
		\begin{equation} \label{F}
			\begin{gathered}
				F = (b_3  R_0 R_1 R_2 R_{01} R_{012} R_{02} R_{12})/(
				R_3 R_{0123} R_{013} R_{023} R_{03} R_{123} R_{13} R_{23}) + \\
				R_0 R_1 R_2 R_3 R_{01} R_{012} R_{0123} R_{013} R_{02} R_{023} R_{03} R_{12} R_{123} R_{13} R_{23},  
			\end{gathered}
		\end{equation}

		Free energy
		
		\begin{equation}\label{Free_energy}
			\begin{split}
				-f/(kT)=  \ln(F).
			\end{split} 
		\end{equation}
		
		All formulas are considered on the set of values of variables for which the denominators of the expressions do not vanish, $s_{02}>0$.
		
		Example. {Case $ A \neq 0,  B \neq 0 , C \neq 0  $}.
		
		We define free parameters.
		
		$	K_3,  K_{13}, K_{23}, K_{03}, K_{012}, K_{123}, K_{013}, 
		K_{023}, K_{0123}, b_3, b_4 .   $
		
		The remaining parameters are calculated.    
		
		\begin{tabular}{ l l }	
			$	K_{012}  =-0.78 $ &{ $K_{0123}  =1.77 $ }\\
			$	K_{013}  = 2.08  $ &{ $ K_{023}  = -0.673 $ }\\
			$	K_{03}  = 0.693 $ &{ $ K_{123}  = 0.64 $ }\\
			$	K_{13}  =0.13  $ &{ $ K_{23}  =0.23 $ }\\
			$b_3  = 0.3$ &{ $K_3  =-2.1714029607603003$ }\\
			$ K_{02} =\ln x_2 /4  =-1.3441741420924833$ & {$	K_{12}  =0.08619047541745381 $ }\\
			$ K_{01}  =0.4877434019297984  $ & {${K_1  =2.2973978914807907} $ }\\
			$	b_4  =1.4  $ &{ $b_2  =3199.070651210929 $ }\\
			${K_0  =-1.8300697949663391} $  &{ $K_2  =1.4298100940137992 $ }\\
			$F= 21.048995184536196$ &{ $- f/ {kT}=3.046852824287685$ 	} \\
		\end{tabular}
		
		The mapping of the set of squares of exponents of parameters of disorder solutions into the set of squares of exponents of parameters of independent parameters of the Hamiltonian (\ref{Hamiltonian_classic})  can be represented  as 
		\begin{equation} \label{Jacob_main_1}
			\begin{gathered}
				\overrightarrow{U}(	u_3,  u_{13}, u_{23}, u_{03}, u_{012}, u_{123}, u_{013}, 
				u_{023}, u_{0123}, b_3, b_4     )=
				(U_1, U_2, ...,U_{10}),	  		
			\end{gathered}
		\end{equation} 
		where 
		\begin{equation} \label{Jacob_main_2}
			\begin{gathered}
				U_1=u_0 u_1 u_2 u_3, \;\;
				U_2=u_{01} u_{23},\;\;
				U_3=u_{02} u_{13},\;\;
				U_4=u_{12},\;\;
				U_5= u_{03}, \\
				U_6=u_{012},\;\;
				U_7=u_{123}, \;\;
				U_8=u_{013},\;\;
				U_9=u_{023},\;\;
				U_{10}= u_{0123},\;\;\\
				U_i=U_i(	u_3,  u_{13}, u_{23}, u_{03}, u_{012}, u_{123}, u_{013}, 
				u_{023}, u_{0123}, b_3, b_4),\\
				i=1,2,...,10. 	
			\end{gathered}
		\end{equation}

		The Jacobi matrix of the mapping (\ref{Jacob_main_1}) - (\ref{Jacob_main_2}) calculated with formulas (\ref{quadratic_equation})- (\ref{Free_energy})  at the point from the example below, can be represented as: 
		
		\begin{equation} \label{Jacob_main_3}
			\begin{gathered} 
				J=(V_1 | V_2 | V_3 | V_4)	,
			\end{gathered}
		\end{equation} 
		where $V_1, V_2, V_3, V_4 $ -  matrices, written in the following form (we divide them into submatrices to fit on the page)

		\[ V_1= \left( \begin{array}{lcr}	    	
			-26.37922682274807 & -0.26441667921872636 & -0.8931098377518154 \\
			-50.48367965221345 & -0.506031775149296 & 2.4358902126486726 \\
			3.0875926895213013 & 0.030949012096159745 & 0.017292171701488446 \\
			-4.309248976985813 & -0.04319449242728979 & 0.603868157811327 \\
			0 & 0 & 0\\
			0 & 0 & 0\\
			0 & 0 & 0\\
			0 & 0 & 0\\
			0 & 0 & 0\\
			0 & 0 & 0\\
		\end{array}  \right ) ,\]

		\[ V_2= \left( \begin{array}{lcr}	    	
			-0.30481921131642054 & -11.147947418521653 & -0.06567164639870171 \\
			0.7593711419673355 & 15.224407536429396 & 0.18201503060311097 \\
			0.007363033786844664 & 0.3230692936714852 & -0.011052153116308805 \\
			-0.049852118655380195 & 4.282311838799124 & 0.015413593512647594 \\
			1 & 0 & 0\\
			0 & 1 & 0\\
			0 & 0 & 1\\
			0 & 0 & 0\\
			0 & 0 & 0\\
			0 & 0 & 0\\
		\end{array} \right ) ,\]

		\[ V_3=\left( \begin{array}{lcr}	    	
			-0.0006926197948509127 & 2.489233563884863 & -0.015519616147763315 \\
			-0.002343806926319303 & -11.655294053003473 & 0.004306538681220218 \\
			-0.0005828524052287065 & -0.11447983119805549 &	0.0010944469497342624 \\
			0.001355972334415867 & -3.7948034735668656 & -0.0025319646379529104 \\
			0 & 0 & 0\\
			0 & 0 & 0\\
			0 & 0 & 0\\
			1 & 0 & 0\\
			0 & 1 & 0\\
			0 & 0 & 1\\
		\end{array} \right ) ,\]

		\[V_4= \left( \begin{array}{lcr}	
			1.1430998222428101 & 0 \\
			2.1876261171449585 & 0\\
			-0.13379568308130585 & 0\\
			0.18673412216507757 & 0\\
			0 & 0\\
			0 & 0\\
			0 & 0\\
			0 & 0\\
			0 & 0\\
			0 & 0\\
		\end{array} \right ) .\]

		The rang of the Jacobi matrix $ J$ with tolerance 0.001 is equal to 8. This fact can be seen if we found the minor of the Jacobi matrix
		
		\[\det \left( \begin{array}{lcr}	
			-0.8931098377518154 & 1.1430998222428101\\
			2.4358902126486726 & 2.1876261171449585\\	  
		\end{array} \right )=-4.7382560756266665, \]	   
		which differs from zero. The last  six lines of the Jacobi matrix  $J$ are linearly independent.
		Hence in the general case the disorder solution form the 8-dimensional subset of 10-dimensional space  of the parameters of the classical Hamiltonian (\ref{Hamiltonian_classic}).

		%%%%%%%%%%%%%%%%%%%%%%%%%%%%%%%%%%%%%%%%
		\subsection{Case 2 }\label{CaseA0Bneq0Cneq0}
		Let us  $ A=0, B \neq 0 , C \neq 0  $.
		
		If $ A=0$, one additional equation appears from (\ref{coefficient_A})
		
		\begin{equation} \label{A=0}
			\begin{gathered}
				(b_3 u_3 u_{012}^4 u_{0123} u_{023}^3 u_{13} - 
				b_3 u_3 u_{0123}^3 u_{013}^2 u_{023} u_{03}^2 u_{13} +\\
				b_3^2 u_{012}^4 u_{013} u_{023}^2 u_{03} u_{123} u_{23} - 
				b_3^2 u_{0123}^4 u_{013} u_{023}^2 u_{03} u_{123} u_{23} - \\
				u_3^2 u_{0123}^2 u_{013} u_{03} u_{123} u_{13}^2 u_{23} + 
				u_3^2 u_{012}^4 u_{0123}^2 u_{013} u_{023}^4 u_{03} u_{123} u_{13}^2 u_{23} - \\
				b_3 u_3 u_{0123}^3 u_{023} u_{123}^2 u_{13} u_{23}^2 + 
				b_3 u_3 u_{012}^4 u_{0123} u_{013}^2 u_{023}^3 u_{03}^2 u_{123}^2 u_{13} u_{23}^2)=0 .  
			\end{gathered}
		\end{equation} 
		From (\ref{A=0})
		
		\begin{equation} \label{u_{012}_A}
			\begin{gathered}
				u_{012}^4 = ((
				b_3 u_3 u_{0123}^3 u_{013}^2 u_{023} u_{03}^2 u_{13} + 
				b_3^2 u_{0123}^4 u_{013} u_{023}^2 u_{03} u_{123} u_{23} + \\
				u_3^2 u_{0123}^2 u_{013} u_{03} u_{123} u_{13}^2 u_{23} + 
				b_3 u_3 u_{0123}^3 u_{023} u_{123}^2 u_{13} u_{23}^2)/ \\
				(u_{023}^2 (b_3 u_3 u_{0123} u_{023} u_{13} + b_3^2 u_{013} u_{03} u_{123} u_{23} + \\
				u_3^2 u_{0123}^2 u_{013} u_{023}^2 u_{03} u_{123} u_{13}^2 u_{23} + 
				b_3 u_3 u_{0123} u_{013}^2 u_{023} u_{03}^2 u_{123}^2 u_{13} u_{23}^2)))   .  
			\end{gathered}
		\end{equation} 
		
		Then $s_{02}=-C/B $, we return to the main part of the solution with (\ref{s12})-
		(\ref{F}).
		
		Example. {Case $ A = 0,  B \neq 0 , C \neq 0  $}.

		We define free parameters:
		
		$ K_{0123}, K_{013}, K_{023}, K_{03}, K_{123}, K_{13}, K_{23}, b_3, K_3, b_4.$ 
		Other parameters are calculated.    
		
		\begin{tabular}{ l l }	
			$ K_{0123}  =1.77 $ & {$K_{013}  =-0.013  $ }\\
			$ K_{023}  =1.5 $ & {$	K_{03}  =0.693   $ }\\
			$ K_{123}  = 0.64 $ & {$	K_{13}  =0.13  $ }\\
			$ K_{23}  =1.23 $ & {$	K_{012}  =-0.291142355720243 $ }\\
			$b_3  =0.003 $ &{ $K_3  =-2.1714029607603003$ }\\
			$ K_{02}  =-0.38497882070925643$ & {$	K_{12}  =-0.3632011114281891 $ }\\
			$ K_{01}  =-0.9097204700380525 $ & {$K_1  =1.7950021671906025 $ }\\
			$	b_4  =1.4 $ &{ $b_2  = 230.66376488481214$ }\\
			${ K_0  =-3.825768655256056 } $  &{ $K_2  =2.401099990054352 $ }\\
			$F=9.02401494007837 $ &{ $- f/ {kT}=2.1998893503481236$ 	}\\
		\end{tabular}

		\subsection{Case 3 }\label{CaseA0B0C0}
		Let us $ A=0, B= 0 , C=0  .$
		
		In this case the equation (\ref{u_{012}_A}) stay remains unchanged. Let us also express $u_{012}^4  $ from the condition 
		$ B=0 $ (\ref{coefficient_B}) и $C=0 $ (\ref{coefficient_C})  (we don't consider the positive multipliers)
		
		\begin{equation} \label{u_{012}_B}
			\begin{gathered}
				u_{012}^4 =  ((b_3 u_3 u_{0123} u_{013}^4 u_{023} u_{03}^2 u_{123}^2 u_{13} + 
				b_3^2 u_{0123}^2 u_{013}^3 u_{023}^2 u_{03} u_{123}^3 u_{23} + \\
				u_3^2 u_{013}^3 u_{03} u_{123}^3 u_{13}^2 u_{23} + 
				b_3 u_3 u_{0123} u_{013}^2 u_{023} u_{03}^4 u_{13} u_{23}^2 + \\
				b_3 u_3 u_{0123} u_{013}^2 u_{023} u_{123}^4 u_{13} u_{23}^2 + 
				b_3^2 u_{0123}^2 u_{013} u_{023}^2 u_{03}^3 u_{123} u_{23}^3 + \\
				u_3^2 u_{013} u_{03}^3 u_{123} u_{13}^2 u_{23}^3 + 
				b_3 u_3 u_{0123} u_{023} u_{03}^2 u_{123}^2 u_{13} u_{23}^4)/\\
				((u_{03}^2 u_{123}^2 + 
				u_{013}^2 u_{23}^2) (b_3 u_3 u_{0123} u_{023} u_{13} + b_3^2 u_{013} u_{03} u_{123} u_{23} + \\
				u_3^2 u_{0123}^2 u_{013} u_{023}^2 u_{03} u_{123} u_{13}^2 u_{23} + 
				b_3 u_3 u_{0123} u_{013}^2 u_{023} u_{03}^2 u_{123}^2 u_{13} u_{23}^2)))         ,  
			\end{gathered}
		\end{equation} 
		
		and 
		\begin{equation} \label{u_{012}_C}
			\begin{gathered}
				u_{012}^4 = ((
				b_3 u_3 u_{0123} u_{013}^2 u_{023}^3 u_{03}^2 u_{13} + 
				b_3^2 u_{0123}^2 u_{013} u_{023}^4 u_{03} u_{123} u_{23} + \\
				u_3^2 u_{013} u_{023}^2 u_{03} u_{123} u_{13}^2 u_{23} + 
				b_3 u_3 u_{0123} u_{023}^3 u_{123}^2 u_{13} u_{23}^2)/\\
				( u_{0123}^2 (b_3 u_3 u_{0123} u_{023} u_{13} + b_3^2 u_{013} u_{03} u_{123} u_{23} + 
				u_3^2 u_{0123}^2 u_{013} u_{023}^2 u_{03} u_{123} u_{13}^2 u_{23} + \\
				b_3 u_3 u_{0123} u_{013}^2 u_{023} u_{03}^2 u_{123}^2 u_{13} u_{23}^2)))   
				.
			\end{gathered}
		\end{equation} 
		
		Equating the right part of the equation (\ref{u_{012}_A}) to the right parts of the equations (\ref{u_{012}_B}), (\ref{u_{012}_C}), simplifying and removing the positive multipliers,  we get two equations

		\begin{equation} \label{u_{012}_AB}
			\begin{gathered}
				(u_{013}^2 u_{023}^2 u_{123}^2 - u_{0123}^2 u_{03}^2 u_{123}^2 - u_{0123}^2 u_{013}^2 u_{23}^2 + 
				u_{023}^2 u_{03}^2 u_{23}^2) = 0      ,  
			\end{gathered}
		\end{equation} 
		
		\begin{equation} \label{u_{012}_AC}
			\begin{gathered}
				(-u_{0123} + u_{023}) = 0    .  
			\end{gathered}
		\end{equation} 
		From  (\ref{u_{012}_AC})
		\begin{equation} \label{u_{0123}}
			\begin{gathered}
				u_{0123}= u_{023}  ,  
			\end{gathered}
		\end{equation}        
		Substituting $u_{0123}$ from (\ref{u_{0123}}) into (\ref{u_{012}_AB}) and simplifying, 
		we have  
		\begin{equation} \label{u_{013}_03_123_23} 
			\begin{gathered}
				(s_{013} - s_{03}) (s_{123} - s_{23})  = 0     .  
			\end{gathered}
		\end{equation} 
		
		We write the formulas in the order of their application:
		
		$u_{03} = u_{013}$ or  $u_{23} = u_{123}$ (\ref{u_{013}_03_123_23}),  $u_{0123} = u_{023}$  (\ref{u_{0123}})   ,   (\ref{u_{012}_A}) , then   (\ref{s12}) -  (\ref{Free_energy})  without changes.
		
		Example. {Case $ A=0, B= 0 , C=0  $}.

		We define free parameters:
		
		$K_{013}, K_{023}, K_{03}, K_{123}, K_{13}, K_{23}, b_3, K_3, K_{02}, b_4.$ 
		Other parameters are calculated.    
		
		\begin{tabular}{ l l }	
			$	K_{013}  =  -0.013 $ &{ $K_{023}  = 1.5$ }\\
			$ 	K_{03}  = -0.013$  &{ $	K_{13}  = 0.13$ }\\		
			$ K_{123}  = 0.64 $ & {$K_{23}  =1.23$ }\\
			$b_3  = 0.003$ &{ $K_{01}  =  -0.8727265471351584$ }\\
			$ K_{0123}  =1.5 $ & {$	K_{012}  = -0.37360059361219283 $ }\\
			$ K_{02}  =0.02 $ & {$	K_{12}  =-0.09327414990922747$ }\\
			$	b_4  =1.4  $ &{ $b_2  =96.6612745738842 $ }\\	
			$ K_0  =-3.551933566990586 $ & {$K_1  =1.5519101075620245 $ }\\
			${K_2  =2.514429232980978} $  &{ $K_3=-2.1714029607603003$ }\\	
			$F=7.36996205928568 $ &{ $- f/ {(kT)}=1.9974125581941924$ 	}\\
		\end{tabular}
		
		%%%%%%%%%%%%%%%%%%%%%%%%%%%%%%%%%%%%%%%%%%%%%%%%%%%%%%%%%%%%%
		%%%%%%%%%%%%%%%%%%%%%%%%%%%%%%%%%%%%%%%%%%%%%%%%%%%%%%%%%%%%%%%%%%%%%%%%%%%%%%%%%%%%%%%%%%%
		
		%%%%%%%%%%%%%%%%%%%%%%%%%%%%%%%%%%%%%%%%%%%%%%%%%%%%%%%%%%%%%%%%%%%%%%%%%%%%%%%%%%%%%%%%%%%
		\section{Finding the largest eigenvalue and eigenvector of the transposed transfer matrix  }\label{largest_eigenvalue_transpon} %max_val}

		The eigenvector of the transposed transfer matrix $ \Theta^T=\Theta_{p,q}^T $, corresponding to
		the largest eigenvalue $ F' $,  we represent in the following form: \\
		
		\begin{equation}\label{transpon_eigenvector1}   
			\overrightarrow{x}=
			(\underbrace{1,d_2,\dots,1,d_2 }_{2^{L_1}};\underbrace{d_3,d_4,\dots,d_3,d_4}_{2^{L_1}})^T
			.
		\end{equation}	
		
		Then by the Perron–Frobenius theorem \cite{Perron} this eigenvector with all positive elements will correspond to the single maximal eigenvalue $F'$ of the transter matrix $ \Theta^T=\Theta_{p,q}^T $. 
		
		From the form of the elementary transposed transfer matrix  $ \Theta^T $ and the form of the eigenvector (\ref{transpon_eigenvector1}), we have 
		the following system of equations 
		\begin{equation}\label{transpon_main_system1}  
			\begin{split} 
				\left\{  
				\begin{array}{rcl}  
					F' &= & a_{00}+ d_2 a_{01}, \\  
					d_2 F' &= & a_{02}+d_2 a_{03}, \\ 
					F' &= & d_3 a_{04}+d_4 a_{05}, \\ 
					d_2 F' &= & d_3 a_{06}+ d_4 a_{07}, \\ 
					d_3 F' &= &  a_{10}+ d_2 a_{11}, \\
					d_4 F' &= &  a_{12}+ d_2 a_{13}, \\
					d_3 F' &= &  d_3 a_{14}+ d_4 a_{15}, \\
					d_4 F' &= &  d_3 a_{16}+ d_4 a_{17}, \\
				\end{array}   
				\right.
			\end{split} 
		\end{equation}
		where $ a_{ij}$ is from (\ref{aij}).
		
		These eight equations (\ref{transpon_main_system1}) will repeat.
		Then , using the form of the Hamiltonian (\ref{Hamiltonian1}), after changing  variable (\ref{R_i1})	, we have

		\begin{multline} \label{transpon_main_system2}
			F' = ((d_2 H_1 H_2 H_3 R_{12} R_{123} R_{13} R_{23})/(
			H_0 R_{01} R_{012} R_{0123} R_{013} R_{02} R_{023} R_{03}) +\\ 
			H_0 H_1 H_2 H_3 R_{01} R_{012} R_{0123} R_{013} R_{02} R_{023} R_{03} R_{12} R_{123} R_{13} R_{23}  ),   \;\; \;\;  \;\; \;\;(a)\\
			d_2 F' = ((d_2 H_2 H_3 R_{01} R_{012} R_{0123} R_{013} R_{23})/(
			H_0 H_1 R_{02} R_{023} R_{03} R_{12} R_{123} R_{13}) +\\ (H_0 H_2 H_3 R_{02} R_{023} R_{03} R_{23})/(
			H_1 R_{01} R_{012} R_{0123} R_{013} R_{12} R_{123} R_{13})  ) ,   \;\; \;\;  \;\; \;\;(b) \\
			F' = ((d_4 H_1 H_3 R_{012} R_{0123} R_{02} R_{023} R_{13})/(
			H_0 H_2 R_{01} R_{013} R_{03} R_{12} R_{123} R_{23}) +\\(d_3 H_0 H_1 H_3 R_{01} R_{013} R_{03} R_{13})/(
			H_2 R_{012} R_{0123} R_{02} R_{023} R_{12} R_{123} R_{23})    ),   \;\; \;\;  \;\; \;\;(c)\\
			d_2 F' = ((d_4 H_3 R_{01} R_{013} R_{02} R_{023} R_{12} R_{123})/(
			H_0 H_1 H_2 R_{012} R_{0123} R_{03} R_{13} R_{23}) +\\ (
			d_3 H_0 H_3 R_{012} R_{0123} R_{03} R_{12} R_{123})/(
			H_1 H_2 R_{01} R_{013} R_{02} R_{023} R_{13} R_{23}) )  ,   \;\; \;\;  \;\; \;\;(d)\\
			d_3 F' = ((H_0 H_1 H_2 R_{01} R_{012} R_{02} R_{12})/(
			H_3 R_{0123} R_{013} R_{023} R_{03} R_{123} R_{13} R_{23}) +\\ (
			d_2 H_1 H_2 R_{0123} R_{013} R_{023} R_{03} R_{12})/(
			H_0 H_3 R_{01} R_{012} R_{02} R_{123} R_{13} R_{23})  ) ,   \;\; \;\;  \;\; \;\;(e)\\
			d_4 F' = ((H_0 H_2 R_{0123} R_{013} R_{02} R_{123} R_{13})/(
			H_1 H_3 R_{01} R_{012} R_{023} R_{03} R_{12} R_{23}) +\\ (
			d_2 H_2 R_{01} R_{012} R_{023} R_{03} R_{123} R_{13})/(
			H_0 H_1 H_3 R_{0123} R_{013} R_{02} R_{12} R_{23})   ),   \;\; \;\;  \;\; \;\;(f)\\
			d_3 F' = ((d_3 H_0 H_1 R_{01} R_{0123} R_{023} R_{123} R_{23})/(
			H_2 H_3 R_{012} R_{013} R_{02} R_{03} R_{12} R_{13}) +\\ (
			d_4 H_1 R_{012} R_{013} R_{02} R_{03} R_{123} R_{23})/(
			H_0 H_2 H_3 R_{01} R_{0123} R_{023} R_{12} R_{13})   ),   \;\; \;\;  \;\; \;\;(g)\\
			d_4 F' = ((d_3 H_0 R_{012} R_{013} R_{023} R_{12} R_{13} R_{23})/(
			H_1 H_2 H_3 R_{01} R_{0123} R_{02} R_{03} R_{123}) +\\ (
			d_4 R_{01} R_{0123} R_{02} R_{03} R_{12} R_{13} R_{23})/(
			H_0 H_1 H_2 H_3 R_{012} R_{013} R_{023} R_{123}) )  .   \;\; \;\;  \;\; \;\;(h)       
		\end{multline}

		The solution scheme for the system of equations (\ref{transpon_main_system2}) is following:
		variable $ F $ from (\ref{transpon_main_system2} a) we substite into other equations (\ref{transpon_main_system2}(b))-(\ref{transpon_main_system2}(h)) of the system (\ref{transpon_main_system2}). 
		\begin{equation}\label{transpon_main_system3}
			\begin{split} 
				\left\{  
				\begin{array}{rcl}  
					Right(a) &= & Right(c),   \; \;\; (a)1\\
					Right(b) &= & Right(d),  \; \;\; (b)2\\
					Right(e) &= & Right(g),   \; \;\; (c)3\\
					Right(f) &= & Right(h),     \; \;\; (d)4\\
					b_2 Right(a) &= & Right(b),    \; \; \;  (e)5\\
					b_3 Right(a) &= & Right(e),   \; \;\; (f)6\\
					b_4 Right(a) &= & Right(f),   \; \;\; (g)7\\
				\end{array}   
				\right.  
			\end{split} 
		\end{equation}
		
		where  $ Right(i) $, $ i=a,\dots,h $ are the right parts of the equation $ i $ of the system (\ref{transpon_main_system2}).
		
		After multiplying by positive denominator and simplifying of the system (\ref{transpon_main_system3}) 
		%D:\Pavel\Ising\models\RootMatrix\equal\K1_K9H\15variables\check\transponir\article_8_transponir2
		\begin{equation}\label{transpon_main_system4new}
			\begin{split} 
				\left\{  
				\begin{array}{rcl}  
					(d_4 u_{012} u_{0123} u_{02} u_{023} + d_3 u_0 u_{01} u_{013} u_{03} - d_2 u_2 u_{12} u_{123} u_{23} - \\
					u_0 u_2 u_{01} u_{012} u_{0123} u_{013} u_{02} u_{023} u_{03} u_{12} u_{123} u_{23})  &= & 0,   \;\; \;\;    (a)\\
					(d_4 u_{01} u_{013} u_{02} u_{023} u_{12} u_{123} + d_3 u_0 u_{012} u_{0123} u_{03} u_{12} u_{123} - \\
					d_2 u_2 u_{01} u_{012} u_{0123} u_{013} u_{23} - 
					u_0 u_2 u_{02} u_{023} u_{03} u_{23})  &= & 0,   \;\; \;\;    (b)\\
					(u_0 u_2 u_{01} u_{012} u_{02} u_{12} + d_2 u_2 u_{0123} u_{013} u_{023} u_{03} u_{12} - \\
					d_3 u_0 u_{01} u_{0123} u_{023} u_{123} u_{23} - 
					d_4 u_{012} u_{013} u_{02} u_{03} u_{123} u_{23})  &= & 0,   \;\; \;\;    (c)\\
					(u_0 u_2 u_{0123} u_{013} u_{02} u_{123} + d_2 u_2 u_{01} u_{012} u_{023} u_{03} u_{123} -\\ 
					d_3 u_0 u_{012} u_{013} u_{023} u_{12} u_{23} - 
					d_4 u_{01} u_{0123} u_{02} u_{03} u_{12} u_{23})  &= & 0,   \;\; \;\;    (d)\\ 
					(-d_2 u_{01} u_{012} u_{0123} u_{013} - u_0 u_{02} u_{023} u_{03} + d_2^2 u_1 u_{12} u_{123} u_{13} + \\
					d_2 u_0 u_1 u_{01} u_{012} u_{0123} u_{013} u_{02} u_{023} u_{03} u_{12} u_{123} u_{13})  &= & 0,   \;\; \;\;    (e)\\
					(-u_0 u_{01} u_{012} u_{02} - d_2 u_{0123} u_{013} u_{023} u_{03} + d_2 d_3 u_3 u_{123} u_{13} u_{23} + \\
					d_3 u_0 u_3 u_{01} u_{012} u_{0123} u_{013} u_{02} u_{023} u_{03} u_{123} u_{13} u_{23}) &= & 0,   \;\; \;\;    (f)\\
					(-u_0 u_{0123} u_{013} u_{02} - d_2 u_{01} u_{012} u_{023} u_{03} + d_2 d_4 u_1 u_3 u_{12} u_{23} + \\
					d_4 u_0 u_1 u_3 u_{01} u_{012} u_{0123} u_{013} u_{02} u_{023} u_{03} u_{12} u_{23})  &= & 0.   \;\; \;\;    (g)\\
				\end{array}   
				\right.  
			\end{split} 
		\end{equation}
		%%%%%%%%%%%%%%%%%%%%%%%%%%%%%%%%%%%%%%%%%%%%%%%%%
		The system (\ref{transpon_main_system4new}) and equation  (\ref{transpon_main_system2}) (a) 
		is equal to the system of equations (\ref{main_system4}):
		
		( (\ref{main_system4})(b) $\leftrightarrow $(\ref{transpon_main_system4new}) (f)),
		(\ref{main_system4})(c)$\leftrightarrow $ (\ref{transpon_main_system4new})(a), (\ref{main_system4})(d)$\leftrightarrow $(\ref{transpon_main_system4new})(c), 
		( (\ref{main_system4})(e) $\leftrightarrow $(\ref{transpon_main_system4new}) (e)),
		( (\ref{main_system4})(f) $\leftrightarrow $(\ref{transpon_main_system4new}) (g)),
		( (\ref{main_system4})(g) $\leftrightarrow $(\ref{transpon_main_system4new}) (f)),
		( (\ref{main_system4})(h) $\leftrightarrow $(\ref{transpon_main_system4new}) (d)),
		%D:\Pavel\Ising\models\RootMatrix\equal\K1_K9H\15variables\check\w1
		
		and to the equation (\ref{main_system2} $\; $a)  
		after the following change of variables:
		$K_{01}\leftrightarrow K_{23}, $ $K_{02}\leftrightarrow K_{13}, $  $K_{012}\leftrightarrow K_{123}, $  $K_{013}\leftrightarrow K_{023}, $  $K_{0}\leftrightarrow K_{3}, $  $K_{2}\leftrightarrow K_{1}, $  $b_3\leftrightarrow d_2, $  $b_2\leftrightarrow d_3, $. The variables $K_{0123},  K_{03}, K_{12}, b_4 $ remain unchanged. This corresponds to changing variables when rotating around the center of the round square cell $\Omega=\{\tau^{0}, ,\tau^{1}, ,\tau^{L_1}, ,\tau^{L_1+1} \}  $ on $\pi.$ Hence there are the bijection between the solutions of the systems (\ref{main_system2}) and (\ref{transpon_main_system2}). 
		
		\begin{remark}\label{transpon_remark2} 
			It is clear that with all possible symmetries of cell $\Omega  $ and the changes of variables corresponding to these symmetries  (not just for the given transposed matrix) the free energy $f$ remains unchanged. This result helps to extend the class of models which disorder solutions are found for. 		
		\end{remark}
		%%%%%%%%%%%%%%%%%%%%%%%%%%%%%%%%%%%%%%%%%%%%%%%%%
		%%%%%%%%%%%%%%%%%%%%%%%%%%%%%%%%%%%%%%%%%%%%%%%%%%%%%%%%%%%%%%%%%%%%%%%%%%%%%%%%%%%%%%%%%%%
		\section{Disoder Solutions for Models with nearest, next-nearest-neighbor and quadruple interactions  in an external field}\label{without_triple} 
		
		Let us write the formulas in the case of absence of the interaction of three spins. Here 
		$K_{012}=0$,  $K_{023}=0$,  $K_{013}=0$,  $K_{123}=0$. We substitute this conditions into (\ref{quadratic_equation}) with coefficients $A$ (\ref{coefficient_A}), $B$ (\ref{coefficient_B}), $C$ (\ref{coefficient_C}), or, what is the same, into the expaneded form of this equation (\ref{for_s02}). After simplifying and factorization we get that the quadratic equation splits into two linear multipliers. 
		Let us consider both cases:
		%%%%%%%%%%%%%%%%%%%%%%%%%%%%%%%%%%%%%%%%%%%%%%%%%%%%%%%%%%%%%
		\subsection{Case 1.   } 
		\begin{equation}\label{without3_s02_1} 
			\begin{array}{rcl}  
				s_{02} = (-((
				u_3 (u_3 u_{03} u_{13} + b_3 u_{0123} u_{23}) (b_3 u_{0123} u_{03} + 
				u_3 u_{13} u_{23}) \\(b_3 u_{0123}^3 - b_3 u_{0123} u_{03}^2 - u_3 u_{03} u_{13} u_{23} + 
				u_3 u_{0123}^4 u_{03} u_{13} u_{23} - \\b_3 u_{0123} u_{23}^2 + 
				b_3 u_{0123}^3 u_{03}^2 u_{23}^2))/\\(
				b_3 u_{13} (u_3 u_{0123} u_{13} + b_3 u_{03} u_{23}) (b_3 + 
				u_3 u_{0123} u_{03} u_{13} u_{23}) \\(-u_3 u_{0123} u_{13} + u_3 u_{0123}^3 u_{03}^2 u_{13} - 
				b_3 u_{03} u_{23} + \\b_3 u_{0123}^4 u_{03} u_{23} + u_3 u_{0123}^3 u_{13} u_{23}^2 - 
				u_3 u_{0123} u_{03}^2 u_{13} u_{23}^2)))). 
			\end{array}  
		\end{equation}
		Let us simplify the following equations (\ref{s12}), (\ref{s01}), (\ref{b_2}), (\ref{u1}), (\ref{u2}), (\ref{u0}) considering the conditions of of absence of the interaction of three spins.
		
		Then
		\begin{equation} \label{without3_s12}
			\begin{gathered}
				s_{12} =  (-(((u_{0123} - u_{03}) (u_{0123} + u_{03}))/((-1 + u_{0123} u_{03}) (1 + 
				u_{0123} u_{03})))),  
			\end{gathered}
		\end{equation} 
		$u_{12} = s_{12}^{1/2}, $ $ R_{12} = s_{12}^{1/4},$
		
		\begin{equation} \label{without3_s01}
			\begin{gathered}
				s_{01} =  -(((u_{0123} - u_{23}) (u_{0123} + u_{23}))/((-1 + u_{0123} u_{23}) (1 + 
				u_{0123} u_{23}))) ,  
			\end{gathered}
		\end{equation} 
		
		$u_{01} = s_{01}^{1/2}, $ $ R_{01} = s_{01}^{1/4},$
		
		\begin{equation} \label{without3_b_2} 
			\begin{gathered}
				b_2 =(b_4 (u_3 u_{03} u_{13} + b_3 u_{0123} u_{23}) (b_3 u_{0123} u_{03} + 
				u_3 u_{13} u_{23}))/\\(b_3 u_{02}^2 (u_3 u_{0123} u_{13} + b_3 u_{03} u_{23}) (b_3 + 
				u_3 u_{0123} u_{03} u_{13} u_{23}))   ,  
			\end{gathered}
		\end{equation} 
		
		\begin{equation} \label{without3_u1} 
			\begin{gathered}
				u_1 = (u_{01} u_{12} (b_2 u_3 + b_4 u_{0123} u_{03} u_{13} u_{23}))/(u_3 u_{0123} u_{13} + b_3 u_{03} u_{23}) ,  
			\end{gathered}
		\end{equation} 
		
		\begin{equation} \label{without3_u2}
			\begin{gathered}
				u_2 =(u_3 u_{03} u_{13} + b_3 u_{0123} u_{23})/(b_3 u_{02} u_{12} (b_3 + u_3 u_{0123} u_{03} u_{13} u_{23})) ,  
			\end{gathered}
		\end{equation}

		\begin{equation} \label{without3_u0}
			\begin{gathered}
				u_0 = (b_3 u_{02} (u_3 u_{0123} u_{13} + b_3 u_{03} u_{23}))/(b_4 u_{01} (u_3 u_{03} u_{13} + 
				b_3 u_{0123} u_{23})).  
			\end{gathered}
		\end{equation} 
		Example. {Case 5.1}.	
		We define free parameters:	
		$b_3,  K_{0123}, K_{03}, K_{13}, K_3,  K_{23},  b_4, K_3.$ 
		Other parameters are calculated.    
		
		\begin{tabular}{ l l }	
			$	K_{013}  =0 $ &{ $ K_{023}  =0 $ }\\
			$	K_{03}  = -3.3 $ &{ $K_{123}  = 0 $ }\\
			{$	K_{13}  = -1.13$} &  {$ b_2  = 3.1331092732504073$ }\\ 
			${b_3  =1.3} $  &{ $b_4  = 1.4$ }\\ 
			$ K_{23}  =-1.23 $ & {$K_{0123}  =-1.123 $ }\\
			$	K_{012}  = 0 $ &{ $K_{02}  =-1.4916727561623406$ }\\
			$	K_{12}  =-1.1230413128435728 $ &{ $ K_{01}  =-1.386732323460238$ }\\
			$ K_0  =-0.039702842314627625$ & {${K_1  =0.309056082086584}$ }\\
			${K_2  =0.1383983223687954} $  &{ $ K_{3}  = 0.13615729766032947 $ }\\	
			$F= 27.53652851020629$ &{ $-f/ {(kT)}=3.3155134327120073$ 	}\\
		\end{tabular}

		%%%%%%%%%%%%%%%%%%%%%%%%%%%%%%%%%%%%%%%%%%%%%%%%%%%%%%%%%%%%%
		\subsection{Case 2.   }
		\begin{equation}\label{without3_s02_2}
			\begin{array}{rcl}  
				s_{02} = (u_3^2 (u_3 u_{03} u_{13} + b_3 u_{0123} u_{23}) (b_3 u_{0123} u_{03} + 
				u_3 u_{13} u_{23}))/\\(
				b_3^2 (u_3 u_{0123} u_{13} + b_3 u_{03} u_{23}) (b_3 + u_3 u_{0123} u_{03} u_{13} u_{23})). 
			\end{array}  
		\end{equation}

		Let us simplify the following equations (\ref{s12}), (\ref{s01}), (\ref{b_2}), (\ref{u1}), (\ref{u2}), (\ref{u0}) considering the conditions of of absence of the interaction of three spins.
		Then
		\begin{equation} \label{without3_s12_2} 
			\begin{gathered}
				s_{12} =   ((u_3 u_{03} u_{13} + b_3 u_{0123} u_{23}) (u_3 u_{0123} u_{13} + 
				b_3 u_{03} u_{23}))/\\((b_3 u_{0123} u_{03} + u_3 u_{13} u_{23}) (b_3 + 
				u_3 u_{0123} u_{03} u_{13} u_{23})),  
			\end{gathered}
		\end{equation} 
		$u_{12} = s_{12}^{1/2}, $ $ R_{12} = s_{12}^{1/4},$
		
		\begin{equation} \label{without3_s01_2}
			\begin{gathered}
				s_{01} =  (u_3 u_{0123} u_{13} + b_3 u_{03} u_{23})^2/\\(
				u_{12}^2 (b_3 + u_3 u_{0123} u_{03} u_{13} u_{23})^2) .  
			\end{gathered}
		\end{equation} 
		
		Then the formulas (\ref{without3_b_2}), (\ref{without3_u1}), (\ref{without3_u2}), (\ref{without3_u0}).

		Example. {Case 5.2}.
		We define free parameters:
		
		$b_3,  K_{0123}, K_{03}, K_{13}, K_3,  K_{23},  b_4.$ 
		
		Other parameters are calculated.    
		
		\begin{tabular}{ l l }	
			$	K_{013}  =0 $ &{ $K_{023}  =0$ }\\
			$	K_{03}  = -3.3 $ &{ $K_{123}  = 0 $ }\\
			{$	K_{13}  = -1.13$} &  {$ b_2  = 1.055703437822838$ }\\ 
			${b_3  =1.3} $  &{ $b_4  = 1.4$ }\\ 
			$ K_{23}  =-1.23 $ & {$K_{0123}  =-1.123 $ }\\
			$	K_{012}  = 0 $ &{ $K_{02}  =-1.2197181113957154 $ }\\
			$	K_{12}  =-1.1204403375810732 $ &{ $K_{01}  =-1.1224017003579632 $ }\\
			$ K_0  =-0.03207882065027702$ & {$K_1  =0.03207882065027688 $ }\\
			${K_2  =-0.1361572976603294 } $  &{ $K_{3}  = 0.13615729766032947 $ }\\	
			$F= 27.398569829889315$ &{ $-f/ {(kT)}=3.3104908160423254$ 	}\\
		\end{tabular}\\
		We notice that $K_0=-K_1$, $K_2=-K_3$ and the external field is equal to zero.
		%%%%%%%%%%%%%%%%%%%%%%%%%%%%%%%%%%%%%%%%%%%%%%%%%%%%%%%%%%%%%%%%%%%%%%%%%%%%%%%%%%%%%%%%%%%
		%%%%%%%%%%%%%%%%%%%%%%%%%%%%%%%%%%%%%%%%%%%%%%%%%%%%%%%%%%%%%%%%%%%%%%%%%%%%%%%%%%%%%%%%%%%
		\section{ 	Disoder Solutions for Models with nearest, next-nearest-neighbor and quadruple interactions}\label{even_model} 
		In this case we consider 	$K_{012}=K_{023}=K_{013}=K_{123}=K_0=K_1=K_2=K_3=0.$  . Then the trasfer matrix 
		$ \Theta=\Theta_{p,q} $ (\ref{Theta1})- (\ref{q}) is centrosymmetric,%centrally symmetrical		
		 and, at the eigenvector (\ref{eigenvector1}), corresponding to the largest eigenvalue $ F $, we assume $ b_3=b_2, b_4=1 .$ From the 8 equations of the system of equations (\ref{main_system1}) remain just the first four equations. Removing $F$ from this eqiations, we get the system of equations (\ref{main_system3} b, c, d ).
		After simplifying,  we have the following system of equations
		\begin{equation}\label{even_main_system1}  
			\begin{split} 
				\left\{  
				\begin{array}{rcl}  
					b_2 u_{01} u_{12} - u_{0123} u_{13} - b_2 u_{03} u_{23} + 
					u_{01} u_{0123} u_{03} u_{12} u_{13} u_{23} &= & 0,   \;\; \;\;   (b) \\					 			
					-b_2 u_{0123} u_{03} u_{12} + u_{01} u_{03} u_{13} +
					b_2 u_{01} u_{0123} u_{23} - 
					u_{12} u_{13} u_{23}  &= & 0,   \;\; \;\;   (c) \\ 			
					b_2^2 u_{01} u_{02} - b_2 u_{0123} u_{03} - u_{13} u_{23} + 
					b_2 u_{01} u_{0123} u_{02} u_{03} u_{13} u_{23} &= & 0 .   \;\; \;\;   (d) \\				
				\end{array}   
				\right.  
			\end{split} 
		\end{equation}
		Let us express $	u_{01}$ from (\ref{even_main_system1} (b)): 
		
		\begin{equation} \label{even_main_system_u01} 
			\begin{gathered} 
				u_{01}= (u_{0123} u_{13} + b_2 u_{03} u_{23})/
				(u_{12} (b_2 + u_{0123} u_{03} u_{13} u_{23}))    
				\;    
			\end{gathered}
		\end{equation}

		and substiture into the other  equations (\ref{even_main_system1} ). After simplifying and removung the positive denominator, we have the system
		
		\begin{equation}\label{even_main_system2}
			\begin{split} 
				\left\{  
				\begin{array}{rcl}  
					b_2^2 u_{0123} u_{03} u_{12}^2 - u_{0123} u_{03} u_{13}^2 - 
					b_2 u_{0123}^2 u_{13} u_{23} - \\
					b_2 u_{03}^2 u_{13} u_{23} + b_2 u_{12}^2 u_{13} u_{23} + 
					b_2 u_{0123}^2 u_{03}^2 u_{12}^2 u_{13} u_{23} - \\
					b_2^2 u_{0123} u_{03} u_{23}^2 + 
					u_{0123} u_{03} u_{12}^2 u_{13}^2 u_{23}^2  &= & 0,   \;\; \;\;   (c) \\	  			
					-b_2 u_{0123} u_{03} u_{12} + b_2 u_{0123} u_{02} u_{13} + \\
					b_2^2 u_{02} u_{03} u_{23} - 
					u_{12} u_{13} u_{23}  &= & 0.   \;\; \;\;   (d) \\	  			 				
				\end{array}   
				\right.  
			\end{split} 
		\end{equation}
		Let us express $u_{12}$  from (\ref{even_main_system2} (d)) $u_{12}$:
		\begin{equation} \label{even_main_system_u12}
			\begin{gathered} 
				u_{12} = (b_2 u_{0123} u_{02} u_{13} + b_2^2 u_{02} u_{03} u_{23})/(
				b_2 u_{0123} u_{03} + u_{13} u_{23})
				\;  ,  
			\end{gathered}
		\end{equation} 
		and substitute it into the   equation (\ref{even_main_system2} (c) ).  Let us express $u_{02}$ from them 
		
		\begin{equation} \label{even_main_system_u02}  
			\begin{gathered} 
				u_{02}^2 = ((u_{03} u_{13} + b_2 u_{0123} u_{23})
				(b_2 u_{0123} u_{03} + u_{13} u_{23}))/ \\
				(b_2^2 (u_{0123} u_{13} + b_2 u_{03} u_{23})
				(b_2 + u_{0123} u_{03} u_{13} u_{23}))			
				\; .  
			\end{gathered}
		\end{equation} 
		We use the following formulas in reserve order to calculate the free energy:
		(\ref{even_main_system_u02} ), 
		(\ref{even_main_system_u12} ),   (\ref{even_main_system_u01} ).

		Example. { The model with the invariant under inversion of all spin values Hamiltonian }.
		
		We define free parameters:
		
		$  K_{03}, K_{13}, K_{23}, K_{0123}, b_2.$ 
		
		Other parameters are calculated.    
		
		\begin{tabular}{ l l }	
			$	K_{03}  = -0.3 $ &{ $ K_{0123}  =0.123$ }\\
			{$	K_{13}  = 1.3$} &  {$ b_2  = 1.2$ }\\ 		
			$ K_{23}  =-2.3 $ & {$K_{02}  = -0.3725283757770392 $ }\\		
			$	K_{12}  = 1.1531594865705397$ &{ $ K_{01}  =0.14078181788637137$ }\\
			$F= 10.558325577247318 $ &{ $-f/ {(kT)}=2.3569147029450415$ 	}\\
		\end{tabular}
		
		%%%%%%%%%%%%%%%
		
		For this case the mapping of the set of squares of exponents of parameters of disorder solutions into theset of squares of exponents of parameters of independent parameters of the Hamiltonian can be written as 
		
		\begin{equation} \label{Jacob_even_1} 
			\begin{gathered}
				\overrightarrow{U}(	 u_{03}, u_{13}, u_{23}, u_{0123}, b_2)=
				(U_1, U_2, ...,U_{5}),	  		
			\end{gathered}
		\end{equation} 
		where 
		\begin{equation} \label{Jacob_even_2} 
			\begin{gathered}
				U_1=u_01 u_{23}, \;\;
				U_2=u_{02} u_{13},\;\;
				U_3=u_{12},\;\;
				U_4=u_{03},\;\;
				U_5= u_{0123}, \\		
				U_i=U_i(	u_3, u_{03}, u_{13}, u_{013}, u_{023}, b_3, b_4), \;\;
				i=1,2,...,5. 	
			\end{gathered}
		\end{equation}

		The Jacobi matrix of the mapping (\ref{Jacob_even_1}) - (\ref{Jacob_even_2}) 	calculated with formulas  (\ref{even_main_system_u02} ), 
		(\ref{even_main_system_u12} ),   (\ref{even_main_system_u01} ) at the point from the example below, can be represented as:
		
		\begin{equation} \label{Jacob_even_3}
			\begin{gathered} 
				J=(V_1 | V_2 )	,
			\end{gathered}
		\end{equation} 
		where $V_1, V_2 $ - matrices, written in the following form (we divide them into submatrices to fit on the page)

		\[ V_1= \left( \begin{array}{lcr}	
			-0.0025403138256135938 & 0.00003303852590813561 & 1.367201532914003 \\
			10.398179391746254 & 0.4897528449454569 & 21.23271967668927 \\
			0.5796669677593513 & 0.6656484883293956 & -104.51051465221184 \\
			1 & 0 & 0 \\
			0 & 0 & 0 \\	
		\end{array} \right ) ,\]

		\[ V_2= \left( \begin{array}{lccr}	
			0.009299925587437363 & -0.0003706850434243014 \\
			-0.5230448625326289 & -5.494920004345261 \\
			0.2620864396618572 & -7.468430724522079 \\
			0 & 0 \\
			1 & 0. \\				
		\end{array} \right ) .\]

		The rang of the Jacobi matrix $ J$ with tolerance 0.001 is equal to 4.
		Hence follows that in the general case the disorder solutions form the 4-dimensional subset of 10-dimensional space  of the parameters of the classical Hamiltonian (\ref{Hamiltonian_classic}), which is invariant under inversion of all spin values.  
		
		%%%%%%%%%%%%%%%

		%%%%%%%%%%%%%%%%%%%%%%%%%%%%%%%%%%%%%%%%%%%%%%%%%%%%%%%%%%%%%%%%%%%%%%%%%%%%%%%%%%%%%%%%%%%
		\section{Disorder solutions  for models on  the triangular lattice}\label{triangles}
		
		In this section we represent our model as the model set on the triangular lattice, considering two possible configurations of sepparating into triangular lattices:	
		
		\begin{multline} \label{triangles_zero}
			K_{0123}=K_{012}=K_{123}=K_{12} =0,   \;\; \;\;  (a)\\
			K_{0123}=K_{013}=K_{023}=K_{03}=0 .   \;\; \;\;  \;\; \;\;(b) \\	      
		\end{multline}
		
		\subsection{Case 1. (\ref{triangles_zero} (a))  }
		Let us consider the case (\ref{triangles_zero} (a)).
		Considering this conditions the formulas will be differ from the main scheme  (\ref{s12} - \ref{F}) in the calculating  $s_{12}$  (\ref{s12}). We substitute this conditions into (\ref{13_57}) and (\ref{1368_1324}). After simplifying and factorization we get the equation

		\begin{multline} \label{triangles_13_57} 
			b_3 u_3 u_{03} (-1 + u_{023} u_{23}) (1 + 
			u_{023} u_{23}) (-b_3 u_3 u_{013}^2 u_{023} u_{03}^2 u_{13} + 
			b_3 u_3 u_{013}^2 u_{02}^2 u_{023} u_{13}^3 -\\ b_3^2 u_{013} u_{023}^2 u_{03} u_{23} - 
			u_3^2 u_{013} u_{03} u_{13}^2 u_{23} + b_3^2 u_{013}^3 u_{02}^2 u_{03} u_{13}^2 u_{23} + \\
			u_3^2 u_{013}^3 u_{02}^2 u_{023}^2 u_{03} u_{13}^4 u_{23} - b_3 u_3 u_{023} u_{13} u_{23}^2 + 
			b_3 u_3 u_{013}^4 u_{02}^2 u_{023} u_{03}^2 u_{13}^3 u_{23}^2) = 0,  
		\end{multline}

		\begin{multline} \label{triangles_1368_1324}
			-b_3 u_3 u_{03} (u_{023} - u_{23}) (u_{023} + u_{23}) (b_3 u_3 u_{013}^4 u_{023} u_{03}^2 u_{13} - 
			b_3 u_3 u_{02}^2 u_{023} u_{13}^3 +\\ b_3^2 u_{013}^3 u_{023}^2 u_{03} u_{23} + 
			u_3^2 u_{013}^3 u_{03} u_{13}^2 u_{23} - b_3^2 u_{013} u_{02}^2 u_{03} u_{13}^2 u_{23} - \\
			u_3^2 u_{013} u_{02}^2 u_{023}^2 u_{03} u_{13}^4 u_{23} + 
			b_3 u_3 u_{013}^2 u_{023} u_{13} u_{23}^2 - 
			b_3 u_3 u_{013}^2 u_{02}^2 u_{023} u_{03}^2 u_{13}^3 u_{23}^2) =  0.  
		\end{multline} 
		
		The solution of the system of the equations (\ref{triangles_13_57}) -  (\ref{triangles_1368_1324}) are divided into four parts. Let us consider them below.
		
		\subsubsection{Case 1.1   }
		Let in the first case from (\ref{triangles_13_57}) -  (\ref{triangles_1368_1324}) be
		\begin{equation}\label{triangles_case1_1}  
			\begin{split} 
				\left\{  
				\begin{array}{rcl}  
					(-1 + u_{023} u_{23})   &= & 0,   \;\; \;\;    (a)\\
					(b_3 u_3 u_{013}^4 u_{023} u_{03}^2 u_{13} - 
					b_3 u_3 u_{02}^2 u_{023} u_{13}^3 + \\b_3^2 u_{013}^3 u_{023}^2 u_{03} u_{23} + 
					u_3^2 u_{013}^3 u_{03} u_{13}^2 u_{23} - \\ b_3^2 u_{013} u_{02}^2 u_{03} u_{13}^2 u_{23} - 
					u_3^2 u_{013} u_{02}^2 u_{023}^2 u_{03} u_{13}^4 u_{23} + \\
					b_3 u_3 u_{013}^2 u_{023} u_{13} u_{23}^2 - 
					b_3 u_3 u_{013}^2 u_{02}^2 u_{023} u_{03}^2 u_{13}^3 u_{23}^2)   &= & 0.\;\; \;\;    (b)
				\end{array}   
				\right.  
			\end{split} 
		\end{equation}
		Hence
		
		\begin{equation}\label{triangles_case1_2}   
			\begin{split} 
				\left\{  
				\begin{array}{rcl}  
					u_{23}   &= & 1/ u_{023},   \;\; \;\;    (a)\\
					u_{02}^2 &= &(b_3 u_{013}^3 u_{023}^2 u_{03} + u_3 u_{013}^2 u_{13})/
					(   b_3 u_{013} u_{03} u_{13}^2 + u_3 u_{023}^2 u_{13}^3)
					.\;\; \;\;    (b)
				\end{array}   
				\right.  
			\end{split} 
		\end{equation}
		Below we write the simplified formulas (from the main solution) considering the conditions (\ref{triangles_zero} (a))
		in the order of their use: firstly (\ref{triangles_case1_2}  $\;\; a, b$) , then 
		
		\begin{equation} \label{triangles_s01} 
			\begin{gathered}
				s_{01} = (b_3 u_{013}^3 + u_3 u_{03} u_{13})/(u_{013} (b_3 + u_3 u_{013} u_{03} u_{13})),  
			\end{gathered}
		\end{equation} 
		
		$u_{01} = s_{01}^{1/2}, $ $ R_{01} = s_{01}^{1/4},$
		
		\begin{equation} \label{triangles_b_2} 
			\begin{gathered}
				b_2 = ((b_4 u_{13}^2)/(b_3 u_{013}^2))   ,  
			\end{gathered}
		\end{equation} 
		
		\begin{equation} \label{triangles_u1}
			\begin{gathered}
				u_1 = (b_4 u_{01} u_{13})/(b_3 u_{013})   ,  
			\end{gathered}
		\end{equation} 
		
		\begin{equation} \label{triangles_u2} 
			\begin{gathered}
				u_2 = (1/(b_3 u_{02}))   ,  
			\end{gathered}
		\end{equation}

		\begin{equation} \label{triangles_u0}
			\begin{gathered}
				u_0 =  (b_3 u_{02} (b_3 u_{013} u_{03} + u_3 u_{023}^2 u_{13}))/(
				b_4 u_{01} u_{023} (b_3 + u_3 u_{013} u_{03} u_{13})),  
			\end{gathered}
		\end{equation} 
		
		$R_i=u_i, i \in I$,
		
		\begin{equation} \label{triangles_F} 
			\begin{gathered}
				F = (b_3  R_0 R_1 R_2 R_{01}  R_{02} )/(
				R_3  R_{013} R_{023} R_{03}  R_{13} R_{23}) + \\
				R_0 R_1 R_2 R_3 R_{01}  R_{013} R_{02} R_{023} R_{03}  R_{13} R_{23}.  
			\end{gathered}
		\end{equation} 
		Example. {Case 1.1}.
		
		We define the free parameters:
		
		$b_3, K_{013}, K_{023}, K_{03}, K_3,  K_{13},  b_4.$ 
		Other parameters are calculated.    
		
		\begin{tabular}{ l l }	
			$	K_{013}  =  2.08 $ &{ $ K_{023}  =-0.673$ }\\
			$	K_{03}  = 0.693 $ &{ $ b_2=0.0019120965685723366 $ }\\
			$ K_{3}  = -2.1714029607603003 $ & {$ K_{123}  = 0  $ }\\
			$	K_{13}  = 0.13 $ &{ $ K_{23}  =0.673 $ }\\
			${b_3  =0.3} $  &{ $b_4  = 1.4$ }\\ 
			$ K_{0123}  =0$ & {$ K_{012}  = 0 $ }\\
			$ K_{02}  =1.2778044500666297 $ & {$ K_{12}  =0 $ }\\
			$ K_{01}  =1.3964226716518735   $ & {$  K_0  =1.1900116040084041  $ }\\
			${K_1  =0.2166451921254478} $  &{ $ K_2  =-0.6758180479036618 $ }\\
			$F= 66.94063582728322$ &{ $- f/ {(kT)}=4.20380619420361$ 	}\\
		\end{tabular}
		
		The mapping of the set of squares of exponents of parameters of disorder solutions into theset of squares of exponents of parameters of independent parameters of the Hamiltonian (\ref{Hamiltonian_classic}) considering the conditions (\ref{triangles_zero} (a)) can be represented as   
		\begin{equation} \label{Jacob_triangle_1} 
			\begin{gathered}
				\overrightarrow{U}(	u_3, u_{03}, u_{13}, u_{013}, u_{023}, b_3, b_4)=
				(U_1, U_2, ...,U_{6}),	  		
			\end{gathered}
		\end{equation} 
		where 
		\begin{equation} \label{Jacob_triangle_2} 
			\begin{gathered}
				U_1=u_0 u_1 u_2 u_3, \;\;
				U_2=u_{01} u_{23},\;\;
				U_3=u_{02} u_{13},\;\;
				U_4=u_{03},\;\;
				U_5= u_{013}, \\
				U_6=u_{023},\;\;\\	
				U_i=U_i(	u_3, u_{03}, u_{13}, u_{013}, u_{023}, b_3, b_4), \;\;
				i=1,2,...,6. 	
			\end{gathered}
		\end{equation}

		The Jacobi matrix of the mapping (\ref{Jacob_triangle_1}) - (\ref{Jacob_triangle_2}), calculated with formulas (\ref{triangles_case1_2}  $\;\; a, b$)- (\ref{triangles_u0}) at the point from the example below, can be represented as:
		
		\begin{equation} \label{Jacob_triangle_main_3} 
			\begin{gathered} 
				J=(V_1 | V_2 )	,
			\end{gathered}
		\end{equation} 
		where $V_1, V_2 $ - matrices, written in the following form (we divide them into submatrices to fit on the page)

		\[ V_1= \left( \begin{array}{lcr}	
			0.28015735153638643 & 0.0009103627227058375 &  0.0028082049984001767 \\
			-2256.0254390207037 & -7.334241246326201 & -22.613655826120294\\
			2.0638785684923278 & -0.0067095768940816924 & 0.020687638269123454 \\
			0 & 1 & 0 \\
			0 & 0 & 0\\
			0  & 0 &  0\\
		\end{array} \right ) ,\]

		\[ V_2= \left( \begin{array}{lccr}	
			-0.0008183210394030471 & -0.21542162059393077 &	-0.012140151842143787 & 0 \\
			0.5213222316058364 & -241.01166704326715 & 97.76110203674193 & 0 \\
			0.26027964139530013 & 63.966616611565996 & -0.08943473872591312 & 0 \\
			0 & 0 & 0 & 0 \\
			1 & 0 & 0 & 0 \\
			0 & 1 & 0 & 0 . \\			
		\end{array} \right ) .\]

		The rang of the Jacobi matrix $ J$ with tolerance 0.001 is equal to 4.
		It follows that in the general case the disorder solutions form the 4-dimensional subset of 6-dimensional space of the parameters of the classical Hamiltonian (\ref{Hamiltonian_classic}) considering the following conditions (\ref{triangles_zero} (a)). At Verhagen's work \cite{Verhagen}, dedicated to an anisotropic triangular Ising model, the triple interactions are absent at the considering Hamiltonian's interaction, i.e. $ K_{013}=K_{023}=0$, and the disorder solutions form two-dimensional subset.

		%%%%%%%%%%%%%%%
		
		Case 1.1.1
		In addition to conditions of the considering case  $K_{023}=0$, let us consider a "checkerboard-triangular" lattice. 
		
		Then from (\ref{triangles_case1_1} a) and  $K_{23}=0$, in the formulas (\ref{triangles_case1_2})-(\ref{triangles_F}) we should take into consideration this conditions.
		Example. {Case 1.1.1}.

		We define the free parameters:
		$b_3, K_{013},  K_{03}, K_3,  K_{13},  b_4.$ 
		Other parameters are calculated. 
		
		\begin{tabular}{ l l }	
			$	K_{013}  =  2.08 $ &{ $K_{023}  =0  $ }\\
			$	K_{03}  = 0.693 $ &{ $ K_{123}  = 0 $ }\\
			$	K_{13}  = 0.13 $ &{ $K_{23}  =0  $ }\\
			${b_3  =0.3} $  &{ $b_4  = 1.4$ }\\ 
			$ K_{0123}  =0$ & {$ 	K_{012}  = 0   $ }\\
			$ K_{02}  =1.95 $ & {$ K_{12}  =0  $ }\\
			$ K_{01}  =1.3964226716518735   $ & {$ b_2  =0.0019120965685723366  $ }\\
			$ K_0  =1.1893093879038046 $ & {$ {K_1  =0.2166451921254478} $ }\\
			${K_2  =-1.3480135978370322 } $  & { $ K_{3}  = -2.1714029607603003 $ }\\
			$F= 66.89364553535397$ &{ $- f/ {(kT)}=4.20310397809901$ 	}\\
		\end{tabular}
		
		At  the "checkerboard-triangular" lattice let us compare the obtained results with the results from \cite{Wu1985}. Below we will add the top or bottom index $W$ to the notations from \cite{Wu1985} in order not to confuse thenwith the notation of this work. Let us write the formalas for calculating $F^W$ from \cite{Wu1985} in a convenient for comparison form.

		\begin{multline} \label{Wu2_case1_1}      
			A_W = \sinh(2 (K_3^W + K^W)), \;\;   C_W = \sinh(2 (K_3^W - K^W)),   \\
			B_W = \exp(2 K_3^W) \cosh(2 (K_1^W + K_2^W)) - \exp(-2 K_3^W) \cosh(2 (K_1^W - K_2^W)),\\  
		\end{multline}
		
		\begin{multline} \label{Wu3_case1_1}        \\
			t_1 = (-B_W + \sqrt{B_W*B_W - 4 A_W C_W})/(2 A_W),\\
			t_2 = (-B_W - \sqrt{B_W*B_W - 4 A_W C_W})/(2 A_W),\\ 
			w=t_1  \;\; or \;\; w=t_2, \\
			L'_W = Log[w]/2,\\
			F_W =( -2 (\sinh(2 K_1^W) \sinh(2 K_2^W) + \sinh(2 K^W) \sinh(2 L'_W))/
			\sinh(2 K_3^W))^{1/2}. \\
		\end{multline}
		This solution is valid along the trajectory
		
		\begin{multline} \label{Wu4_case1_1}        
			\sinh( L_W)=(\exp(-K_1^W-K_2^W) \sinh(2 L'_W+K_W)-\exp(K_1^W+K_2^W) \sinh(K^W))/\\
			(2 \cosh(2 (K_1^W + K_2^W)) + 2 \cosh(2 (K^W + L'_W)))^{1/2}.
		\end{multline}
		In the example for the case 1.1.1
		\begin{multline} \label{Wu1_case1_1}
			K_1^W = K_{02} + K_{13}, \; \;  
			K_2^W =K_{03} ,  \; \;\\	
			K_3^W = K_{01} + K_{23},   \; \; 
			K^W = K_{013} ,
			w=t_1.
		\end{multline}
		
		Then in the numerical example above for the "checkerboard-triangular" lattice we have $F_W=F=66.89364553534551$  and $\sinh( L_W)=\sinh( K_0+K_1+K_2+K_3)=-4.078013518091266$ . I.e. the parameters from the case1.1 match the parameters of the example, calculated at the article \cite{Wu1985}. This also means indirect confirmation of the coincidence in the thermodynamic limit of the considered model and the model from the work \cite{Wu1985}, where a periodic boundary condition is imposed in the horizontal direction only.

		Case 1.1.2. 
		In addition to conditions of the Case 1.1 $K_{013}=0$ let us consider a "checkerboard-triangular" lattice.
		In the formulas (\ref{triangles_case1_2})-(\ref{triangles_F}) let us additionally take this condition into account.
		
		Example. {Case 1.1.2}.

		We define the free parameters:
		$b_3, K_{023},  K_{03}, K_3,  K_{13},  b_4.$ 
		Other parameters are calculated.   
		
		\begin{tabular}{ l l }	
			$	K_{013}  =  0 $ &{ $K_{023}  =0.0673 $ }\\
			$	K_{03}  = -1 $ &{ $ b_2=0.2286251174348$ }\\
			$ K_{123}  = 0 $ & {$	K_{13}  = 0.13 $ }\\
			${b_3  =10.3} $  &{ $b_4  = 1.4$ }\\  	 
			$ K_{23}  =-0.0673 $ & {$K_{0123}  =0 $ }\\
			$	K_{012}  = 0 $ &{ $ K_{02}  =-0.06432734857193605$ }\\
			$	K_{12}  =0 $ &{ $ K_{01}  =0 $ } \\
			$ K_0  =-0.12604849793430253 $ & {$ K_1  =-0.8678358293071887$ }\\
			${K_2  =-1.101744599045859 } $  &{ $ K_{3}  = -2.1714029607603003$ }\\
			$F=24.873726536087936$ &{ $- f /{(kT)}=3.2138120872050444$ 	}\\
		\end{tabular}
		
		At  the "checkerboard-triangular" lattice let us compare the obtained results with the results from \cite{Wu1985}.	In the example for the case 1.1.2 
		\begin{multline} \label{Wu1_case1_2}
			K_1^W = K_{01} + K_{23}, \; \;  
			K_2^W =K_{03} ,  \; \;\\
			K_3^W = K_{02} + K_{13},   \; \;
			K^W = K_{023} ,
			w=t_1.
		\end{multline}
		
		Then in the numerical example above for the "checkerboard-triangular" lattice we have $F_W=F=24.873726536089315$ and the condition on the trajectory $\sinh( L_W)=\sinh( K_0+K_1+K_2+K_3)=-35.64782135613634$ . I.e. the parameters from the case1.1.2 match the parameters of the example, calculated at the article \cite{Wu1985}. 
		%%%%%%%%%%%%%%%%%%%%%%%%%%%%%%%%%%%%%%%%%%%%%%%%%%%%%%%%%%%%%
		
		\subsubsection{Case 2.    }
		Let in the second case from (\ref{triangles_13_57}) - (\ref{triangles_1368_1324}) be
		\begin{equation}\label{triangles_case2_1}
			\begin{split} 
				\left\{  
				\begin{array}{rcl}  
					(u_{023} - u_{23})  &= & 0,   \;\; \;\;    (a)\\
					(-b_3 u_3 u_{013}^2 u_{023} u_{03}^2 u_{13} + 
					b_3 u_3 u_{013}^2 u_{02}^2 u_{023} u_{13}^3 -\\ b_3^2 u_{013} u_{023}^2 u_{03} u_{23} - 
					u_3^2 u_{013} u_{03} u_{13}^2 u_{23} +\\ b_3^2 u_{013}^3 u_{02}^2 u_{03} u_{13}^2 u_{23} + 
					u_3^2 u_{013}^3 u_{02}^2 u_{023}^2 u_{03} u_{13}^4 u_{23} -\\ b_3 u_3 u_{023} u_{13} u_{23}^2 + 
					b_3 u_3 u_{013}^4 u_{02}^2 u_{023} u_{03}^2 u_{13}^3 u_{23}^2)  &= & 0.\;\; \;\;    (b)
				\end{array}   
				\right.  
			\end{split} 
		\end{equation}
		Hence
		
		\begin{equation}\label{triangles_case2_2} %bbb}  
			\begin{split} 
				\left\{  
				\begin{array}{rcl}  
					u_{23}   &= &  u_{023},   \;\; \;\;    (a)\\
					u_{02}^2 &= &((b_3 u_{023}^2 + u_3 u_{013} u_{03} u_{13})/(
					u_{013}^2 u_{13}^2 (b_3 + u_3 u_{013} u_{023}^2 u_{03} u_{13})))
					.\;\; \;\;    (b)
				\end{array}   
				\right.  
			\end{split} 
		\end{equation}
		Below we write the simplified formulas (from the main solution) considering the conditions ( $K_{0123}=K_{012}=K_{123}=K_{12}=0$)
		in the order of their use: firstly (\ref{triangles_case2_2}  $\;\; a, b$) , then 
		
		\begin{equation} \label{triangles_case2_2_s01} 
			\begin{gathered}
				s_{01} =  (u_{013} (b_3 u_{013} u_{03} + u_3 u_{13}))/(b_3 u_{03} + u_3 u_{013}^3 u_{13}),  
			\end{gathered}
		\end{equation} 
		
		$u_{01} = s_{01}^{1/2}, $ $ R_{01} = s_{01}^{1/4},$
		
		\begin{equation} \label{triangles_case2_2_b_2} 
			\begin{gathered}
				b_2 = (b_4 u_{013}^2 u_{13}^2)/b_3   ,  
			\end{gathered}
		\end{equation} 
		
		\begin{equation} \label{triangles_case2_2_u1} 
			\begin{gathered}
				u_1 = (b_4 u_{01} u_{13} (b_3 u_{03} + u_3 u_{013}^3 u_{13}))/(b_3 (b_3 u_{013} u_{03} + u_3 u_{13}))   ,  
			\end{gathered}
		\end{equation} 
		
		\begin{equation} \label{triangles_case2_2_u2} 
			\begin{gathered}
				u_2 = (b_3 u_{023}^2 + u_3 u_{013} u_{03} u_{13})/(b_3 u_{02} (b_3 + u_3 u_{013} u_{023}^2 u_{03} u_{13}))   ,  
			\end{gathered}
		\end{equation}

		\begin{equation} \label{triangles_case2_2_u0} 
			\begin{gathered}
				u_0 = (b_3 u_{02} u_{023} (b_3 u_{013} u_{03} + u_3 u_{13}))/(b_4 u_{01} (b_3 u_{023}^2 + 
				u_3 u_{013} u_{03} u_{13})),  
			\end{gathered}
		\end{equation} 
		
		$R_i=u_i, i \in I$,
		
		\begin{equation} \label{triangles_case2_2_F} 
			\begin{gathered}
				F = (b_3  R_0 R_1 R_2 R_{01}  R_{02} )/(
				R_3  R_{013} R_{023} R_{03}  R_{13} R_{23}) + \\
				R_0 R_1 R_2 R_3 R_{01}  R_{013} R_{02} R_{023} R_{03}  R_{13} R_{23}.  
			\end{gathered}
		\end{equation} 
		Example. {Case 1.2 }.
		
		We define the free parameters:
		
		$b_3, K_{013}, K_{023}, K_{03}, K_3,  K_{13},  b_4.$ 
		Other parameters are calculated.   
		
		\begin{tabular}{ l l }	
			$	K_{013}  =  2.08 $ &{ $K_{023}  =-0.673 $ }\\
			$	K_{03}  = 0.693 $ &{ $ b_2  =32223.298988324717 $ }\\
			$ K_{123}  = 0 $ & {$	K_{13}  = 0.13 $ }\\
			${b_3  =0.3} $  &{ $b_4  = 1.4$ }\\ 
			$ K_{23}  =-0.673 $ & {$  K_{0123}  =0 $ }\\
			$	K_{012}  = 0 $ &{ $K_{02}  =-1.7122367578969977 $ }\\
			$	K_{12}  =0 $ &{ $ K_{01}  =0.026195490573812423 $ }\\
			$ K_0  =-1.744475441933697 $ & {${K_1  =2.954027029899762}  $ }\\
			${K_2  =3.3097496442659704} $  &{ $  K_{3}  = -2.1714029607603003 $ }\\
			$F= 18.62487078994398$ &{ $-f/ {(kT)}=2.924497826788887 $ 	}\\
		\end{tabular}

		%%%%%%%%%%%%%%%%%%%%%%%%%%%%%%%%%%%%%%%%%%%%%%%%%%%%%%%%%%%%%
		
		\subsubsection{Case 3.   }
		Let in the third case from (\ref{triangles_13_57}) - (\ref{triangles_1368_1324}) be
		\begin{equation}\label{triangles_case3_1}  
			\begin{split} 
				\left\{  
				\begin{array}{rcl}  
					(-b_3 u_3 u_{013}^2 u_{023} u_{03}^2 u_{13} + b_3 u_3 u_{013}^2 u_{02}^2 u_{023} u_{13}^3 -\\ 
					b_3^2 u_{013} u_{023}^2 u_{03} u_{23} - u_3^2 u_{013} u_{03} u_{13}^2 u_{23} + \\
					b_3^2 u_{013}^3 u_{02}^2 u_{03} u_{13}^2 u_{23} + 
					u_3^2 u_{013}^3 u_{02}^2 u_{023}^2 u_{03} u_{13}^4 u_{23} - \\b_3 u_3 u_{023} u_{13} u_{23}^2 + 
					b_3 u_3 u_{013}^4 u_{02}^2 u_{023} u_{03}^2 u_{13}^3 u_{23}^2) &= & 0,   \;\; \;\;    (a)\\   
					(b_3 u_3 u_{013}^4 u_{023} u_{03}^2 u_{13} - b_3 u_3 u_{02}^2 u_{023} u_{13}^3 + \\
					b_3^2 u_{013}^3 u_{023}^2 u_{03} u_{23} + u_3^2 u_{013}^3 u_{03} u_{13}^2 u_{23} - \\
					b_3^2 u_{013} u_{02}^2 u_{03} u_{13}^2 u_{23} - 
					u_3^2 u_{013} u_{02}^2 u_{023}^2 u_{03} u_{13}^4 u_{23} + \\
					b_3 u_3 u_{013}^2 u_{023} u_{13} u_{23}^2 - 
					b_3 u_3 u_{013}^2 u_{02}^2 u_{023} u_{03}^2 u_{13}^3 u_{23}^2) &= & 0.   \;\; \;\;    (b)\\ 
				\end{array}   
				\right.  
			\end{split} 
		\end{equation}
		Hence
		
		\begin{equation}\label{triangles_case3_2} 
			\begin{split} 
				\left\{  
				\begin{array}{l}  
					(u_{02}^2) = ((
					b_3 u_3 u_{013}^2 u_{023} u_{03}^2 u_{13} + b_3^2 u_{013} u_{023}^2 u_{03} u_{23} +  \\
					u_3^2 u_{013} u_{03} u_{13}^2 u_{23} + b_3 u_3 u_{023} u_{13} u_{23}^2)/\\(
					b_3 u_3 u_{013}^2 u_{023} u_{13}^3 + b_3^2 u_{013}^3 u_{03} u_{13}^2 u_{23} + \\
					u_3^2 u_{013}^3 u_{023}^2 u_{03} u_{13}^4 u_{23} + 
					b_3 u_3 u_{013}^4 u_{023} u_{03}^2 u_{13}^3 u_{23}^2))  
					% (*1\[Equal]3&&5\[Equal]7*)
					,   \;\; \;\;    (a)\\
					(u_{02}^2) = ((
					b_3 u_3 u_{013}^4 u_{023} u_{03}^2 u_{13} + b_3^2 u_{013}^3 u_{023}^2 u_{03} u_{23} + \\
					u_3^2 u_{013}^3 u_{03} u_{13}^2 u_{23} + b_3 u_3 u_{013}^2 u_{023} u_{13} u_{23}^2)/\\(
					b_3 u_3 u_{023} u_{13}^3 + b_3^2 u_{013} u_{03} u_{13}^2 u_{23} + \\
					u_3^2 u_{013} u_{023}^2 u_{03} u_{13}^4 u_{23} + 
					b_3 u_3 u_{013}^2 u_{023} u_{03}^2 u_{13}^3 u_{23}^2))
					.\;\; \;\;    (b)
				\end{array}   
				\right.  
			\end{split} 
		\end{equation}
		Equating the rigth parts of the equation (\ref{triangles_case3_2}  $\;\; a, b$), we get
		\begin{equation} \label{triangles_case3_u_{013}} 
			\begin{gathered}
				u_{013} = 1.  
			\end{gathered}
		\end{equation} 
		This means a solution at a "checkerboard-triangular" lattice (\cite{Wu1985}).
		Below we write the simplified formulas (from the main solution) considering the conditions ( $K_{0123}=K_{012}=K_{123}=K_{12}=K_{013}=0$) и  (\ref{triangles_case3_u_{013}}), (\ref{triangles_case3_2}  $\;\; a, b$)
		in the order of their use:  
		
		\begin{equation} \label{triangles_case3_u_{02}}
			\begin{gathered}
				u_{02}^2 = (((u_3 u_{03} u_{13} + b_3 u_{023} u_{23}) (b_3 u_{023} u_{03} + u_3 u_{13} u_{23}))/\\(
				u_{13}^2 (u_3 u_{023} u_{13} + b_3 u_{03} u_{23}) (b_3 + u_3 u_{023} u_{03} u_{13} u_{23}))).  
			\end{gathered}
		\end{equation} 
		
		After simplifying we have:
		
		\begin{equation} \label{triangles_case3_s01}  
			\begin{gathered}
				s_{01} = 1   ,  
			\end{gathered}
		\end{equation} 
		
		\begin{equation} \label{triangles_case3_b_2}  
			\begin{gathered}
				b_2 = ((b_4 u_{13}^2)/b_3)  ,  
			\end{gathered}
		\end{equation} 
		
		\begin{equation} \label{triangles_case3_u1} 
			\begin{gathered}
				u_1 = (b_4 u_{13})/b_3  ,  
			\end{gathered}
		\end{equation} 
		
		\begin{equation} \label{triangles_case3_u2}
			\begin{gathered}
				u_2 =  ((u_3 u_{03} u_{13} + b_3 u_{023} u_{23})/(b_3 u_{02} (b_3 + u_3 u_{023} u_{03} u_{13} u_{23})))   ,  
			\end{gathered}
		\end{equation}

		\begin{equation} \label{triangles_case3_u0}
			\begin{gathered}
				u_0 =  (b_3 u_{02} (u_3 u_{023} u_{13} + b_3 u_{03} u_{23}))/(
				b_4 (u_3 u_{03} u_{13} + b_3 u_{023} u_{23})),  
			\end{gathered}
		\end{equation}

		\begin{equation} \label{triangles_case3_F}
			\begin{gathered}
				F = (b_3  R_0 R_1 R_2 R_{01}  R_{02} )/(
				R_3  R_{013} R_{023} R_{03}  R_{13} R_{23}) + \\
				R_0 R_1 R_2 R_3 R_{01}  R_{013} R_{02} R_{023} R_{03}  R_{13} R_{23}.  
			\end{gathered}
		\end{equation} 
		Example. {Case 1.3}.
		
		We define the free parameters:
		
		$b_3,  K_{023}, K_{23}, K_{03}, K_3,  K_{13},  b_4.$ 
		Other parameters are calculated. 
		
		\begin{tabular}{ l l }
			
			$	K_{013}  = 0 $ &{ $ K_{023}  =-0.673$ }\\
			$	K_{03}  = 0.693 $ &{ $K_{01}  =0 $ }\\
			$ K_{123}  = 0 $ & {$b_2  = 7.849462365261467$ }\\
			$	K_{13}  = 0.13 $ &{ $K_{23}  =0.23 $ }\\
			${b_3  =0.3} $  &{ $b_4  = 1.4$ }\\ 
			$ K_{0123}  =0$ & {$	K_{012}  = 0 $ }\\
			$ K_{02}  =-0.6964439335507264 $ & {$	K_{12}  =0$ }\\
			$ K_0  =-0.3170491689085145 $ & {${K_1  =0.9002225204735743} $ }\\
			${K_2  =1.028659422643789} $  &{ $K_{3}  = -2.1714029607603003 $ }\\
			$F= 4.9107802487716$ &{ $-f/ {(kT)}=1.5914328393183328$ 	}\\
		\end{tabular}

		Let us compare the obtained results with the results from \cite{Wu1985}.

		\begin{multline} \label{Wu2_case3}      
			A_W = \sinh(2 (K_3^W + K^W)), \;\;   C_W = \sinh(2 (K_3^W - K^W)),   \\
			B_W = \exp(2 K_3^W) \cosh(2 (K_1^W + K_2^W)) - \exp(-2 K_3^W) \cosh(2 (K_1^W - K_2^W)),\\  
		\end{multline}
		
		\begin{multline} \label{Wu3_case3}        
			t_1 = (-B_W + \sqrt{B_W*B_W - 4 A_W C_W})/(2 A_W),\\
			t_2 = (-B_W - \sqrt{B_W*B_W - 4 A_W C_W})/(2 A_W),\\ 
			w=t_1  \;\; or \;\; w=t_2, \\
			L'_W = Log[w]/2,\\  
			F_W =( -2 (\sinh(2 K_1^W) \sinh(2 K_2^W) + \sinh(2 K^W) \sinh(2 L'_W))/
			\sinh(2 K_3^W))^{1/2}. \\ 
		\end{multline}
		This solution is valid along the trajectory
		\begin{multline} \label{Wu4_case3}        
			\sinh( L_W)=(\exp(-K_1^W-K_2^W) \sinh(2 L'_W+K_W)-\exp(K_1^W+K_2^W) \sinh(K^W))/\\
			(2 \cosh(2 (K_1^W + K_2^W)) + 2 \cosh(2 (K^W + L'_W)))^{1/2}. 
		\end{multline}
		In the example for the case 1.3
		\begin{multline} \label{Wu1_case3}
			K_1^W = K_{01} + K_{23}, \; \; 
			K_2^W =K_{03} ,  \; \;\\
			% (*  Log[K3]   *)
			K_3^W = K_{02} + K_{13},   \; \;
			% (*     Log[K1]   *)    
			K^W = K_{023} ,
			w=t_2.
		\end{multline}
		
		Then in the numerical example above for the "checkerboard-triangular" lattice we have $F_W=F=4.910780248771596$ and $\sinh( L_W)=\sinh( K_0+K_1+K_2+K_3)=-0.589232784720577$ . I.e. the parameters from the case3 match the parameters of the example, calculated at the article \cite{Wu1985}.

		%%%%%%%%%%%%%%%%%%%%%%%%%%%%%%%%%%%%%%%%%%%%%%%%%%%%%%%%%%%%%

		\subsubsection{Case 4.   }
		Let in the fourth case from (\ref{triangles_13_57}) -  (\ref{triangles_1368_1324}) be
		\begin{equation}\label{triangles_case1_4_1} 
			\begin{split} 
				\left\{  
				\begin{array}{rcl}  
					(-1 + u_{023} u_{23})   &= & 0,   \;\; \;\;    (a)\\ 
					(u_{023} - u_{23})  &= & 0.     \;\; \;\;    (b)\\    
				\end{array}   
				\right.  
			\end{split} 
		\end{equation}
		Hence
		
		$K23 =0,  K023 = 0$.
		
		This means a solution at a "checkerboard-triangular" lattice (\cite{Wu1985}).
		Below we write the simplified formulas (from the main solution) considering the conditions ( $K_{0123}=K_{012}=K_{123}=K_{12}=K_{013}=0$) и  (\ref{triangles_case3_u_{013}}), (\ref{triangles_case3_2}  $\;\; a, b$)
		in the order of their use:   
		
		\begin{equation} \label{triangles_case1_3_u_{02}} 
			\begin{gathered}
				s_{01} = ((b_3 u_{013} u_{03} + u_3 u_{13}) (u_3 u_{03} + b_3 u_{013} u_{02}^2 u_{13}))/\\
				((b_3 + 
				u_3 u_{013} u_{03} u_{13}) (u_3 u_{013} + 
				b_3 u_{02}^2 u_{03} u_{13})) 	,  
			\end{gathered}
		\end{equation}

		\begin{equation} \label{triangles_case1_4_b_2} 
			\begin{gathered}
				b_2 = b_4/(b_3 u_{02}^2)  ,  
			\end{gathered}
		\end{equation} 
		
		\begin{equation} \label{triangles_case1_4_u1} 
			\begin{gathered}
				u_1 = ((b_4 u_{01} (u_3 u_{013} + b_3 u_{02}^2 u_{03} u_{13}))/\\(
				b_3 u_{02}^2 (b_3 u_{013} u_{03} + u_3 u_{13}))) 
				%  (*   6\[Equal]8   *)  
				,  
			\end{gathered}
		\end{equation} 
		
		\begin{equation} \label{triangles_case1_4_u2}
			\begin{gathered}
				u_2 = (1/(b_3 u_{02}))	,  
			\end{gathered}
		\end{equation}

		\begin{equation} \label{triangles_case1_4_u0} 
			\begin{gathered}
				u_0 = ((b_3 u_{02} (b_3 u_{013} u_{03} + u_3 u_{13}))/\\
				(
				b_4 u_{01} (b_3 + u_3 u_{013} u_{03} u_{13}))) . 
			\end{gathered}
		\end{equation}

		Example. {Case 1.4}.
		
		We define the free parameters:
		
		$b_3,  K_{123}, K_{02}, K_{13}, K_3,  K_{23},  b_4.$ 
		Other parameters are calculated.   
		
		\begin{tabular}{ l l }	
			$	K_{013}  =2.08 $ &{ $K_{023}  =0 $ }\\
			$	K_{03}  = 0.693 $ &{ $K_{01}  =1.3428230030246964 $ }\\
			{$K_{123}  = 0 $} &  {$ b_2  = 2.0968684992137008$ }\\ 
			$	K_{13}  = 0.13 $ &{ $K_{23}  =0  $ }\\
			${b_3  =0.3} $  &{ $b_4  = 1.4$ }\\ 
			$ K_{0123}  =0$ & {$	K_{012}  = 0 $ }\\
			$ K_{02}  =0.2 $ & {$K_{12}  =0$ }\\
			$ K_0  =-0.5070909434690185$ & {$K_1  =0.270712711449207 $ }\\
			${K_2  =0.40198640216296805} $  &{ $K_{3}  = -2.1714029607603003 $ }\\	
			$F= 12.2701747100922$ &{ $- f /{(kT)}=2.5071714974227697$ 	}\\
		\end{tabular}
		
	Let us compare the obtained results with the results from \cite{Wu1985}.
		
		\begin{multline} \label{Wu1_case1_4}
			K_1^W = K_{02} + K_{13}, \; \;   % (*  Log[K2]   *)
			K_2^W =K_{03} ,  \; \;
			K_3^W = K_{01} + K_{23},   \; \;   
			K^W = K_{013} ,\\
		\end{multline}
		
		\begin{multline} \label{Wu2}      
			A_W = \sinh(2 (K_3^W + K^W)), \;\;   C_W = \sinh(2 (K_3^W - K^W)) ,  \\
			B_W = \exp(2 K_3^W) \cosh(2 (K_1^W + K_2^W)) - \exp(-2 K_3^W) \cosh(2 (K_1^W - K_2^W)) ,\\  
		\end{multline}
		
		\begin{multline} \label{Wu3}        
			t_1 = (-B_W + \sqrt{B_W*B_W - 4 A_W C_W})/(2 A_W)\,\
			t_2 = (-B_W - \sqrt{B_W*B_W - 4 A_W C_W})/(2 A_W),\\ 
			w=t_1  \;\; or \;\; w=t_2, \\
			L'_W = Log[w]/2,\\ 
			F_W =( -2 (\sinh(2 K_1^W) \sinh(2 K_2^W) + \sinh(2 K^W) \sinh(2 L'_W))/
			\sinh(2 K_3^W))^{1/2}. \\
		\end{multline}
		This solution is valid along the trajectory
		
		\begin{multline} \label{Wu4}        
			\sinh( L_W)=(\exp(-K_1^W-K_2^W) \sinh(2 L'_W+K_W)-\exp(K_1^W+K_2^W) \sinh(K^W))/\\
			(2 \cosh(2 (K_1^W + K_2^W)) + 2 \cosh(2 (K^W + L'_W)))^{1/2}. 
		\end{multline}

		Then in the numerical example above for the "checkerboard-triangular" lattice we have  $F_W=F$ (we assume $w=t_1$ in the example), а $\sinh( L_W)=\sinh( K_0+K_1+K_2+K_3)$. I.e. the parameters from the case1.4 match the parameters of the example, calculated at the article \cite{Wu1985}.  
		
		In the example for case 1.4 we use (\ref{Wu2_case3} -\ref{Wu4_case3}) with the following parameters
		\begin{multline} \label{Wu1_case4}
			K_1^W = K_{02} + K_{13}, \; \;  
			K_2^W =K_{03} ,  \; \;
			K_3^W = K_{01} + K_{23},   \; \;    
			K^W = K_{013} ,\\
			w=t_1.  
		\end{multline}
		
		Then in the numerical example above (case1.4) for the "checkerboard - triangular" lattice we have  $F_W=F=12.270174710092201$ and $\sinh( L_W)=\sinh( K_0+K_1+K_2+K_3)=-3.648722560578193$ . Hence the parameters from the example founded in case2 are equal to the parameters from the example, founded in the article \cite{Wu1985}.

		%%%%%%%%%%%%%%%%%%%%%%%%%%%%%%%%%%%%%%%%%%%%%%%%%%%%%%%%%%%%%
		\subsection{Case 2. (\ref{triangles_zero} (b)) }
		
		Let us assume $K_{0123}=K_{013}=K_{023}=K_{03}=0$
		Let us substitute this conditions into (\ref{quadratic_equation}) with coefficients $A$ (\ref{coefficient_A}), $B$ (\ref{coefficient_B}), $C$ (\ref{coefficient_C}), or, that is the same, into the expanded form of this equation (\ref{for_s02}) . After simplifying and factorization we get  the only solution accepteble to us.
		
		\begin{equation}\label{triangles_case2_u_{012}} 
			\begin{array}{rcl}  
				u_{012}=1.
			\end{array}  
		\end{equation}

		This means a solution at a "checkerboard-triangular" lattice (\cite{Wu1985}).
		Below we write the simplified formulas (from the main solution) considering the conditions
		(\ref{triangles_case2_u_{012}})  
		in the order of their use:  
		
		\begin{equation} \label{triangles_case2_s12}
			\begin{gathered}
				s12 = (((b_3 u_{02}^2 u_{123} u_{13} + u_3 u_{23}) (u_3 u_{13} + 
				b_3 u_{123} u_{23}))/\\
				((u_3 u_{123} + b_3 u_{02}^2 u_{13} u_{23}) (b_3 + 
				u_3 u_{123} u_{13} u_{23})) )    ,  
			\end{gathered}
		\end{equation}

		\begin{equation} \label{triangles_case2_s01}
			\begin{gathered}
				s01 = 1  ,  
			\end{gathered}
		\end{equation} 
		
		\begin{equation} \label{triangles_case2_b_2}
			\begin{gathered}
				b_2 = b_4/(b_3 u_{02}^2)  ,  
			\end{gathered}
		\end{equation} 
		
		\begin{equation} \label{triangles_case2_u_1}
			\begin{gathered}
				u_1 = (b_4 u_{12} (u_3 u_{123} + b_3 u_{02}^2 u_{13} u_{23}))/(
				b_3 u_{02}^2 (u_3 u_{13} + b_3 u_{123} u_{23}))  ,  
			\end{gathered}
		\end{equation}

		\begin{equation} \label{triangles_case2_u_2} 
			\begin{gathered}
				u_2 = (u_3 u_{13} + b_3 u_{123} u_{23})/(b_3 u_{02} u_{12} (b_3 + u_3 u_{123} u_{13} u_{23}))  ,  
			\end{gathered}
		\end{equation}

		\begin{equation} \label{triangles_case2_u_0}
			\begin{gathered}
				u_0 = (b_3 u_{02})/b_4  .  
			\end{gathered}
		\end{equation}

		We define the free parameters:
		
		$b_3,  K_{123}, K_{02}, K_{13}, K_3,  K_{23},  b_4.$ 
		Other parameters are calculated.    
		
		\begin{tabular}{ l l }	
			$	K_{013}  =0 $ &{ $K_{023}  =0$ }\\
			$	K_{03}  = 0 $ &{ $K_{123}  = 0.123$ }\\
			{$K_{23}  =0.23 $} &  {$ b_2  = 0.061073839782923314$ }\\ 
			$	K_{13}  = 0.13 $ &{ $K_{01}  =0  $ }\\
			${b_3  =10.3} $  &{ $b_4  = 1.4$ }\\ 
			$ K_{0123}  =0$ & {$	K_{012}  = 0 $ }\\
			$ K_{02}  =0.2 $ & {$K_{12}  =0.10973601177580669$ }\\
			$ K_0  =1.1978358293071885$ & {$K_1  =-0.8865014650838456$ }\\
			${K_2  =-1.1515820240539831} $  &{ $K_{3}  = -0.5807760442209919 $ }\\		
			$F= 7.2154440545359195$ &{ $-f/ {(kT)}=1.97622373635218$ 	}\\
		\end{tabular}
		
		At  the "checkerboard-triangular" lattice let us compare the obtained results with the results from \cite{Wu1985}. 
		\begin{multline} \label{Wu1}
			K_2^W = K_{02} + K_{13}, \; \;  
			K_3^W =K_{12} ,  \; \;
			K_1^W = K_{01} + K_{23},   \; \;  
			K^W = K_{123} .\\ 
		\end{multline}

		Then in the numerical example above for the "checkerboard-triangular" lattice we have  $F_W=F$ (we assume $w=t_1$ in the example), а $\sinh( L_W)=\sinh( K_0+K_1+K_2+K_3)$. I.e. the parameters from the case2 match the parameters of the example, calculated at the article \cite{Wu1985}.  
		In the example for case 1.4 we use (\ref{Wu2_case3} -\ref{Wu4_case3}) with the following parameters
		
		\begin{multline} \label{Wu1_case2}
			K_1^W = K_{01} + K_{23}, \; \;  
			K_2^W = K_{02} + K_{13},  \; \;
			K_3^W =K_{12} ,    \; \;
			K^W = K_{123} ,\\
			w=t_1.
		\end{multline}
		
		Then in the numerical example above (case2) for the "checkerboard-triangular" lattice we have  $F_W=F=7.2154440545359195$ and $\sinh( L_W)=\sinh( K_0+K_1+K_2+K_3)=-1.9499455479235897$ . Hence the parameters from the example founded in case2 are equal to the parameters from the example, founded in the article \cite{Wu1985}.

		\section{ Solving a system of equations for finding the maximum eigenvalue of a transfer matrix}\label{solution_main_system}
		
		The solution scheme for the system of equations (\ref{main_system2}) is following:
		Variable $ F $ from (\ref{main_system2} a) we substite into other equations (\ref{main_system2}(b))-(\ref{main_system2}(h)) of the system (\ref{main_system2}). 
		\begin{equation}\label{main_system3} 
			\begin{split} 
				\left\{  
				\begin{array}{rcl}  
					b_2 Right(a) &= & Right(b),   \; \;\; (b)2\\
					Right(a) &= & Right(c),  \; \;\; (c)3\\
					Right(b) &= & Right(d),   \; \;\; (d)4\\
					b_3 Right(a) &= & Right(e),     \; \;\; (e)5\\
					b_4 Right(a) &= & Right(f),    \; \; \;  (f)6\\
					Right(e) &= & Right(g),   \; \;\; (g)7\\
					Right(f) &= & Right(h),   \; \;\; (h)8\\
				\end{array}   
				\right.  
			\end{split} 
		\end{equation}
		
		where  $ Right(i) $, $ i=a,\dots,h $ are the right parts of the equation $ i $ of the system (\ref{main_system2}).
		
		After multiplying by positive denominator and simplifying of the system (\ref{main_system3}) 
		
		\begin{equation}\label{main_system4}  
			\begin{split} 
				\left\{  
				\begin{array}{rcl}  
					b_2 b_3  h_0 u_{01} u_{012} u_{02} - b_3  u_{0123} u_{013} u_{023} u_{03} -  h_3 u_{123} u_{13} u_{23} + \\
					b_2 h_0 h_3 u_{01} u_{012} u_{0123} u_{013} u_{02} u_{023} u_{03} u_{123} u_{13} u_{23} &= & 0,   \;\; \;\;   (b) \\
					b_3  h_1 u_{01} u_{012} u_{12} - b_4  u_{0123} u_{013} u_{123} u_{13} - b_2 h_3 u_{023} u_{03} u_{23} + \\
					h_1 h_3 u_{01} u_{012} u_{0123} u_{013} u_{023} u_{03} u_{12} u_{123} u_{13} u_{23} &= & 0,   \;\; \;\;   (c)   \\
					b_3  h_1 u_{0123} u_{013} u_{023} u_{03} u_{12} - b_4  u_{01} u_{012} u_{023} u_{03} u_{123} u_{13} - \\
					b_2 h_3 u_{01} u_{012} u_{0123} u_{013} u_{23} + 
					h_1 h_3 u_{12} u_{123} u_{13} u_{23} &= & 0 ,   \;\; \;\;   (d)\\
					b_3^2  h_2 u_{012} u_{02} u_{12} - h_3 u_{013} u_{03} u_{13} - b_3  u_{0123} u_{023} u_{123} u_{23} + \\
					b_3 h_2 h_3 u_{012} u_{0123} u_{013} u_{02} u_{023} u_{03} u_{12} u_{123} u_{13} u_{23} &= & 0,   \;\; \;\;    (e)\\
					b_3 b_4  h_0 h_2 u_{01} u_{12} - h_3 u_{0123} u_{023} u_{13} - b_3  u_{013} u_{03} u_{123} u_{23} + \\
					b_4 h_0 h_2 h_3 u_{01} u_{0123} u_{013} u_{023} u_{03} u_{12} u_{123} u_{13} u_{23} &= & 0  ,   \;\; \;\;   (f)\\
					-b_2 h_3 u_{012} u_{0123} u_{03} u_{12} u_{123} + h_1 h_3 u_{01} u_{013} u_{03} u_{13} + \\
					b_3  h_1 u_{01} u_{0123} u_{023} u_{123} u_{23} - 
					b_4  u_{012} u_{013} u_{023} u_{12} u_{13} u_{23} &= & 0,   \;\; \;\;    (g)\\
					-b_2 h_3 u_{01} u_{013} u_{023} u_{12} u_{123} + h_1 h_3 u_{012} u_{0123} u_{023} u_{13} + \\
					b_3  h_1 u_{012} u_{013} u_{03} u_{123} u_{23} - 
					b_4  u_{01} u_{0123} u_{03} u_{12} u_{13} u_{23} &= & 0,   \;\; \;\;     (h)
				\end{array}   
				\right.  
			\end{split} 
		\end{equation}

		Let us solve the equation (\ref{main_system4} $\;\;b , f$) with respect to $ u_0 $ and equate their rigth sides. 
		
		Сonsidering that denominators are positive in the obtained equations, we get the following equations after simplifying. 
		\begin{equation} \label{main_system4_u0_b_f}  
			\begin{gathered} 
				u_0 = ((b_3 u_{0123} u_{013} u_{023} u_{03} + u_3 u_{123} u_{13} u_{23})/\\
				( b_2 u_{01} u_{012} u_{02} (b_3 + 
				u_3 u_{0123} u_{013} u_{023} u_{03} u_{123} u_{13} u_{23}))) \;\;\;  (b)\\
				u_0 = ((u_3 u_{0123} u_{023} u_{13} + b_3 u_{013} u_{03} u_{123} u_{23})/\\
				( b_4 u_2 u_{01} u_{12} (b_3 + 
				u_3 u_{0123} u_{013} u_{023} u_{03} u_{123} u_{13} u_{23})))  \;\;\;  (f)\\
				b_3 b_4 u_2 u_{0123} u_{013} u_{023} u_{03} u_{12} - b_2 u_3 u_{012} u_{0123} u_{02} u_{023} u_{13} - \\
				b_2 b_3 u_{012} u_{013} u_{02} u_{03} u_{123} u_{23} + 
				b_4 u_2 u_3 u_{12} u_{123} u_{13} u_{23} = 0   \;  (bf),  
			\end{gathered}
		\end{equation}

		After substituting $u_2 $ from the equations (\ref{main_system4_u0_b_f} $\;\; bf$) and  (\ref{main_system4} $\;\; e$), we equate the obtained rigth sides.
		
		Сonsidering that denominators are positive in the obtained equations, we get the following equations after simplifying.  
		\begin{equation} \label{main_system4_u2_bfe} 
			\begin{gathered} 
				u_2 = ((b_2 u_3 u_{012} u_{0123} u_{02} u_{023} u_{13} + 
				b_2 b_3 u_{012} u_{013} u_{02} u_{03} u_{123} u_{23})/\\
				(b_4 u_{12} (b_3 u_{0123} u_{013} u_{023} u_{03} + 
				u_3 u_{123} u_{13} u_{23})))  \; \; \; \; \; \; \; \; \;  (bf)\\
				u_2 == ((u_3 u_{013} u_{03} u_{13} + b_3 u_{0123} u_{023} u_{123} u_{23})/\\
				(b_3 u_{012} u_{02} u_{12} (b_3 + 
				u_3 u_{0123} u_{013} u_{023} u_{03} u_{123} u_{13} u_{23})))\; \; \;  \; \; \;   (g)\\
				(-b_2 b_3^2 u_3 u_{012}^2 u_{0123} u_{02}^2 u_{023} u_{13} + 
				b_3 b_4 u_3 u_{0123} u_{013}^2 u_{023} u_{03}^2 u_{13} - \\
				b_2 b_3^3 u_{012}^2 u_{013} u_{02}^2 u_{03} u_{123} u_{23} + 
				b_3^2 b_4 u_{0123}^2 u_{013} u_{023}^2 u_{03} u_{123} u_{23} + \\
				b_4 u_3^2 u_{013} u_{03} u_{123} u_{13}^2 u_{23} - 
				b_2 b_3 u_3^2 u_{012}^2 u_{0123}^2 u_{013} u_{02}^2 u_{023}^2 u_{03} u_{123} u_{13}^2 u_{23} + \\
				b_3 b_4 u_3 u_{0123} u_{023} u_{123}^2 u_{13} u_{23}^2 - 
				b_2 b_3^2 u_3 u_{012}^2 u_{0123} u_{013}^2 u_{02}^2 u_{023} u_{03}^2 u_{123}^2 u_{13} u_{23}^2)= 0 \; \; \; (bfe).  
			\end{gathered}
		\end{equation} 
		
		Let us solve the equation (\ref{main_system4} $\;\;c , d, g, h$) with respect to $ u_1 $ .	
		After simplifying we have the following equations. 
		
		\begin{equation} \label{main_system4_u1_cdgh} 
			\begin{gathered} 
				u_1 = ((b_4 u_{0123} u_{013} u_{123} u_{13} + b_2 u_3 u_{023} u_{03} u_{23})/\\
				(u_{01} u_{012} u_{12} (b_3 + 
				u_3 u_{0123} u_{013} u_{023} u_{03} u_{123} u_{13} u_{23})))  \; \; \;  \; \; \;   ( c)\\
				u_1 = ((b_4 u_{01} u_{012} u_{023} u_{03} u_{123} u_{13} + 
				b_2 u_3 u_{01} u_{012} u_{0123} u_{013} u_{23})/\\
				(u_{12} (b_3 u_{0123} u_{013} u_{023} u_{03} + 
				u_3 u_{123} u_{13} u_{23}))) \; \; \;\; \; \;  (d)\\
				u_1 = ((b_2 u_3 u_{012} u_{0123} u_{03} u_{12} u_{123} + 
				b_4 u_{012} u_{013} u_{023} u_{12} u_{13} u_{23})/\\
				(u_{01} (u_3 u_{013} u_{03} u_{13} + 
				b_3 u_{0123} u_{023} u_{123} u_{23}))) \; \; \; \; \; \; (g)\\
				u_1 = ((b_2 u_3 u_{01} u_{013} u_{023} u_{12} u_{123} + b_4 u_{01} u_{0123} u_{03} u_{12} u_{13} u_{23})/\\
				(u_{012} (u_3 u_{0123} u_{023} u_{13} + 
				b_3 u_{013} u_{03} u_{123} u_{23})))  \; \; \;\; \; \; (h).  
			\end{gathered}
		\end{equation} 
		Equating the rigth part of the equation (\ref{main_system4_u1_cdgh} $\;\;c $) to the rigth part of the equation (\ref{main_system4_u1_cdgh} $\;\; d, g, h$). After simplifying the equations and removing the positive denominators we have the following three equations

		\begin{equation} \label{main_system4_cdgh} 
			\begin{gathered} 
				(b_3 b_4 u_{01}^2 u_{012}^2 u_{023} u_{03} u_{123} u_{13} - 
				b_3 b_4 u_{0123}^2 u_{013}^2 u_{023} u_{03} u_{123} u_{13} + \\
				b_2 b_3 u_3 u_{01}^2 u_{012}^2 u_{0123} u_{013} u_{23} - 
				b_2 b_3 u_3 u_{0123} u_{013} u_{023}^2 u_{03}^2 u_{23} - \\
				b_4 u_3 u_{0123} u_{013} u_{123}^2 u_{13}^2 u_{23} + 
				b_4 u_3 u_{01}^2 u_{012}^2 u_{0123} u_{013} u_{023}^2 u_{03}^2 u_{123}^2 u_{13}^2 u_{23} - \\
				b_2 u_3^2 u_{023} u_{03} u_{123} u_{13} u_{23}^2 + 
				b_2 u_3^2 u_{01}^2 u_{012}^2 u_{0123}^2 u_{013}^2 u_{023} u_{03} u_{123} u_{13} u_{23}^2) = 0  \; \; \; (cd)\\
				(b_2 b_3 u_3 u_{012}^2 u_{0123} u_{03} u_{12}^2 u_{123} - 
				b_4 u_3 u_{0123} u_{013}^2 u_{03} u_{123} u_{13}^2 - \\
				b_2 u_3^2 u_{013} u_{023} u_{03}^2 u_{13} u_{23} + 
				b_3 b_4 u_{012}^2 u_{013} u_{023} u_{12}^2 u_{13} u_{23} - \\
				b_3 b_4 u_{0123}^2 u_{013} u_{023} u_{123}^2 u_{13} u_{23} + 
				b_2 u_3^2 u_{012}^2 u_{0123}^2 u_{013} u_{023} u_{03}^2 u_{12}^2 u_{123}^2 u_{13} u_{23} - \\
				b_2 b_3 u_3 u_{0123} u_{023}^2 u_{03} u_{123} u_{23}^2 + 
				b_4 u_3 u_{012}^2 u_{0123} u_{013}^2 u_{023}^2 u_{03} u_{12}^2 u_{123} u_{13}^2 u_{23}^2) 
				= 0  \; \; \; (cg) \\
				(b_2 b_3 u_3 u_{01}^2 u_{013} u_{023} u_{12}^2 u_{123} - 
				b_4 u_3 u_{0123}^2 u_{013} u_{023} u_{123} u_{13}^2 - \\
				b_2 u_3^2 u_{0123} u_{023}^2 u_{03} u_{13} u_{23} + 
				b_3 b_4 u_{01}^2 u_{0123} u_{03} u_{12}^2 u_{13} u_{23} - \\
				b_3 b_4 u_{0123} u_{013}^2 u_{03} u_{123}^2 u_{13} u_{23} + 
				b_2 u_3^2 u_{01}^2 u_{0123} u_{013}^2 u_{023}^2 u_{03} u_{12}^2 u_{123}^2 u_{13} u_{23} - \\
				b_2 b_3 u_3 u_{013} u_{023} u_{03}^2 u_{123} u_{23}^2 + 
				b_4 u_3 u_{01}^2 u_{0123}^2 u_{013} u_{023} u_{03}^2 u_{12}^2 u_{123} u_{13}^2 u_{23}^2) 
				= 0   \; \; \; (ch).  
			\end{gathered}
		\end{equation} 
		Substituting $b_2 $ from (\ref{main_system4_u2_bfe} $\;\;(bfe) $) 
		\begin{equation} \label{main_system4_b_2} 
			\begin{gathered} 
				b_2 = ((b_3 b_4 u_3 u_{0123} u_{013}^2 u_{023} u_{03}^2 u_{13} + 
				b_3^2 b_4 u_{0123}^2 u_{013} u_{023}^2 u_{03} u_{123} u_{23} + \\
				b_4 u_3^2 u_{013} u_{03} u_{123} u_{13}^2 u_{23} + 
				b_3 b_4 u_3 u_{0123} u_{023} u_{123}^2 u_{13} u_{23}^2)/\\
				(b_3 u_{012}^2 u_{02}^2 (b_3 u_3 u_{0123} u_{023} u_{13} + b_3^2 u_{013} u_{03} u_{123} u_{23} + \\
				u_3^2 u_{0123}^2 u_{013} u_{023}^2 u_{03} u_{123} u_{13}^2 u_{23} + 
				b_3 u_3 u_{0123} u_{013}^2 u_{023} u_{03}^2 u_{123}^2 u_{13} u_{23}^2))).  
			\end{gathered}
		\end{equation} 
		We substitute $b_2$  from  (\ref{main_system4_b_2} ) into all equations (\ref{main_system4_cdgh} ). After simplifying the equations and removing the positive denominators we have the following equations (let us write them separately because of their size)
		
		1. Equation (\ref{main_system4_cdgh}$\;\;cd$):
		
		\begin{multline} \label{main_system4_cd}  
			(b_3^3 u_3 u_{01}^2 u_{012}^4 u_{0123} u_{02}^2 u_{023}^2 u_{03} u_{123} u_{13}^2 - 
			b_3^3 u_3 u_{012}^2 u_{0123}^3 u_{013}^2 u_{02}^2 u_{023}^2 u_{03} u_{123} u_{13}^2 + \\
			b_3^2 u_3^2 u_{01}^2 u_{012}^2 u_{0123}^2 u_{013}^3 u_{023} u_{03}^2 u_{13} u_{23} - 
			b_3^2 u_3^2 u_{0123}^2 u_{013}^3 u_{023}^3 u_{03}^4 u_{13} u_{23} + \\
			b_3^4 u_{01}^2 u_{012}^4 u_{013} u_{02}^2 u_{023} u_{03}^2 u_{123}^2 u_{13} u_{23} - 
			b_3^4 u_{012}^2 u_{0123}^2 u_{013}^3 u_{02}^2 u_{023} u_{03}^2 u_{123}^2 u_{13} u_{23} -\\ 
			b_3^2 u_3^2 u_{012}^2 u_{0123}^2 u_{013} u_{02}^2 u_{023} u_{123}^2 u_{13}^3 u_{23} + 
			2 b_3^2 u_3^2 u_{01}^2 u_{012}^4 u_{0123}^2 u_{013} u_{02}^2 u_{023}^3 u_{03}^2 u_{123}^2 \
			u_{13}^3 u_{23} -\\ 
			b_3^2 u_3^2 u_{012}^2 u_{0123}^4 u_{013}^3 u_{02}^2 u_{023}^3 u_{03}^2 u_{123}^2 u_{13}^3 \
			u_{23} + b_3^3 u_3 u_{01}^2 u_{012}^2 u_{0123}^3 u_{013}^2 u_{023}^2 u_{03} u_{123} u_{23}^2 - \\
			b_3^3 u_3 u_{0123}^3 u_{013}^2 u_{023}^4 u_{03}^3 u_{123} u_{23}^2 + 
			b_3 u_3^3 u_{01}^2 u_{012}^2 u_{0123} u_{013}^2 u_{03} u_{123} u_{13}^2 u_{23}^2 -\\ 
			2 b_3 u_3^3 u_{0123} u_{013}^2 u_{023}^2 u_{03}^3 u_{123} u_{13}^2 u_{23}^2 + 
			b_3 u_3^3 u_{01}^2 u_{012}^2 u_{0123}^3 u_{013}^4 u_{023}^2 u_{03}^3 u_{123} u_{13}^2 u_{23}^2 -\\
			b_3^3 u_3 u_{012}^2 u_{0123} u_{013}^2 u_{02}^2 u_{03} u_{123}^3 u_{13}^2 u_{23}^2 + 
			2 b_3^3 u_3 u_{01}^2 u_{012}^4 u_{0123} u_{013}^2 u_{02}^2 u_{023}^2 u_{03}^3 u_{123}^3 \
			u_{13}^2 u_{23}^2 - \\
			b_3^3 u_3 u_{012}^2 u_{0123}^3 u_{013}^4 u_{02}^2 u_{023}^2 u_{03}^3 u_{123}^3 u_{13}^2 \
			u_{23}^2 - b_3 u_3^3 u_{012}^2 u_{0123}^3 u_{013}^2 u_{02}^2 u_{023}^2 u_{03} u_{123}^3 u_{13}^4 \
			u_{23}^2 +\\ b_3 u_3^3 u_{01}^2 u_{012}^4 u_{0123}^3 u_{013}^2 u_{02}^2 u_{023}^4 u_{03}^3 u_{123}^3 \
			u_{13}^4 u_{23}^2 + 
			b_3^2 u_3^2 u_{01}^2 u_{012}^2 u_{0123}^2 u_{013} u_{023} u_{123}^2 u_{13} u_{23}^3 - \\
			2 b_3^2 u_3^2 u_{0123}^2 u_{013} u_{023}^3 u_{03}^2 u_{123}^2 u_{13} u_{23}^3 + 
			b_3^2 u_3^2 u_{01}^2 u_{012}^2 u_{0123}^4 u_{013}^3 u_{023}^3 u_{03}^2 u_{123}^2 u_{13} \
			u_{23}^3 -\\ u_3^4 u_{013} u_{023} u_{03}^2 u_{123}^2 u_{13}^3 u_{23}^3 + 
			u_3^4 u_{01}^2 u_{012}^2 u_{0123}^2 u_{013}^3 u_{023} u_{03}^2 u_{123}^2 u_{13}^3 u_{23}^3 -\\ 
			b_3^2 u_3^2 u_{012}^2 u_{0123}^2 u_{013}^3 u_{02}^2 u_{023} u_{03}^2 u_{123}^4 u_{13}^3 \
			u_{23}^3 + b_3^2 u_3^2 u_{01}^2 u_{012}^4 u_{0123}^2 u_{013}^3 u_{02}^2 u_{023}^3 u_{03}^4 \
			u_{123}^4 u_{13}^3 u_{23}^3 - \\b_3 u_3^3 u_{0123} u_{023}^2 u_{03} u_{123}^3 u_{13}^2 u_{23}^4 + 
			b_3 u_3^3 u_{01}^2 u_{012}^2 u_{0123}^3 u_{013}^2 u_{023}^2 u_{03} u_{123}^3 u_{13}^2 u_{23}^4) \
			= 0      ,  
		\end{multline}

		2. Equation (\ref{main_system4_cdgh}$\;\;cg$):
		
		\begin{multline} \label{main_system4_cg}  
			(b_3^2 u_3^2 u_{012}^2 u_{0123}^2 u_{013}^2 u_{023} u_{03}^3 u_{12}^2 u_{123} u_{13} - 
			b_3^2 u_3^2 u_{012}^2 u_{0123}^2 u_{013}^2 u_{02}^2 u_{023} u_{03} u_{123} u_{13}^3 + \\
			b_3^3 u_3 u_{012}^2 u_{0123}^3 u_{013} u_{023}^2 u_{03}^2 u_{12}^2 u_{123}^2 u_{23} - 
			b_3 u_3^3 u_{0123} u_{013}^3 u_{023}^2 u_{03}^4 u_{13}^2 u_{23} + \\
			b_3^3 u_3 u_{012}^4 u_{0123} u_{013} u_{02}^2 u_{023}^2 u_{12}^2 u_{13}^2 u_{23} - 
			b_3^3 u_3 u_{012}^2 u_{0123}^3 u_{013} u_{02}^2 u_{023}^2 u_{123}^2 u_{13}^2 u_{23} - \\
			b_3^3 u_3 u_{012}^2 u_{0123} u_{013}^3 u_{02}^2 u_{03}^2 u_{123}^2 u_{13}^2 u_{23} + 
			b_3 u_3^3 u_{012}^2 u_{0123} u_{013} u_{03}^2 u_{12}^2 u_{123}^2 u_{13}^2 u_{23} + \\
			b_3 u_3^3 u_{012}^2 u_{0123}^3 u_{013}^3 u_{023}^2 u_{03}^4 u_{12}^2 u_{123}^2 u_{13}^2 u_{23} -
			b_3 u_3^3 u_{012}^2 u_{0123}^3 u_{013}^3 u_{02}^2 u_{023}^2 u_{03}^2 u_{123}^2 u_{13}^4 u_{23} \
			-\\ 2 b_3^2 u_3^2 u_{0123}^2 u_{013}^2 u_{023}^3 u_{03}^3 u_{123} u_{13} u_{23}^2 + 
			b_3^4 u_{012}^4 u_{013}^2 u_{02}^2 u_{023} u_{03} u_{12}^2 u_{123} u_{13} u_{23}^2 - \\
			b_3^4 u_{012}^2 u_{0123}^2 u_{013}^2 u_{02}^2 u_{023} u_{03} u_{123}^3 u_{13} u_{23}^2 + 
			b_3^2 u_3^2 u_{012}^2 u_{0123}^2 u_{023} u_{03} u_{12}^2 u_{123}^3 u_{13} u_{23}^2 + \\
			b_3^2 u_3^2 u_{012}^2 u_{0123}^4 u_{013}^2 u_{023}^3 u_{03}^3 u_{12}^2 u_{123}^3 u_{13} \
			u_{23}^2 - u_3^4 u_{013}^2 u_{023} u_{03}^3 u_{123} u_{13}^3 u_{23}^2 + \\
			2 b_3^2 u_3^2 u_{012}^4 u_{0123}^2 u_{013}^2 u_{02}^2 u_{023}^3 u_{03} u_{12}^2 u_{123} \
			u_{13}^3 u_{23}^2 - 
			b_3^2 u_3^2 u_{012}^2 u_{0123}^4 u_{013}^2 u_{02}^2 u_{023}^3 u_{03} u_{123}^3 u_{13}^3 \
			u_{23}^2 -\\ b_3^2 u_3^2 u_{012}^2 u_{0123}^2 u_{013}^4 u_{02}^2 u_{023} u_{03}^3 u_{123}^3 u_{13}^3 \
			u_{23}^2 + u_3^4 u_{012}^2 u_{0123}^2 u_{013}^2 u_{023} u_{03}^3 u_{12}^2 u_{123}^3 u_{13}^3 \
			u_{23}^2 -\\ b_3^3 u_3 u_{0123}^3 u_{013} u_{023}^4 u_{03}^2 u_{123}^2 u_{23}^3 - 
			2 b_3 u_3^3 u_{0123} u_{013} u_{023}^2 u_{03}^2 u_{123}^2 u_{13}^2 u_{23}^3 + \\
			2 b_3^3 u_3 u_{012}^4 u_{0123} u_{013}^3 u_{02}^2 u_{023}^2 u_{03}^2 u_{12}^2 u_{123}^2 \
			u_{13}^2 u_{23}^3 - 
			b_3^3 u_3 u_{012}^2 u_{0123}^3 u_{013}^3 u_{02}^2 u_{023}^2 u_{03}^2 u_{123}^4 u_{13}^2 \
			u_{23}^3 + \\b_3 u_3^3 u_{012}^2 u_{0123}^3 u_{013} u_{023}^2 u_{03}^2 u_{12}^2 u_{123}^4 u_{13}^2 \
			u_{23}^3 + b_3 u_3^3 u_{012}^4 u_{0123}^3 u_{013}^3 u_{02}^2 u_{023}^4 u_{03}^2 u_{12}^2 u_{123}^2 \
			u_{13}^4 u_{23}^3 -\\ b_3^2 u_3^2 u_{0123}^2 u_{023}^3 u_{03} u_{123}^3 u_{13} u_{23}^4 + 
			b_3^2 u_3^2 u_{012}^4 u_{0123}^2 u_{013}^4 u_{02}^2 u_{023}^3 u_{03}^3 u_{12}^2 u_{123}^3 \
			u_{13}^3 u_{23}^4) = 0    ,  
		\end{multline} 
		
		3. Equation (\ref{main_system4_cdgh}$\;\;ch$):
		
		\begin{multline} \label{main_system4_ch}  
			(b_3^2 u_3^2 u_{01}^2 u_{0123} u_{013}^3 u_{023}^2 u_{03}^2 u_{12}^2 u_{123} u_{13} - 
			b_3^2 u_3^2 u_{012}^2 u_{0123}^3 u_{013} u_{02}^2 u_{023}^2 u_{123} u_{13}^3 + \\
			b_3^3 u_3 u_{01}^2 u_{0123}^2 u_{013}^2 u_{023}^3 u_{03} u_{12}^2 u_{123}^2 u_{23} - 
			b_3 u_3^3 u_{0123}^2 u_{013}^2 u_{023}^3 u_{03}^3 u_{13}^2 u_{23} + \\
			b_3^3 u_3 u_{01}^2 u_{012}^2 u_{0123}^2 u_{02}^2 u_{023} u_{03} u_{12}^2 u_{13}^2 u_{23} - 
			2 b_3^3 u_3 u_{012}^2 u_{0123}^2 u_{013}^2 u_{02}^2 u_{023} u_{03} u_{123}^2 u_{13}^2 u_{23} +\\ 
			b_3 u_3^3 u_{01}^2 u_{013}^2 u_{023} u_{03} u_{12}^2 u_{123}^2 u_{13}^2 u_{23} + 
			b_3 u_3^3 u_{01}^2 u_{0123}^2 u_{013}^4 u_{023}^3 u_{03}^3 u_{12}^2 u_{123}^2 u_{13}^2 u_{23} -\\ 
			b_3 u_3^3 u_{012}^2 u_{0123}^4 u_{013}^2 u_{02}^2 u_{023}^3 u_{03} u_{123}^2 u_{13}^4 u_{23} - 
			b_3^2 u_3^2 u_{0123}^3 u_{013} u_{023}^4 u_{03}^2 u_{123} u_{13} u_{23}^2 - \\
			b_3^2 u_3^2 u_{0123} u_{013}^3 u_{023}^2 u_{03}^4 u_{123} u_{13} u_{23}^2 + 
			b_3^4 u_{01}^2 u_{012}^2 u_{0123} u_{013} u_{02}^2 u_{03}^2 u_{12}^2 u_{123} u_{13} u_{23}^2 -\\ 
			b_3^4 u_{012}^2 u_{0123} u_{013}^3 u_{02}^2 u_{03}^2 u_{123}^3 u_{13} u_{23}^2 + 
			b_3^2 u_3^2 u_{01}^2 u_{0123} u_{013} u_{023}^2 u_{12}^2 u_{123}^3 u_{13} u_{23}^2 + \\
			b_3^2 u_3^2 u_{01}^2 u_{0123}^3 u_{013}^3 u_{023}^4 u_{03}^2 u_{12}^2 u_{123}^3 u_{13} u_{23}^2 \
			- u_3^4 u_{0123} u_{013} u_{023}^2 u_{03}^2 u_{123} u_{13}^3 u_{23}^2 + \\
			2 b_3^2 u_3^2 u_{01}^2 u_{012}^2 u_{0123}^3 u_{013} u_{02}^2 u_{023}^2 u_{03}^2 u_{12}^2 \
			u_{123} u_{13}^3 u_{23}^2 - \\
			2 b_3^2 u_3^2 u_{012}^2 u_{0123}^3 u_{013}^3 u_{02}^2 u_{023}^2 u_{03}^2 u_{123}^3 u_{13}^3 \
			u_{23}^2 + \\u_3^4 u_{01}^2 u_{0123} u_{013}^3 u_{023}^2 u_{03}^2 u_{12}^2 u_{123}^3 u_{13}^3 u_{23}^2 \
			- b_3^3 u_3 u_{0123}^2 u_{013}^2 u_{023}^3 u_{03}^3 u_{123}^2 u_{23}^3 - \\
			b_3 u_3^3 u_{0123}^2 u_{023}^3 u_{03} u_{123}^2 u_{13}^2 u_{23}^3 - 
			b_3 u_3^3 u_{013}^2 u_{023} u_{03}^3 u_{123}^2 u_{13}^2 u_{23}^3 + \\
			2 b_3^3 u_3 u_{01}^2 u_{012}^2 u_{0123}^2 u_{013}^2 u_{02}^2 u_{023} u_{03}^3 u_{12}^2 \
			u_{123}^2 u_{13}^2 u_{23}^3 - \\
			b_3^3 u_3 u_{012}^2 u_{0123}^2 u_{013}^4 u_{02}^2 u_{023} u_{03}^3 u_{123}^4 u_{13}^2 u_{23}^3 +\\
			b_3 u_3^3 u_{01}^2 u_{0123}^2 u_{013}^2 u_{023}^3 u_{03} u_{12}^2 u_{123}^4 u_{13}^2 u_{23}^3 +\\
			b_3 u_3^3 u_{01}^2 u_{012}^2 u_{0123}^4 u_{013}^2 u_{02}^2 u_{023}^3 u_{03}^3 u_{12}^2 \
			u_{123}^2 u_{13}^4 u_{23}^3 - \\
			b_3^2 u_3^2 u_{0123} u_{013} u_{023}^2 u_{03}^2 u_{123}^3 u_{13} u_{23}^4 +\\ 
			b_3^2 u_3^2 u_{01}^2 u_{012}^2 u_{0123}^3 u_{013}^3 u_{02}^2 u_{023}^2 u_{03}^4 u_{12}^2 \
			u_{123}^3 u_{13}^3 u_{23}^4) = 0     ,  
		\end{multline} 
		
		We substitute $s_{01}$ from the equation (\ref{main_system4_cd})
		\begin{multline} \label{main_system4_s01}  
			s_{01} = ((b_3 u_{0123} u_{013} u_{023} u_{03} + 
			u_3 u_{123} u_{13} u_{23}) (b_3^2 u_3 u_{012}^2 u_{0123}^2 u_{013} u_{02}^2 u_{023} u_{123} \
			u_{13}^2 +\\ b_3 u_3^2 u_{0123} u_{013}^2 u_{023}^2 u_{03}^3 u_{13} u_{23} + 
			b_3^3 u_{012}^2 u_{0123} u_{013}^2 u_{02}^2 u_{03} u_{123}^2 u_{13} u_{23} + \\
			b_3 u_3^2 u_{012}^2 u_{0123}^3 u_{013}^2 u_{02}^2 u_{023}^2 u_{03} u_{123}^2 u_{13}^3 u_{23} \
			+ b_3^2 u_3 u_{0123}^2 u_{013} u_{023}^3 u_{03}^2 u_{123} u_{23}^2 + \\
			u_3^3 u_{013} u_{023} u_{03}^2 u_{123} u_{13}^2 u_{23}^2 + 
			b_3^2 u_3 u_{012}^2 u_{0123}^2 u_{013}^3 u_{02}^2 u_{023} u_{03}^2 u_{123}^3 u_{13}^2 \
			u_{23}^2 +\\ b_3 u_3^2 u_{0123} u_{023}^2 u_{03} u_{123}^2 u_{13} u_{23}^3))/(u_{012}^2 (b_3 +
			u_3 u_{0123} u_{013} u_{023} u_{03} u_{123} u_{13} u_{23}) \\(b_3^2 u_3 u_{012}^2 u_{0123} \
			u_{02}^2 u_{023}^2 u_{03} u_{123} u_{13}^2 + 
			b_3 u_3^2 u_{0123}^2 u_{013}^3 u_{023} u_{03}^2 u_{13} u_{23} + \\
			b_3^3 u_{012}^2 u_{013} u_{02}^2 u_{023} u_{03}^2 u_{123}^2 u_{13} u_{23} + 
			b_3 u_3^2 u_{012}^2 u_{0123}^2 u_{013} u_{02}^2 u_{023}^3 u_{03}^2 u_{123}^2 u_{13}^3 u_{23} \
			+\\ b_3^2 u_3 u_{0123}^3 u_{013}^2 u_{023}^2 u_{03} u_{123} u_{23}^2 + 
			u_3^3 u_{0123} u_{013}^2 u_{03} u_{123} u_{13}^2 u_{23}^2 + \\
			b_3^2 u_3 u_{012}^2 u_{0123} u_{013}^2 u_{02}^2 u_{023}^2 u_{03}^3 u_{123}^3 u_{13}^2 \
			u_{23}^2 + b_3 u_3^2 u_{0123}^2 u_{013} u_{023} u_{123}^2 u_{13} u_{23}^3))     ,  
		\end{multline} 
		Substituting $s_{01}$  from  (\ref{main_system4_s01}) into  (\ref{main_system4_cg}) and (\ref{main_system4_ch}). We get two equations:

		\begin{multline} \label{13_57}
			(b_3^2 u_3^2 u_{012}^2 u_{0123}^2 u_{013}^2 u_{023} u_{03}^3 u_{12}^2 u_{123} u_{13} - 
			b_3^2 u_3^2 u_{012}^2 u_{0123}^2 u_{013}^2 u_{02}^2 u_{023} u_{03} u_{123} u_{13}^3 + \\
			b_3^3 u_3 u_{012}^2 u_{0123}^3 u_{013} u_{023}^2 u_{03}^2 u_{12}^2 u_{123}^2 u_{23} - 
			b_3 u_3^3 u_{0123} u_{013}^3 u_{023}^2 u_{03}^4 u_{13}^2 u_{23} + \\
			b_3^3 u_3 u_{012}^4 u_{0123} u_{013} u_{02}^2 u_{023}^2 u_{12}^2 u_{13}^2 u_{23} - 
			b_3^3 u_3 u_{012}^2 u_{0123}^3 u_{013} u_{02}^2 u_{023}^2 u_{123}^2 u_{13}^2 u_{23} - \\
			b_3^3 u_3 u_{012}^2 u_{0123} u_{013}^3 u_{02}^2 u_{03}^2 u_{123}^2 u_{13}^2 u_{23} + 
			b_3 u_3^3 u_{012}^2 u_{0123} u_{013} u_{03}^2 u_{12}^2 u_{123}^2 u_{13}^2 u_{23} + \\
			b_3 u_3^3 u_{012}^2 u_{0123}^3 u_{013}^3 u_{023}^2 u_{03}^4 u_{12}^2 u_{123}^2 u_{13}^2 u_{23} -
			b_3 u_3^3 u_{012}^2 u_{0123}^3 u_{013}^3 u_{02}^2 u_{023}^2 u_{03}^2 u_{123}^2 u_{13}^4 u_{23} \
			- \\2 b_3^2 u_3^2 u_{0123}^2 u_{013}^2 u_{023}^3 u_{03}^3 u_{123} u_{13} u_{23}^2 + 
			b_3^4 u_{012}^4 u_{013}^2 u_{02}^2 u_{023} u_{03} u_{12}^2 u_{123} u_{13} u_{23}^2 -\\ 
			b_3^4 u_{012}^2 u_{0123}^2 u_{013}^2 u_{02}^2 u_{023} u_{03} u_{123}^3 u_{13} u_{23}^2 + 
			b_3^2 u_3^2 u_{012}^2 u_{0123}^2 u_{023} u_{03} u_{12}^2 u_{123}^3 u_{13} u_{23}^2 + \\
			b_3^2 u_3^2 u_{012}^2 u_{0123}^4 u_{013}^2 u_{023}^3 u_{03}^3 u_{12}^2 u_{123}^3 u_{13} \
			u_{23}^2 - u_3^4 u_{013}^2 u_{023} u_{03}^3 u_{123} u_{13}^3 u_{23}^2 + \\
			2 b_3^2 u_3^2 u_{012}^4 u_{0123}^2 u_{013}^2 u_{02}^2 u_{023}^3 u_{03} u_{12}^2 u_{123} \
			u_{13}^3 u_{23}^2 - \\
			b_3^2 u_3^2 u_{012}^2 u_{0123}^4 u_{013}^2 u_{02}^2 u_{023}^3 u_{03} u_{123}^3 u_{13}^3 \
			u_{23}^2 - b_3^2 u_3^2 u_{012}^2 u_{0123}^2 u_{013}^4 u_{02}^2 u_{023} u_{03}^3 u_{123}^3 u_{13}^3 \
			u_{23}^2 +\\ u_3^4 u_{012}^2 u_{0123}^2 u_{013}^2 u_{023} u_{03}^3 u_{12}^2 u_{123}^3 u_{13}^3 \
			u_{23}^2 - b_3^3 u_3 u_{0123}^3 u_{013} u_{023}^4 u_{03}^2 u_{123}^2 u_{23}^3 - \\
			2 b_3 u_3^3 u_{0123} u_{013} u_{023}^2 u_{03}^2 u_{123}^2 u_{13}^2 u_{23}^3 + 
			2 b_3^3 u_3 u_{012}^4 u_{0123} u_{013}^3 u_{02}^2 u_{023}^2 u_{03}^2 u_{12}^2 u_{123}^2 \
			u_{13}^2 u_{23}^3 - \\
			b_3^3 u_3 u_{012}^2 u_{0123}^3 u_{013}^3 u_{02}^2 u_{023}^2 u_{03}^2 u_{123}^4 u_{13}^2 \
			u_{23}^3 + b_3 u_3^3 u_{012}^2 u_{0123}^3 u_{013} u_{023}^2 u_{03}^2 u_{12}^2 u_{123}^4 u_{13}^2 \
			u_{23}^3 +\\ b_3 u_3^3 u_{012}^4 u_{0123}^3 u_{013}^3 u_{02}^2 u_{023}^4 u_{03}^2 u_{12}^2 u_{123}^2 \
			u_{13}^4 u_{23}^3 - b_3^2 u_3^2 u_{0123}^2 u_{023}^3 u_{03} u_{123}^3 u_{13} u_{23}^4 + \\
			b_3^2 u_3^2 u_{012}^4 u_{0123}^2 u_{013}^4 u_{02}^2 u_{023}^3 u_{03}^3 u_{12}^2 u_{123}^3 \
			u_{13}^3 u_{23}^4) = 0  ,
		\end{multline}

		\begin{multline} \label{1368_1324}
			(-b_3^2 u_3^2 u_{0123}^2 u_{013}^4 u_{023}^3 u_{03}^3 u_{12}^2 u_{123} u_{13} + 
			b_3^2 u_3^2 u_{012}^4 u_{0123}^2 u_{02}^2 u_{023}^3 u_{03} u_{123} u_{13}^3 - \\
			b_3^3 u_3 u_{0123}^3 u_{013}^3 u_{023}^4 u_{03}^2 u_{12}^2 u_{123}^2 u_{23} + 
			b_3 u_3^3 u_{012}^2 u_{0123}^3 u_{013}^3 u_{023}^2 u_{03}^2 u_{13}^2 u_{23} - \\
			b_3^3 u_3 u_{012}^2 u_{0123}^3 u_{013} u_{02}^2 u_{023}^2 u_{03}^2 u_{12}^2 u_{13}^2 u_{23} + 
			2 b_3^3 u_3 u_{012}^4 u_{0123} u_{013} u_{02}^2 u_{023}^2 u_{03}^2 u_{123}^2 u_{13}^2 u_{23} - \\
			2 b_3 u_3^3 u_{0123} u_{013}^3 u_{023}^2 u_{03}^2 u_{12}^2 u_{123}^2 u_{13}^2 u_{23} + 
			b_3 u_3^3 u_{012}^4 u_{0123}^3 u_{013} u_{02}^2 u_{023}^4 u_{03}^2 u_{123}^2 u_{13}^4 u_{23} +\\ 
			b_3^2 u_3^2 u_{012}^2 u_{0123}^4 u_{013}^2 u_{023}^3 u_{03} u_{123} u_{13} u_{23}^2 + 
			b_3^2 u_3^2 u_{012}^2 u_{0123}^2 u_{013}^4 u_{023} u_{03}^3 u_{123} u_{13} u_{23}^2 -\\ 
			b_3^4 u_{012}^2 u_{0123}^2 u_{013}^2 u_{02}^2 u_{023} u_{03}^3 u_{12}^2 u_{123} u_{13} u_{23}^2 + 
			b_3^4 u_{012}^4 u_{013}^2 u_{02}^2 u_{023} u_{03}^3 u_{123}^3 u_{13} u_{23}^2 -\\ 
			2 b_3^2 u_3^2 u_{0123}^2 u_{013}^2 u_{023}^3 u_{03} u_{12}^2 u_{123}^3 u_{13} u_{23}^2 + 
			u_3^4 u_{012}^2 u_{0123}^2 u_{013}^2 u_{023} u_{03} u_{123} u_{13}^3 u_{23}^2 - \\
			b_3^2 u_3^2 u_{012}^2 u_{0123}^2 u_{02}^2 u_{023} u_{03} u_{12}^2 u_{123} u_{13}^3 u_{23}^2 - 
			b_3^2 u_3^2 u_{012}^2 u_{0123}^4 u_{013}^2 u_{02}^2 u_{023}^3 u_{03}^3 u_{12}^2 u_{123} \
			u_{13}^3 u_{23}^2 + \\
			2 b_3^2 u_3^2 u_{012}^4 u_{0123}^2 u_{013}^2 u_{02}^2 u_{023}^3 u_{03}^3 u_{123}^3 u_{13}^3 \
			u_{23}^2 - u_3^4 u_{013}^2 u_{023} u_{03} u_{12}^2 u_{123}^3 u_{13}^3 u_{23}^2 + \\
			b_3^3 u_3 u_{012}^2 u_{0123}^3 u_{013}^3 u_{023}^2 u_{03}^2 u_{123}^2 u_{23}^3 + 
			b_3 u_3^3 u_{012}^2 u_{0123}^3 u_{013} u_{023}^2 u_{123}^2 u_{13}^2 u_{23}^3 + \\
			b_3 u_3^3 u_{012}^2 u_{0123} u_{013}^3 u_{03}^2 u_{123}^2 u_{13}^2 u_{23}^3 - 
			b_3^3 u_3 u_{012}^2 u_{0123} u_{013} u_{02}^2 u_{03}^2 u_{12}^2 u_{123}^2 u_{13}^2 u_{23}^3 - \\
			b_3^3 u_3 u_{012}^2 u_{0123}^3 u_{013}^3 u_{02}^2 u_{023}^2 u_{03}^4 u_{12}^2 u_{123}^2 \
			u_{13}^2 u_{23}^3 + \\
			b_3^3 u_3 u_{012}^4 u_{0123} u_{013}^3 u_{02}^2 u_{023}^2 u_{03}^4 u_{123}^4 u_{13}^2 u_{23}^3 -
			b_3 u_3^3 u_{0123} u_{013} u_{023}^2 u_{12}^2 u_{123}^4 u_{13}^2 u_{23}^3 -\\ 
			b_3 u_3^3 u_{012}^2 u_{0123}^3 u_{013} u_{02}^2 u_{023}^2 u_{03}^2 u_{12}^2 u_{123}^2 u_{13}^4 \
			u_{23}^3 + b_3^2 u_3^2 u_{012}^2 u_{0123}^2 u_{013}^2 u_{023} u_{03} u_{123}^3 u_{13} u_{23}^4 - \\
			b_3^2 u_3^2 u_{012}^2 u_{0123}^2 u_{013}^2 u_{02}^2 u_{023} u_{03}^3 u_{12}^2 u_{123}^3 \
			u_{13}^3 u_{23}^4) = 0       .	
		\end{multline} 
		
		Substituting $s_{12}$  from equations (\ref{13_57}) , (\ref{1368_1324}):
		
		\begin{multline} \label{13_57_s12} 
			s_{12} = ((u_3 u_{013} u_{03} u_{13} + 
			b_3 u_{0123} u_{023} u_{123} u_{23}) (b_3^2 u_3 u_{012}^2 u_{0123}^2 u_{013} u_{02}^2 u_{023} \
			u_{123} u_{13}^2 +\\ b_3 u_3^2 u_{0123} u_{013}^2 u_{023}^2 u_{03}^3 u_{13} u_{23} + 
			b_3^3 u_{012}^2 u_{0123} u_{013}^2 u_{02}^2 u_{03} u_{123}^2 u_{13} u_{23} + \\
			b_3 u_3^2 u_{012}^2 u_{0123}^3 u_{013}^2 u_{02}^2 u_{023}^2 u_{03} u_{123}^2 u_{13}^3 u_{23} \
			+ b_3^2 u_3 u_{0123}^2 u_{013} u_{023}^3 u_{03}^2 u_{123} u_{23}^2 + \\
			u_3^3 u_{013} u_{023} u_{03}^2 u_{123} u_{13}^2 u_{23}^2 + 
			b_3^2 u_3 u_{012}^2 u_{0123}^2 u_{013}^3 u_{02}^2 u_{023} u_{03}^2 u_{123}^3 u_{13}^2 \
			u_{23}^2 + \\b_3 u_3^2 u_{0123} u_{023}^2 u_{03} u_{123}^2 u_{13} u_{23}^3))/(u_{012}^2 (b_3 + 
			u_3 u_{0123} u_{013} u_{023} u_{03} u_{123} u_{13} u_{23})\\ (b_3 u_3^2 u_{0123}^2 u_{013}^2 \
			u_{023} u_{03}^3 u_{123} u_{13} + b_3^2 u_3 u_{0123}^3 u_{013} u_{023}^2 u_{03}^2 u_{123}^2 u_{23} + \\
			b_3^2 u_3 u_{012}^2 u_{0123} u_{013} u_{02}^2 u_{023}^2 u_{13}^2 u_{23} + 
			u_3^3 u_{0123} u_{013} u_{03}^2 u_{123}^2 u_{13}^2 u_{23} + \\
			b_3^3 u_{012}^2 u_{013}^2 u_{02}^2 u_{023} u_{03} u_{123} u_{13} u_{23}^2 + 
			b_3 u_3^2 u_{0123}^2 u_{023} u_{03} u_{123}^3 u_{13} u_{23}^2 + \\
			b_3 u_3^2 u_{012}^2 u_{0123}^2 u_{013}^2 u_{02}^2 u_{023}^3 u_{03} u_{123} u_{13}^3 u_{23}^2 \
			+ b_3^2 u_3 u_{012}^2 u_{0123} u_{013}^3 u_{02}^2 u_{023}^2 u_{03}^2 u_{123}^2 u_{13}^2 u_{23}^3)) .	
		\end{multline}

		\begin{multline} \label{1368_1324_s12} 
			s_{12} = (u_{012}^2 (u_3 u_{0123} u_{023} u_{13} + 
			b_3 u_{013} u_{03} u_{123} u_{23}) (b_3^2 u_3 u_{012}^2 u_{0123} u_{02}^2 u_{023}^2 u_{03} \
			u_{123} u_{13}^2 +\\ b_3 u_3^2 u_{0123}^2 u_{013}^3 u_{023} u_{03}^2 u_{13} u_{23} + 
			b_3^3 u_{012}^2 u_{013} u_{02}^2 u_{023} u_{03}^2 u_{123}^2 u_{13} u_{23} + \\
			b_3 u_3^2 u_{012}^2 u_{0123}^2 u_{013} u_{02}^2 u_{023}^3 u_{03}^2 u_{123}^2 u_{13}^3 u_{23} \
			+ b_3^2 u_3 u_{0123}^3 u_{013}^2 u_{023}^2 u_{03} u_{123} u_{23}^2 + \\
			u_3^3 u_{0123} u_{013}^2 u_{03} u_{123} u_{13}^2 u_{23}^2 + 
			b_3^2 u_3 u_{012}^2 u_{0123} u_{013}^2 u_{02}^2 u_{023}^2 u_{03}^3 u_{123}^3 u_{13}^2 \
			u_{23}^2 +\\ b_3 u_3^2 u_{0123}^2 u_{013} u_{023} u_{123}^2 u_{13} u_{23}^3))/((b_3 u_{0123} u_{013} \
			u_{023} u_{03} + 
			u_3 u_{123} u_{13} u_{23}) \\(b_3 u_3^2 u_{0123} u_{013}^3 u_{023}^2 u_{03}^2 u_{123} u_{13} + 
			b_3^2 u_3 u_{0123}^2 u_{013}^2 u_{023}^3 u_{03} u_{123}^2 u_{23} + \\
			b_3^2 u_3 u_{012}^2 u_{0123}^2 u_{02}^2 u_{023} u_{03} u_{13}^2 u_{23} + 
			u_3^3 u_{013}^2 u_{023} u_{03} u_{123}^2 u_{13}^2 u_{23} + \\
			b_3^3 u_{012}^2 u_{0123} u_{013} u_{02}^2 u_{03}^2 u_{123} u_{13} u_{23}^2 + 
			b_3 u_3^2 u_{0123} u_{013} u_{023}^2 u_{123}^3 u_{13} u_{23}^2 + \\
			b_3 u_3^2 u_{012}^2 u_{0123}^3 u_{013} u_{02}^2 u_{023}^2 u_{03}^2 u_{123} u_{13}^3 u_{23}^2 \
			+ b_3^2 u_3 u_{012}^2 u_{0123}^2 u_{013}^2 u_{02}^2 u_{023} u_{03}^3 u_{123}^2 u_{13}^2 u_{23}^3)) .
		\end{multline} 
		
		From equality of the rigth parts of (\ref{13_57_s12}) and (\ref{1368_1324_s12}) we get the following equation
		\begin{multline} \label{for_s02}    
			(-b_3^4 u_3^4 u_{012}^6 u_{0123}^4 u_{013}^2 u_{02}^2 u_{023}^4 u_{03}^4 u_{123}^2 u_{13}^4 + 
			b_3^4 u_3^4 u_{012}^2 u_{0123}^4 u_{013}^6 u_{02}^2 u_{023}^4 u_{03}^4 u_{123}^2 u_{13}^4 - \\
			b_3^3 u_3^5 u_{012}^4 u_{0123}^5 u_{013}^5 u_{023}^3 u_{03}^5 u_{123} u_{13}^3 u_{23} + 
			b_3^3 u_3^5 u_{0123}^3 u_{013}^7 u_{023}^5 u_{03}^7 u_{123} u_{13}^3 u_{23} -\\ 
			b_3^5 u_3^3 u_{012}^6 u_{0123}^5 u_{013} u_{02}^2 u_{023}^5 u_{03}^3 u_{123}^3 u_{13}^3 u_{23} \
			+ 2 b_3^5 u_3^3 u_{012}^2 u_{0123}^5 u_{013}^5 u_{02}^2 u_{023}^5 u_{03}^3 u_{123}^3 u_{13}^3 \
			u_{23} -\\ 2 b_3^5 u_3^3 u_{012}^6 u_{0123}^3 u_{013}^3 u_{02}^2 u_{023}^3 u_{03}^5 u_{123}^3 \
			u_{13}^3 u_{23} + 
			b_3^5 u_3^3 u_{012}^2 u_{0123}^3 u_{013}^7 u_{02}^2 u_{023}^3 u_{03}^5 u_{123}^3 u_{13}^3 \
			u_{23} -\\ b_3^5 u_3^3 u_{012}^8 u_{0123}^3 u_{013} u_{02}^4 u_{023}^5 u_{03} u_{123} u_{13}^5 u_{23} + 
			b_3^5 u_3^3 u_{012}^4 u_{0123}^5 u_{013}^3 u_{02}^4 u_{023}^3 u_{03}^3 u_{123} u_{13}^5 u_{23} -\\
			b_3^3 u_3^5 u_{012}^6 u_{0123}^3 u_{013} u_{02}^2 u_{023}^3 u_{03}^3 u_{123}^3 u_{13}^5 u_{23} \
			+ 2 b_3^3 u_3^5 u_{012}^2 u_{0123}^3 u_{013}^5 u_{02}^2 u_{023}^3 u_{03}^3 u_{123}^3 u_{13}^5 \
			u_{23} -\\ 2 b_3^3 u_3^5 u_{012}^6 u_{0123}^5 u_{013}^3 u_{02}^2 u_{023}^5 u_{03}^5 u_{123}^3 \
			u_{13}^5 u_{23} + 
			b_3^3 u_3^5 u_{012}^2 u_{0123}^5 u_{013}^7 u_{02}^2 u_{023}^5 u_{03}^5 u_{123}^3 u_{13}^5 \
			u_{23} -\\ 2 b_3^4 u_3^4 u_{012}^4 u_{0123}^6 u_{013}^4 u_{023}^4 u_{03}^4 u_{123}^2 u_{13}^2 \
			u_{23}^2 - b_3^4 u_3^4 u_{012}^4 u_{0123}^4 u_{013}^6 u_{023}^2 u_{03}^6 u_{123}^2 u_{13}^2 \
			u_{23}^2 +\\ 3 b_3^4 u_3^4 u_{0123}^4 u_{013}^6 u_{023}^6 u_{03}^6 u_{123}^2 u_{13}^2 u_{23}^2 + 
			b_3^6 u_3^2 u_{012}^2 u_{0123}^6 u_{013}^4 u_{02}^2 u_{023}^6 u_{03}^2 u_{123}^4 u_{13}^2 \
			u_{23}^2 - \\2 b_3^6 u_3^2 u_{012}^6 u_{0123}^4 u_{013}^2 u_{02}^2 u_{023}^4 u_{03}^4 u_{123}^4 \
			u_{13}^2 u_{23}^2 + 
			2 b_3^6 u_3^2 u_{012}^2 u_{0123}^4 u_{013}^6 u_{02}^2 u_{023}^4 u_{03}^4 u_{123}^4 u_{13}^2 \
			u_{23}^2 -\\ b_3^6 u_3^2 u_{012}^6 u_{0123}^2 u_{013}^4 u_{02}^2 u_{023}^2 u_{03}^6 u_{123}^4 \
			u_{13}^2 u_{23}^2 - 
			b_3^4 u_3^4 u_{012}^6 u_{0123}^4 u_{013}^4 u_{02}^2 u_{023}^4 u_{03}^2 u_{13}^4 u_{23}^2 + \\
			b_3^4 u_3^4 u_{012}^2 u_{0123}^4 u_{013}^4 u_{02}^2 u_{023}^4 u_{03}^6 u_{13}^4 u_{23}^2 - 
			3 b_3^6 u_3^2 u_{012}^8 u_{0123}^2 u_{013}^2 u_{02}^4 u_{023}^4 u_{03}^2 u_{123}^2 u_{13}^4 \
			u_{23}^2 +\\ b_3^6 u_3^2 u_{012}^4 u_{0123}^6 u_{013}^2 u_{02}^4 u_{023}^4 u_{03}^2 u_{123}^2 \
			u_{13}^4 u_{23}^2 - 
			2 b_3^2 u_3^6 u_{012}^4 u_{0123}^4 u_{013}^4 u_{023}^2 u_{03}^4 u_{123}^2 u_{13}^4 u_{23}^2 \
			+ \\2 b_3^6 u_3^2 u_{012}^4 u_{0123}^4 u_{013}^4 u_{02}^4 u_{023}^2 u_{03}^4 u_{123}^2 u_{13}^4 \
			u_{23}^2 + 3 b_3^2 u_3^6 u_{0123}^2 u_{013}^6 u_{023}^4 u_{03}^6 u_{123}^2 u_{13}^4 u_{23}^2 - \\
			b_3^2 u_3^6 u_{012}^4 u_{0123}^6 u_{013}^6 u_{023}^4 u_{03}^6 u_{123}^2 u_{13}^4 u_{23}^2 - 
			b_3^4 u_3^4 u_{012}^6 u_{0123}^4 u_{02}^2 u_{023}^4 u_{03}^2 u_{123}^4 u_{13}^4 u_{23}^2 + \\
			4 b_3^4 u_3^4 u_{012}^2 u_{0123}^4 u_{013}^4 u_{02}^2 u_{023}^4 u_{03}^2 u_{123}^4 u_{13}^4 \
			u_{23}^2 - 2 b_3^4 u_3^4 u_{012}^6 u_{0123}^2 u_{013}^2 u_{02}^2 u_{023}^2 u_{03}^4 u_{123}^4 \
			u_{13}^4 u_{23}^2 + \\
			2 b_3^4 u_3^4 u_{012}^2 u_{0123}^2 u_{013}^6 u_{02}^2 u_{023}^2 u_{03}^4 u_{123}^4 u_{13}^4 \
			u_{23}^2 - 2 b_3^4 u_3^4 u_{012}^6 u_{0123}^6 u_{013}^2 u_{02}^2 u_{023}^6 u_{03}^4 u_{123}^4 \
			u_{13}^4 u_{23}^2 + \\
			2 b_3^4 u_3^4 u_{012}^2 u_{0123}^6 u_{013}^6 u_{02}^2 u_{023}^6 u_{03}^4 u_{123}^4 u_{13}^4 \
			u_{23}^2 - 4 b_3^4 u_3^4 u_{012}^6 u_{0123}^4 u_{013}^4 u_{02}^2 u_{023}^4 u_{03}^6 u_{123}^4 \
			u_{13}^4 u_{23}^2 + \\
			b_3^4 u_3^4 u_{012}^2 u_{0123}^4 u_{013}^8 u_{02}^2 u_{023}^4 u_{03}^6 u_{123}^4 u_{13}^4 \
			u_{23}^2 + b_3^4 u_3^4 u_{012}^4 u_{0123}^4 u_{013}^2 u_{02}^4 u_{023}^2 u_{03}^2 u_{123}^2 \
			u_{13}^6 u_{23}^2 - \\
			3 b_3^4 u_3^4 u_{012}^8 u_{0123}^4 u_{013}^2 u_{02}^4 u_{023}^6 u_{03}^2 u_{123}^2 u_{13}^6 \
			u_{23}^2 + 2 b_3^4 u_3^4 u_{012}^4 u_{0123}^6 u_{013}^4 u_{02}^4 u_{023}^4 u_{03}^4 u_{123}^2 \
			u_{13}^6 u_{23}^2 + \\
			b_3^2 u_3^6 u_{012}^2 u_{0123}^2 u_{013}^4 u_{02}^2 u_{023}^2 u_{03}^2 u_{123}^4 u_{13}^6 \
			u_{23}^2 - 2 b_3^2 u_3^6 u_{012}^6 u_{0123}^4 u_{013}^2 u_{02}^2 u_{023}^4 u_{03}^4 u_{123}^4 \
			u_{13}^6 u_{23}^2 + \\
			2 b_3^2 u_3^6 u_{012}^2 u_{0123}^4 u_{013}^6 u_{02}^2 u_{023}^4 u_{03}^4 u_{123}^4 u_{13}^6 \
			u_{23}^2 - b_3^2 u_3^6 u_{012}^6 u_{0123}^6 u_{013}^4 u_{02}^2 u_{023}^6 u_{03}^6 u_{123}^4 \
			u_{13}^6 u_{23}^2 - \\
			b_3^5 u_3^3 u_{012}^4 u_{0123}^7 u_{013}^3 u_{023}^5 u_{03}^3 u_{123}^3 u_{13} u_{23}^3 - 
			2 b_3^5 u_3^3 u_{012}^4 u_{0123}^5 u_{013}^5 u_{023}^3 u_{03}^5 u_{123}^3 u_{13} u_{23}^3 + \\
			3 b_3^5 u_3^3 u_{0123}^5 u_{013}^5 u_{023}^7 u_{03}^5 u_{123}^3 u_{13} u_{23}^3 + 
			b_3^7 u_3 u_{012}^2 u_{0123}^5 u_{013}^5 u_{02}^2 u_{023}^5 u_{03}^3 u_{123}^5 u_{13} u_{23}^3 -\\
			b_3^7 u_3 u_{012}^6 u_{0123}^3 u_{013}^3 u_{02}^2 u_{023}^3 u_{03}^5 u_{123}^5 u_{13} u_{23}^3 \
			- b_3^5 u_3^3 u_{012}^6 u_{0123}^5 u_{013}^3 u_{02}^2 u_{023}^5 u_{03} u_{123} u_{13}^3 u_{23}^3 - \\
			2 b_3^5 u_3^3 u_{012}^6 u_{0123}^3 u_{013}^5 u_{02}^2 u_{023}^3 u_{03}^3 u_{123} u_{13}^3 \
			u_{23}^3 + 2 b_3^5 u_3^3 u_{012}^2 u_{0123}^5 u_{013}^3 u_{02}^2 u_{023}^5 u_{03}^5 u_{123} \
			u_{13}^3 u_{23}^3 + \\
			b_3^5 u_3^3 u_{012}^2 u_{0123}^3 u_{013}^5 u_{02}^2 u_{023}^3 u_{03}^7 u_{123} u_{13}^3 \
			u_{23}^3 - 4 b_3^3 u_3^5 u_{012}^4 u_{0123}^5 u_{013}^3 u_{023}^3 u_{03}^3 u_{123}^3 u_{13}^3 \
			u_{23}^3 -\\ 3 b_3^7 u_3 u_{012}^8 u_{0123} u_{013}^3 u_{02}^4 u_{023}^3 u_{03}^3 u_{123}^3 u_{13}^3 \
			u_{23}^3 + 2 b_3^7 u_3 u_{012}^4 u_{0123}^5 u_{013}^3 u_{02}^4 u_{023}^3 u_{03}^3 u_{123}^3 \
			u_{13}^3 u_{23}^3 - \\
			2 b_3^3 u_3^5 u_{012}^4 u_{0123}^3 u_{013}^5 u_{023} u_{03}^5 u_{123}^3 u_{13}^3 u_{23}^3 + 
			b_3^7 u_3 u_{012}^4 u_{0123}^3 u_{013}^5 u_{02}^4 u_{023} u_{03}^5 u_{123}^3 u_{13}^3 u_{23}^3 \
			+\\ 9 b_3^3 u_3^5 u_{0123}^3 u_{013}^5 u_{023}^5 u_{03}^5 u_{123}^3 u_{13}^3 u_{23}^3 - 
			2 b_3^3 u_3^5 u_{012}^4 u_{0123}^7 u_{013}^5 u_{023}^5 u_{03}^5 u_{123}^3 u_{13}^3 u_{23}^3 -\\
			b_3^3 u_3^5 u_{012}^4 u_{0123}^5 u_{013}^7 u_{023}^3 u_{03}^7 u_{123}^3 u_{13}^3 u_{23}^3 + 
			2 b_3^5 u_3^3 u_{012}^2 u_{0123}^5 u_{013}^3 u_{02}^2 u_{023}^5 u_{03} u_{123}^5 u_{13}^3 \
			u_{23}^3 -\\ 2 b_3^5 u_3^3 u_{012}^6 u_{0123}^3 u_{013} u_{02}^2 u_{023}^3 u_{03}^3 u_{123}^5 \
			u_{13}^3 u_{23}^3 + 
			4 b_3^5 u_3^3 u_{012}^2 u_{0123}^3 u_{013}^5 u_{02}^2 u_{023}^3 u_{03}^3 u_{123}^5 u_{13}^3 \
			u_{23}^3 + \\b_3^5 u_3^3 u_{012}^2 u_{0123}^7 u_{013}^5 u_{02}^2 u_{023}^7 u_{03}^3 u_{123}^5 \
			u_{13}^3 u_{23}^3 - 
			b_3^5 u_3^3 u_{012}^6 u_{0123} u_{013}^3 u_{02}^2 u_{023} u_{03}^5 u_{123}^5 u_{13}^3 u_{23}^3 \
			-\\ 4 b_3^5 u_3^3 u_{012}^6 u_{0123}^5 u_{013}^3 u_{02}^2 u_{023}^5 u_{03}^5 u_{123}^5 u_{13}^3 \
			u_{23}^3 + 2 b_3^5 u_3^3 u_{012}^2 u_{0123}^5 u_{013}^7 u_{02}^2 u_{023}^5 u_{03}^5 u_{123}^5 \
			u_{13}^3 u_{23}^3 - \\
			2 b_3^5 u_3^3 u_{012}^6 u_{0123}^3 u_{013}^5 u_{02}^2 u_{023}^3 u_{03}^7 u_{123}^5 u_{13}^3 \
			u_{23}^3 - b_3^3 u_3^5 u_{012}^6 u_{0123}^3 u_{013}^3 u_{02}^2 u_{023}^3 u_{03} u_{123} u_{13}^5 \
			u_{23}^3 - \\2 b_3^3 u_3^5 u_{012}^6 u_{0123}^5 u_{013}^5 u_{02}^2 u_{023}^5 u_{03}^3 u_{123} \
			u_{13}^5 u_{23}^3 + 
			2 b_3^3 u_3^5 u_{012}^2 u_{0123}^3 u_{013}^3 u_{02}^2 u_{023}^3 u_{03}^5 u_{123} u_{13}^5 \
			u_{23}^3 + \\b_3^3 u_3^5 u_{012}^2 u_{0123}^5 u_{013}^5 u_{02}^2 u_{023}^5 u_{03}^7 u_{123} u_{13}^5 \
			u_{23}^3 + b_3^5 u_3^3 u_{012}^4 u_{0123}^5 u_{013} u_{02}^4 u_{023}^3 u_{03} u_{123}^3 u_{13}^5 \
			u_{23}^3 -\\ b_3 u_3^7 u_{012}^4 u_{0123}^3 u_{013}^3 u_{023} u_{03}^3 u_{123}^3 u_{13}^5 u_{23}^3 + 
			2 b_3^5 u_3^3 u_{012}^4 u_{0123}^3 u_{013}^3 u_{02}^4 u_{023} u_{03}^3 u_{123}^3 u_{13}^5 \
			u_{23}^3 -\\ 9 b_3^5 u_3^3 u_{012}^8 u_{0123}^3 u_{013}^3 u_{02}^4 u_{023}^5 u_{03}^3 u_{123}^3 \
			u_{13}^5 u_{23}^3 + 
			2 b_3^5 u_3^3 u_{012}^4 u_{0123}^7 u_{013}^3 u_{02}^4 u_{023}^5 u_{03}^3 u_{123}^3 u_{13}^5 \
			u_{23}^3 + \\3 b_3 u_3^7 u_{0123} u_{013}^5 u_{023}^3 u_{03}^5 u_{123}^3 u_{13}^5 u_{23}^3 - 
			2 b_3 u_3^7 u_{012}^4 u_{0123}^5 u_{013}^5 u_{023}^3 u_{03}^5 u_{123}^3 u_{13}^5 u_{23}^3 + \\
			4 b_3^5 u_3^3 u_{012}^4 u_{0123}^5 u_{013}^5 u_{02}^4 u_{023}^3 u_{03}^5 u_{123}^3 u_{13}^5 \
			u_{23}^3 + 2 b_3^3 u_3^5 u_{012}^2 u_{0123}^3 u_{013}^3 u_{02}^2 u_{023}^3 u_{03} u_{123}^5 \
			u_{13}^5 u_{23}^3 + \\
			b_3^3 u_3^5 u_{012}^2 u_{0123} u_{013}^5 u_{02}^2 u_{023} u_{03}^3 u_{123}^5 u_{13}^5 u_{23}^3 \
			- 2 b_3^3 u_3^5 u_{012}^6 u_{0123}^5 u_{013} u_{02}^2 u_{023}^5 u_{03}^3 u_{123}^5 u_{13}^5 \
			u_{23}^3 +\\ 4 b_3^3 u_3^5 u_{012}^2 u_{0123}^5 u_{013}^5 u_{02}^2 u_{023}^5 u_{03}^3 u_{123}^5 \
			u_{13}^5 u_{23}^3 - 
			4 b_3^3 u_3^5 u_{012}^6 u_{0123}^3 u_{013}^3 u_{02}^2 u_{023}^3 u_{03}^5 u_{123}^5 u_{13}^5 \
			u_{23}^3 + \\2 b_3^3 u_3^5 u_{012}^2 u_{0123}^3 u_{013}^7 u_{02}^2 u_{023}^3 u_{03}^5 u_{123}^5 \
			u_{13}^5 u_{23}^3 - 
			b_3^3 u_3^5 u_{012}^6 u_{0123}^7 u_{013}^3 u_{02}^2 u_{023}^7 u_{03}^5 u_{123}^5 u_{13}^5 \
			u_{23}^3 - \\2 b_3^3 u_3^5 u_{012}^6 u_{0123}^5 u_{013}^5 u_{02}^2 u_{023}^5 u_{03}^7 u_{123}^5 \
			u_{13}^5 u_{23}^3 + 
			2 b_3^3 u_3^5 u_{012}^4 u_{0123}^5 u_{013}^3 u_{02}^4 u_{023}^3 u_{03}^3 u_{123}^3 u_{13}^7 \
			u_{23}^3 - \\3 b_3^3 u_3^5 u_{012}^8 u_{0123}^5 u_{013}^3 u_{02}^4 u_{023}^7 u_{03}^3 u_{123}^3 \
			u_{13}^7 u_{23}^3 + 
			b_3^3 u_3^5 u_{012}^4 u_{0123}^7 u_{013}^5 u_{02}^4 u_{023}^5 u_{03}^5 u_{123}^3 u_{13}^7 \
			u_{23}^3 +\\ b_3 u_3^7 u_{012}^2 u_{0123}^3 u_{013}^5 u_{02}^2 u_{023}^3 u_{03}^3 u_{123}^5 u_{13}^7 \
			u_{23}^3 - b_3 u_3^7 u_{012}^6 u_{0123}^5 u_{013}^3 u_{02}^2 u_{023}^5 u_{03}^5 u_{123}^5 u_{13}^7 \
			u_{23}^3 -\\ b_3^6 u_3^2 u_{012}^4 u_{0123}^6 u_{013}^4 u_{023}^4 u_{03}^4 u_{123}^4 u_{23}^4 + 
			b_3^6 u_3^2 u_{0123}^6 u_{013}^4 u_{023}^8 u_{03}^4 u_{123}^4 u_{23}^4 - \\
			2 b_3^6 u_3^2 u_{012}^6 u_{0123}^4 u_{013}^4 u_{02}^2 u_{023}^4 u_{03}^2 u_{123}^2 u_{13}^2 \
			u_{23}^4 - b_3^6 u_3^2 u_{012}^6 u_{0123}^2 u_{013}^6 u_{02}^2 u_{023}^2 u_{03}^4 u_{123}^2 \
			u_{13}^2 u_{23}^4 + \\
			b_3^6 u_3^2 u_{012}^2 u_{0123}^6 u_{013}^2 u_{02}^2 u_{023}^6 u_{03}^4 u_{123}^2 u_{13}^2 \
			u_{23}^4 + 2 b_3^6 u_3^2 u_{012}^2 u_{0123}^4 u_{013}^4 u_{02}^2 u_{023}^4 u_{03}^6 u_{123}^2 \
			u_{13}^2 u_{23}^4 - \\
			2 b_3^4 u_3^4 u_{012}^4 u_{0123}^6 u_{013}^2 u_{023}^4 u_{03}^2 u_{123}^4 u_{13}^2 u_{23}^4 \
			- 4 b_3^4 u_3^4 u_{012}^4 u_{0123}^4 u_{013}^4 u_{023}^2 u_{03}^4 u_{123}^4 u_{13}^2 u_{23}^4 - \\
			b_3^8 u_{012}^8 u_{013}^4 u_{02}^4 u_{023}^2 u_{03}^4 u_{123}^4 u_{13}^2 u_{23}^4 + 
			b_3^8 u_{012}^4 u_{0123}^4 u_{013}^4 u_{02}^4 u_{023}^2 u_{03}^4 u_{123}^4 u_{13}^2 u_{23}^4 + \\
			9 b_3^4 u_3^4 u_{0123}^4 u_{013}^4 u_{023}^6 u_{03}^4 u_{123}^4 u_{13}^2 u_{23}^4 - 
			b_3^4 u_3^4 u_{012}^4 u_{0123}^8 u_{013}^4 u_{023}^6 u_{03}^4 u_{123}^4 u_{13}^2 u_{23}^4 - \\
			2 b_3^4 u_3^4 u_{012}^4 u_{0123}^6 u_{013}^6 u_{023}^4 u_{03}^6 u_{123}^4 u_{13}^2 u_{23}^4 \
			+ 2 b_3^6 u_3^2 u_{012}^2 u_{0123}^4 u_{013}^4 u_{02}^2 u_{023}^4 u_{03}^2 u_{123}^6 u_{13}^2 \
			u_{23}^4 - \\b_3^6 u_3^2 u_{012}^6 u_{0123}^2 u_{013}^2 u_{02}^2 u_{023}^2 u_{03}^4 u_{123}^6 \
			u_{13}^2 u_{23}^4 + 
			b_3^6 u_3^2 u_{012}^2 u_{0123}^6 u_{013}^6 u_{02}^2 u_{023}^6 u_{03}^4 u_{123}^6 u_{13}^2 \
			u_{23}^4 - \\2 b_3^6 u_3^2 u_{012}^6 u_{0123}^4 u_{013}^4 u_{02}^2 u_{023}^4 u_{03}^6 u_{123}^6 \
			u_{13}^2 u_{23}^4 - 
			b_3^4 u_3^4 u_{012}^6 u_{0123}^4 u_{013}^2 u_{02}^2 u_{023}^4 u_{123}^2 u_{13}^4 u_{23}^4 - \\
			2 b_3^4 u_3^4 u_{012}^6 u_{0123}^2 u_{013}^4 u_{02}^2 u_{023}^2 u_{03}^2 u_{123}^2 u_{13}^4 \
			u_{23}^4 - 2 b_3^4 u_3^4 u_{012}^6 u_{0123}^6 u_{013}^4 u_{02}^2 u_{023}^6 u_{03}^2 u_{123}^2 \
			u_{13}^4 u_{23}^4 + \\
			4 b_3^4 u_3^4 u_{012}^2 u_{0123}^4 u_{013}^2 u_{02}^2 u_{023}^4 u_{03}^4 u_{123}^2 u_{13}^4 \
			u_{23}^4 - 4 b_3^4 u_3^4 u_{012}^6 u_{0123}^4 u_{013}^6 u_{02}^2 u_{023}^4 u_{03}^4 u_{123}^2 \
			u_{13}^4 u_{23}^4 + \\
			2 b_3^4 u_3^4 u_{012}^2 u_{0123}^2 u_{013}^4 u_{02}^2 u_{023}^2 u_{03}^6 u_{123}^2 u_{13}^4 \
			u_{23}^4 + 2 b_3^4 u_3^4 u_{012}^2 u_{0123}^6 u_{013}^4 u_{02}^2 u_{023}^6 u_{03}^6 u_{123}^2 \
			u_{13}^4 u_{23}^4 + \\
			b_3^4 u_3^4 u_{012}^2 u_{0123}^4 u_{013}^6 u_{02}^2 u_{023}^4 u_{03}^8 u_{123}^2 u_{13}^4 \
			u_{23}^4 - 2 b_3^2 u_3^6 u_{012}^4 u_{0123}^4 u_{013}^2 u_{023}^2 u_{03}^2 u_{123}^4 u_{13}^4 \
			u_{23}^4 + \\2 b_3^6 u_3^2 u_{012}^4 u_{0123}^4 u_{013}^2 u_{02}^4 u_{023}^2 u_{03}^2 u_{123}^4 \
			u_{13}^4 u_{23}^4 - 
			b_3^2 u_3^6 u_{012}^4 u_{0123}^2 u_{013}^4 u_{03}^4 u_{123}^4 u_{13}^4 u_{23}^4 + \\
			b_3^6 u_3^2 u_{012}^4 u_{0123}^2 u_{013}^4 u_{02}^4 u_{03}^4 u_{123}^4 u_{13}^4 u_{23}^4 + 
			9 b_3^2 u_3^6 u_{0123}^2 u_{013}^4 u_{023}^4 u_{03}^4 u_{123}^4 u_{13}^4 u_{23}^4 -\\ 
			4 b_3^2 u_3^6 u_{012}^4 u_{0123}^6 u_{013}^4 u_{023}^4 u_{03}^4 u_{123}^4 u_{13}^4 u_{23}^4 \
			- 9 b_3^6 u_3^2 u_{012}^8 u_{0123}^2 u_{013}^4 u_{02}^4 u_{023}^4 u_{03}^4 u_{123}^4 u_{13}^4 \
			u_{23}^4 + \\4 b_3^6 u_3^2 u_{012}^4 u_{0123}^6 u_{013}^4 u_{02}^4 u_{023}^4 u_{03}^4 u_{123}^4 \
			u_{13}^4 u_{23}^4 - 
			2 b_3^2 u_3^6 u_{012}^4 u_{0123}^4 u_{013}^6 u_{023}^2 u_{03}^6 u_{123}^4 u_{13}^4 u_{23}^4 \
			+ \\2 b_3^6 u_3^2 u_{012}^4 u_{0123}^4 u_{013}^6 u_{02}^4 u_{023}^2 u_{03}^6 u_{123}^4 u_{13}^4 \
			u_{23}^4 + b_3^4 u_3^4 u_{012}^2 u_{0123}^4 u_{013}^2 u_{02}^2 u_{023}^4 u_{123}^6 u_{13}^4 \
			u_{23}^4 +\\ 2 b_3^4 u_3^4 u_{012}^2 u_{0123}^2 u_{013}^4 u_{02}^2 u_{023}^2 u_{03}^2 u_{123}^6 \
			u_{13}^4 u_{23}^4 + 
			2 b_3^4 u_3^4 u_{012}^2 u_{0123}^6 u_{013}^4 u_{02}^2 u_{023}^6 u_{03}^2 u_{123}^6 u_{13}^4 \
			u_{23}^4 -\\ 4 b_3^4 u_3^4 u_{012}^6 u_{0123}^4 u_{013}^2 u_{02}^2 u_{023}^4 u_{03}^4 u_{123}^6 \
			u_{13}^4 u_{23}^4 + 
			4 b_3^4 u_3^4 u_{012}^2 u_{0123}^4 u_{013}^6 u_{02}^2 u_{023}^4 u_{03}^4 u_{123}^6 u_{13}^4 \
			u_{23}^4 -\\ 2 b_3^4 u_3^4 u_{012}^6 u_{0123}^2 u_{013}^4 u_{02}^2 u_{023}^2 u_{03}^6 u_{123}^6 \
			u_{13}^4 u_{23}^4 - 
			2 b_3^4 u_3^4 u_{012}^6 u_{0123}^6 u_{013}^4 u_{02}^2 u_{023}^6 u_{03}^6 u_{123}^6 u_{13}^4 \
			u_{23}^4 -\\ b_3^4 u_3^4 u_{012}^6 u_{0123}^4 u_{013}^6 u_{02}^2 u_{023}^4 u_{03}^8 u_{123}^6 \
			u_{13}^4 u_{23}^4 - 
			2 b_3^2 u_3^6 u_{012}^6 u_{0123}^4 u_{013}^4 u_{02}^2 u_{023}^4 u_{03}^2 u_{123}^2 u_{13}^6 \
			u_{23}^4 + \\b_3^2 u_3^6 u_{012}^2 u_{0123}^2 u_{013}^2 u_{02}^2 u_{023}^2 u_{03}^4 u_{123}^2 \
			u_{13}^6 u_{23}^4 - 
			b_3^2 u_3^6 u_{012}^6 u_{0123}^6 u_{013}^6 u_{02}^2 u_{023}^6 u_{03}^4 u_{123}^2 u_{13}^6 \
			u_{23}^4 + \\2 b_3^2 u_3^6 u_{012}^2 u_{0123}^4 u_{013}^4 u_{02}^2 u_{023}^4 u_{03}^6 u_{123}^2 \
			u_{13}^6 u_{23}^4 + 
			2 b_3^4 u_3^4 u_{012}^4 u_{0123}^6 u_{013}^2 u_{02}^4 u_{023}^4 u_{03}^2 u_{123}^4 u_{13}^6 \
			u_{23}^4 +\\ u_3^8 u_{013}^4 u_{023}^2 u_{03}^4 u_{123}^4 u_{13}^6 u_{23}^4 - 
			u_3^8 u_{012}^4 u_{0123}^4 u_{013}^4 u_{023}^2 u_{03}^4 u_{123}^4 u_{13}^6 u_{23}^4 + \\
			4 b_3^4 u_3^4 u_{012}^4 u_{0123}^4 u_{013}^4 u_{02}^4 u_{023}^2 u_{03}^4 u_{123}^4 u_{13}^6 \
			u_{23}^4 - 9 b_3^4 u_3^4 u_{012}^8 u_{0123}^4 u_{013}^4 u_{02}^4 u_{023}^6 u_{03}^4 u_{123}^4 \
			u_{13}^6 u_{23}^4 + \\
			b_3^4 u_3^4 u_{012}^4 u_{0123}^8 u_{013}^4 u_{02}^4 u_{023}^6 u_{03}^4 u_{123}^4 u_{13}^6 \
			u_{23}^4 + 2 b_3^4 u_3^4 u_{012}^4 u_{0123}^6 u_{013}^6 u_{02}^4 u_{023}^4 u_{03}^6 u_{123}^4 \
			u_{13}^6 u_{23}^4 + \\
			2 b_3^2 u_3^6 u_{012}^2 u_{0123}^4 u_{013}^4 u_{02}^2 u_{023}^4 u_{03}^2 u_{123}^6 u_{13}^6 \
			u_{23}^4 + b_3^2 u_3^6 u_{012}^2 u_{0123}^2 u_{013}^6 u_{02}^2 u_{023}^2 u_{03}^4 u_{123}^6 \
			u_{13}^6 u_{23}^4 - \\
			b_3^2 u_3^6 u_{012}^6 u_{0123}^6 u_{013}^2 u_{02}^2 u_{023}^6 u_{03}^4 u_{123}^6 u_{13}^6 \
			u_{23}^4 - 2 b_3^2 u_3^6 u_{012}^6 u_{0123}^4 u_{013}^4 u_{02}^2 u_{023}^4 u_{03}^6 u_{123}^6 \
			u_{13}^6 u_{23}^4 + \\
			b_3^2 u_3^6 u_{012}^4 u_{0123}^6 u_{013}^4 u_{02}^4 u_{023}^4 u_{03}^4 u_{123}^4 u_{13}^8 \
			u_{23}^4 - b_3^2 u_3^6 u_{012}^8 u_{0123}^6 u_{013}^4 u_{02}^4 u_{023}^8 u_{03}^4 u_{123}^4 \
			u_{13}^8 u_{23}^4 - \\
			b_3^7 u_3 u_{012}^6 u_{0123}^3 u_{013}^5 u_{02}^2 u_{023}^3 u_{03}^3 u_{123}^3 u_{13} u_{23}^5 +
			b_3^7 u_3 u_{012}^2 u_{0123}^5 u_{013}^3 u_{02}^2 u_{023}^5 u_{03}^5 u_{123}^3 u_{13} u_{23}^5 \
			- \\2 b_3^5 u_3^3 u_{012}^4 u_{0123}^5 u_{013}^3 u_{023}^3 u_{03}^3 u_{123}^5 u_{13} u_{23}^5 + 
			3 b_3^5 u_3^3 u_{0123}^5 u_{013}^3 u_{023}^7 u_{03}^3 u_{123}^5 u_{13} u_{23}^5 - \\
			b_3^5 u_3^3 u_{012}^4 u_{0123}^7 u_{013}^5 u_{023}^5 u_{03}^5 u_{123}^5 u_{13} u_{23}^5 - 
			2 b_3^5 u_3^3 u_{012}^6 u_{0123}^3 u_{013}^3 u_{02}^2 u_{023}^3 u_{03} u_{123}^3 u_{13}^3 \
			u_{23}^5 - \\b_3^5 u_3^3 u_{012}^6 u_{0123} u_{013}^5 u_{02}^2 u_{023} u_{03}^3 u_{123}^3 u_{13}^3 \
			u_{23}^5 + 2 b_3^5 u_3^3 u_{012}^2 u_{0123}^5 u_{013} u_{02}^2 u_{023}^5 u_{03}^3 u_{123}^3 \
			u_{13}^3 u_{23}^5 - \\
			4 b_3^5 u_3^3 u_{012}^6 u_{0123}^5 u_{013}^5 u_{02}^2 u_{023}^5 u_{03}^3 u_{123}^3 u_{13}^3 \
			u_{23}^5 + 4 b_3^5 u_3^3 u_{012}^2 u_{0123}^3 u_{013}^3 u_{02}^2 u_{023}^3 u_{03}^5 u_{123}^3 \
			u_{13}^3 u_{23}^5 - \\
			2 b_3^5 u_3^3 u_{012}^6 u_{0123}^3 u_{013}^7 u_{02}^2 u_{023}^3 u_{03}^5 u_{123}^3 u_{13}^3 \
			u_{23}^5 + b_3^5 u_3^3 u_{012}^2 u_{0123}^7 u_{013}^3 u_{02}^2 u_{023}^7 u_{03}^5 u_{123}^3 \
			u_{13}^3 u_{23}^5 + \\
			2 b_3^5 u_3^3 u_{012}^2 u_{0123}^5 u_{013}^5 u_{02}^2 u_{023}^5 u_{03}^7 u_{123}^3 u_{13}^3 \
			u_{23}^5 - b_3^3 u_3^5 u_{012}^4 u_{0123}^5 u_{013} u_{023}^3 u_{03} u_{123}^5 u_{13}^3 u_{23}^5 - \\
			2 b_3^3 u_3^5 u_{012}^4 u_{0123}^3 u_{013}^3 u_{023} u_{03}^3 u_{123}^5 u_{13}^3 u_{23}^5 + 
			b_3^7 u_3 u_{012}^4 u_{0123}^3 u_{013}^3 u_{02}^4 u_{023} u_{03}^3 u_{123}^5 u_{13}^3 u_{23}^5 \
			+\\ 9 b_3^3 u_3^5 u_{0123}^3 u_{013}^3 u_{023}^5 u_{03}^3 u_{123}^5 u_{13}^3 u_{23}^5 - 
			2 b_3^3 u_3^5 u_{012}^4 u_{0123}^7 u_{013}^3 u_{023}^5 u_{03}^3 u_{123}^5 u_{13}^3 u_{23}^5 \
			- \\4 b_3^3 u_3^5 u_{012}^4 u_{0123}^5 u_{013}^5 u_{023}^3 u_{03}^5 u_{123}^5 u_{13}^3 u_{23}^5 - 
			3 b_3^7 u_3 u_{012}^8 u_{0123} u_{013}^5 u_{02}^4 u_{023}^3 u_{03}^5 u_{123}^5 u_{13}^3 \
			u_{23}^5 + \\2 b_3^7 u_3 u_{012}^4 u_{0123}^5 u_{013}^5 u_{02}^4 u_{023}^3 u_{03}^5 u_{123}^5 \
			u_{13}^3 u_{23}^5 + 
			b_3^5 u_3^3 u_{012}^2 u_{0123}^3 u_{013}^3 u_{02}^2 u_{023}^3 u_{03} u_{123}^7 u_{13}^3 \
			u_{23}^5 +\\ 2 b_3^5 u_3^3 u_{012}^2 u_{0123}^5 u_{013}^5 u_{02}^2 u_{023}^5 u_{03}^3 u_{123}^7 \
			u_{13}^3 u_{23}^5 - 
			2 b_3^5 u_3^3 u_{012}^6 u_{0123}^3 u_{013}^3 u_{02}^2 u_{023}^3 u_{03}^5 u_{123}^7 u_{13}^3 \
			u_{23}^5 - \\b_3^5 u_3^3 u_{012}^6 u_{0123}^5 u_{013}^5 u_{02}^2 u_{023}^5 u_{03}^7 u_{123}^7 \
			u_{13}^3 u_{23}^5 - 
			2 b_3^3 u_3^5 u_{012}^6 u_{0123}^5 u_{013}^3 u_{02}^2 u_{023}^5 u_{03} u_{123}^3 u_{13}^5 \
			u_{23}^5 + \\2 b_3^3 u_3^5 u_{012}^2 u_{0123}^3 u_{013} u_{02}^2 u_{023}^3 u_{03}^3 u_{123}^3 \
			u_{13}^5 u_{23}^5 - 
			4 b_3^3 u_3^5 u_{012}^6 u_{0123}^3 u_{013}^5 u_{02}^2 u_{023}^3 u_{03}^3 u_{123}^3 u_{13}^5 \
			u_{23}^5 - \\b_3^3 u_3^5 u_{012}^6 u_{0123}^7 u_{013}^5 u_{02}^2 u_{023}^7 u_{03}^3 u_{123}^3 \
			u_{13}^5 u_{23}^5 + 
			b_3^3 u_3^5 u_{012}^2 u_{0123} u_{013}^3 u_{02}^2 u_{023} u_{03}^5 u_{123}^3 u_{13}^5 u_{23}^5 \
			+\\ 4 b_3^3 u_3^5 u_{012}^2 u_{0123}^5 u_{013}^3 u_{02}^2 u_{023}^5 u_{03}^5 u_{123}^3 u_{13}^5 \
			u_{23}^5 - 2 b_3^3 u_3^5 u_{012}^6 u_{0123}^5 u_{013}^7 u_{02}^2 u_{023}^5 u_{03}^5 u_{123}^3 \
			u_{13}^5 u_{23}^5 + \\
			2 b_3^3 u_3^5 u_{012}^2 u_{0123}^3 u_{013}^5 u_{02}^2 u_{023}^3 u_{03}^7 u_{123}^3 u_{13}^5 \
			u_{23}^5 + 3 b_3 u_3^7 u_{0123} u_{013}^3 u_{023}^3 u_{03}^3 u_{123}^5 u_{13}^5 u_{23}^5 - \\
			2 b_3 u_3^7 u_{012}^4 u_{0123}^5 u_{013}^3 u_{023}^3 u_{03}^3 u_{123}^5 u_{13}^5 u_{23}^5 + 
			4 b_3^5 u_3^3 u_{012}^4 u_{0123}^5 u_{013}^3 u_{02}^4 u_{023}^3 u_{03}^3 u_{123}^5 u_{13}^5 \
			u_{23}^5 - \\b_3 u_3^7 u_{012}^4 u_{0123}^3 u_{013}^5 u_{023} u_{03}^5 u_{123}^5 u_{13}^5 u_{23}^5 + 
			2 b_3^5 u_3^3 u_{012}^4 u_{0123}^3 u_{013}^5 u_{02}^4 u_{023} u_{03}^5 u_{123}^5 u_{13}^5 \
			u_{23}^5 -\\ 9 b_3^5 u_3^3 u_{012}^8 u_{0123}^3 u_{013}^5 u_{02}^4 u_{023}^5 u_{03}^5 u_{123}^5 \
			u_{13}^5 u_{23}^5 + 
			2 b_3^5 u_3^3 u_{012}^4 u_{0123}^7 u_{013}^5 u_{02}^4 u_{023}^5 u_{03}^5 u_{123}^5 u_{13}^5 \
			u_{23}^5 +\\ b_3^5 u_3^3 u_{012}^4 u_{0123}^5 u_{013}^7 u_{02}^4 u_{023}^3 u_{03}^7 u_{123}^5 \
			u_{13}^5 u_{23}^5 + 
			b_3^3 u_3^5 u_{012}^2 u_{0123}^5 u_{013}^3 u_{02}^2 u_{023}^5 u_{03} u_{123}^7 u_{13}^5 \
			u_{23}^5 +\\ 2 b_3^3 u_3^5 u_{012}^2 u_{0123}^3 u_{013}^5 u_{02}^2 u_{023}^3 u_{03}^3 u_{123}^7 \
			u_{13}^5 u_{23}^5 - 
			2 b_3^3 u_3^5 u_{012}^6 u_{0123}^5 u_{013}^3 u_{02}^2 u_{023}^5 u_{03}^5 u_{123}^7 u_{13}^5 \
			u_{23}^5 -\\ b_3^3 u_3^5 u_{012}^6 u_{0123}^3 u_{013}^5 u_{02}^2 u_{023}^3 u_{03}^7 u_{123}^7 \
			u_{13}^5 u_{23}^5 - 
			b_3 u_3^7 u_{012}^6 u_{0123}^5 u_{013}^5 u_{02}^2 u_{023}^5 u_{03}^3 u_{123}^3 u_{13}^7 \
			u_{23}^5 + \\b_3 u_3^7 u_{012}^2 u_{0123}^3 u_{013}^3 u_{02}^2 u_{023}^3 u_{03}^5 u_{123}^3 u_{13}^7 \
			u_{23}^5 + b_3^3 u_3^5 u_{012}^4 u_{0123}^7 u_{013}^3 u_{02}^4 u_{023}^5 u_{03}^3 u_{123}^5 \
			u_{13}^7 u_{23}^5 + \\
			2 b_3^3 u_3^5 u_{012}^4 u_{0123}^5 u_{013}^5 u_{02}^4 u_{023}^3 u_{03}^5 u_{123}^5 u_{13}^7 \
			u_{23}^5 - 3 b_3^3 u_3^5 u_{012}^8 u_{0123}^5 u_{013}^5 u_{02}^4 u_{023}^7 u_{03}^5 u_{123}^5 \
			u_{13}^7 u_{23}^5 - \\
			b_3^6 u_3^2 u_{012}^6 u_{0123}^2 u_{013}^4 u_{02}^2 u_{023}^2 u_{03}^2 u_{123}^4 u_{13}^2 \
			u_{23}^6 + 2 b_3^6 u_3^2 u_{012}^2 u_{0123}^4 u_{013}^2 u_{02}^2 u_{023}^4 u_{03}^4 u_{123}^4 \
			u_{13}^2 u_{23}^6 - \\
			2 b_3^6 u_3^2 u_{012}^6 u_{0123}^4 u_{013}^6 u_{02}^2 u_{023}^4 u_{03}^4 u_{123}^4 u_{13}^2 \
			u_{23}^6 + b_3^6 u_3^2 u_{012}^2 u_{0123}^6 u_{013}^4 u_{02}^2 u_{023}^6 u_{03}^6 u_{123}^4 \
			u_{13}^2 u_{23}^6 - \\
			b_3^4 u_3^4 u_{012}^4 u_{0123}^4 u_{013}^2 u_{023}^2 u_{03}^2 u_{123}^6 u_{13}^2 u_{23}^6 + 
			3 b_3^4 u_3^4 u_{0123}^4 u_{013}^2 u_{023}^6 u_{03}^2 u_{123}^6 u_{13}^2 u_{23}^6 - \\
			2 b_3^4 u_3^4 u_{012}^4 u_{0123}^6 u_{013}^4 u_{023}^4 u_{03}^4 u_{123}^6 u_{13}^2 u_{23}^6 +
			b_3^4 u_3^4 u_{012}^2 u_{0123}^4 u_{02}^2 u_{023}^4 u_{03}^2 u_{123}^4 u_{13}^4 u_{23}^6 - \\
			4 b_3^4 u_3^4 u_{012}^6 u_{0123}^4 u_{013}^4 u_{02}^2 u_{023}^4 u_{03}^2 u_{123}^4 u_{13}^4 \
			u_{23}^6 + 2 b_3^4 u_3^4 u_{012}^2 u_{0123}^2 u_{013}^2 u_{02}^2 u_{023}^2 u_{03}^4 u_{123}^4 \
			u_{13}^4 u_{23}^6 - \\
			2 b_3^4 u_3^4 u_{012}^6 u_{0123}^2 u_{013}^6 u_{02}^2 u_{023}^2 u_{03}^4 u_{123}^4 u_{13}^4 \
			u_{23}^6 + 2 b_3^4 u_3^4 u_{012}^2 u_{0123}^6 u_{013}^2 u_{02}^2 u_{023}^6 u_{03}^4 u_{123}^4 \
			u_{13}^4 u_{23}^6 - \\
			2 b_3^4 u_3^4 u_{012}^6 u_{0123}^6 u_{013}^6 u_{02}^2 u_{023}^6 u_{03}^4 u_{123}^4 u_{13}^4 \
			u_{23}^6 + 4 b_3^4 u_3^4 u_{012}^2 u_{0123}^4 u_{013}^4 u_{02}^2 u_{023}^4 u_{03}^6 u_{123}^4 \
			u_{13}^4 u_{23}^6 - \\
			b_3^4 u_3^4 u_{012}^6 u_{0123}^4 u_{013}^8 u_{02}^2 u_{023}^4 u_{03}^6 u_{123}^4 u_{13}^4 \
			u_{23}^6 + 3 b_3^2 u_3^6 u_{0123}^2 u_{013}^2 u_{023}^4 u_{03}^2 u_{123}^6 u_{13}^4 u_{23}^6 - \\
			b_3^2 u_3^6 u_{012}^4 u_{0123}^6 u_{013}^2 u_{023}^4 u_{03}^2 u_{123}^6 u_{13}^4 u_{23}^6 - 
			2 b_3^2 u_3^6 u_{012}^4 u_{0123}^4 u_{013}^4 u_{023}^2 u_{03}^4 u_{123}^6 u_{13}^4 u_{23}^6 \
			+ \\2 b_3^6 u_3^2 u_{012}^4 u_{0123}^4 u_{013}^4 u_{02}^4 u_{023}^2 u_{03}^4 u_{123}^6 u_{13}^4 \
			u_{23}^6 - 3 b_3^6 u_3^2 u_{012}^8 u_{0123}^2 u_{013}^6 u_{02}^4 u_{023}^4 u_{03}^6 u_{123}^6 \
			u_{13}^4 u_{23}^6 + \\
			b_3^6 u_3^2 u_{012}^4 u_{0123}^6 u_{013}^6 u_{02}^4 u_{023}^4 u_{03}^6 u_{123}^6 u_{13}^4 \
			u_{23}^6 + b_3^4 u_3^4 u_{012}^2 u_{0123}^4 u_{013}^4 u_{02}^2 u_{023}^4 u_{03}^2 u_{123}^8 \
			u_{13}^4 u_{23}^6 - \\
			b_3^4 u_3^4 u_{012}^6 u_{0123}^4 u_{013}^4 u_{02}^2 u_{023}^4 u_{03}^6 u_{123}^8 u_{13}^4 \
			u_{23}^6 - b_3^2 u_3^6 u_{012}^6 u_{0123}^6 u_{013}^4 u_{02}^2 u_{023}^6 u_{03}^2 u_{123}^4 \
			u_{13}^6 u_{23}^6 + \\
			2 b_3^2 u_3^6 u_{012}^2 u_{0123}^4 u_{013}^2 u_{02}^2 u_{023}^4 u_{03}^4 u_{123}^4 u_{13}^6 \
			u_{23}^6 - 2 b_3^2 u_3^6 u_{012}^6 u_{0123}^4 u_{013}^6 u_{02}^2 u_{023}^4 u_{03}^4 u_{123}^4 \
			u_{13}^6 u_{23}^6 + \\
			b_3^2 u_3^6 u_{012}^2 u_{0123}^2 u_{013}^4 u_{02}^2 u_{023}^2 u_{03}^6 u_{123}^4 u_{13}^6 \
			u_{23}^6 + 2 b_3^4 u_3^4 u_{012}^4 u_{0123}^6 u_{013}^4 u_{02}^4 u_{023}^4 u_{03}^4 u_{123}^6 \
			u_{13}^6 u_{23}^6 + \\
			b_3^4 u_3^4 u_{012}^4 u_{0123}^4 u_{013}^6 u_{02}^4 u_{023}^2 u_{03}^6 u_{123}^6 u_{13}^6 \
			u_{23}^6 - 3 b_3^4 u_3^4 u_{012}^8 u_{0123}^4 u_{013}^6 u_{02}^4 u_{023}^6 u_{03}^6 u_{123}^6 \
			u_{13}^6 u_{23}^6 + \\
			b_3^5 u_3^3 u_{012}^2 u_{0123}^3 u_{013} u_{02}^2 u_{023}^3 u_{03}^3 u_{123}^5 u_{13}^3 \
			u_{23}^7 - 2 b_3^5 u_3^3 u_{012}^6 u_{0123}^3 u_{013}^5 u_{02}^2 u_{023}^3 u_{03}^3 u_{123}^5 \
			u_{13}^3 u_{23}^7 + \\
			2 b_3^5 u_3^3 u_{012}^2 u_{0123}^5 u_{013}^3 u_{02}^2 u_{023}^5 u_{03}^5 u_{123}^5 u_{13}^3 \
			u_{23}^7 - b_3^5 u_3^3 u_{012}^6 u_{0123}^5 u_{013}^7 u_{02}^2 u_{023}^5 u_{03}^5 u_{123}^5 \
			u_{13}^3 u_{23}^7 + \\b_3^3 u_3^5 u_{0123}^3 u_{013} u_{023}^5 u_{03} u_{123}^7 u_{13}^3 u_{23}^7 - 
			b_3^3 u_3^5 u_{012}^4 u_{0123}^5 u_{013}^3 u_{023}^3 u_{03}^3 u_{123}^7 u_{13}^3 u_{23}^7 + \\
			b_3^3 u_3^5 u_{012}^2 u_{0123}^5 u_{013} u_{02}^2 u_{023}^5 u_{03}^3 u_{123}^5 u_{13}^5 \
			u_{23}^7 - 2 b_3^3 u_3^5 u_{012}^6 u_{0123}^5 u_{013}^5 u_{02}^2 u_{023}^5 u_{03}^3 u_{123}^5 \
			u_{13}^5 u_{23}^7 + \\
			2 b_3^3 u_3^5 u_{012}^2 u_{0123}^3 u_{013}^3 u_{02}^2 u_{023}^3 u_{03}^5 u_{123}^5 u_{13}^5 \
			u_{23}^7 - b_3^3 u_3^5 u_{012}^6 u_{0123}^3 u_{013}^7 u_{02}^2 u_{023}^3 u_{03}^5 u_{123}^5 \
			u_{13}^5 u_{23}^7 + \\
			b_3^5 u_3^3 u_{012}^4 u_{0123}^5 u_{013}^5 u_{02}^4 u_{023}^3 u_{03}^5 u_{123}^7 u_{13}^5 \
			u_{23}^7 - b_3^5 u_3^3 u_{012}^8 u_{0123}^3 u_{013}^7 u_{02}^4 u_{023}^5 u_{03}^7 u_{123}^7 \
			u_{13}^5 u_{23}^7 + \\
			b_3^4 u_3^4 u_{012}^2 u_{0123}^4 u_{013}^2 u_{02}^2 u_{023}^4 u_{03}^4 u_{123}^6 u_{13}^4 \
			u_{23}^8 - b_3^4 u_3^4 u_{012}^6 u_{0123}^4 u_{013}^6 u_{02}^2 u_{023}^4 u_{03}^4 u_{123}^6 \
			u_{13}^4 u_{23}^8) = 0.
		\end{multline} 
		
		The equation (\ref{for_s02}) can be written as quadratic equation  (\ref{quadratic_equation}) with respect to  
		$ x=s_{02} $  with coefficients $A$ (\ref{coefficient_A}) , $B$ (\ref{coefficient_B})  , $C$ (\ref{coefficient_C})  .

		%%%%%%%%%%%%%%%%%%%%%%%%%%%%%%%%%%%%%%%%%%%%%%%%%%%%%%%%%%%%%%%%% 

		%\begin{acknowledgements}
		%\end{acknowledgements}


		

	\end{document}